\documentclass[fleqn,usenatbib]{mnras}
\usepackage{newtxtext,newtxmath}
\usepackage[T1]{fontenc}

\DeclareRobustCommand{\VAN}[3]{#2}
\let\VANthebibliography\thebibliography
\def\thebibliography{\DeclareRobustCommand{\VAN}[3]{##3}\VANthebibliography}

\usepackage{graphicx}	% Including figure files
\usepackage{amsmath}	% Advanced maths commands
\usepackage{amssymb}	% Extra maths symbols
\usepackage{pdflscape}    % PDFlscape

\defcitealias{2015ApJS..218....6B}{B15}
\defcitealias{2018MNRAS.480.3562D}{D18}
\title[The AGN contribution of interacting galaxies]{The AGN contribution to the UV-FIR luminosities of interacting galaxies and its role in identifying the Main Sequence}

\author[Andr\'{e}s F. Ramos P.\ et al.]{Andr\'{e}s F. Ramos P.,$^{1,2,3}$\thanks{ramos@astro.rug.nl (AFRP)} M.\,L.\,N.\,Ashby$^{3}$, Howard  A.\,Smith$^{3}$,  Juan R.\,Mart\'{i}nez-Galarza$^{3}$, \newauthor{Aliza G.\,Beverage$^{3,4}$,  Jeremy Dietrich$^{3,5}$, Mario-A.\,Higuera-G.$^{6}$,} and Aaron\,S.\,Weiner$^{3}$
  \\
  $^{1}$Kapteyn Astronomical Institute, University of Groningen, Landleven 12, 9747 AD Groningen, The Netherlands\\
  $^{2}$SRON Netherlands Institute for Space Research, Landleven 12, 9747 AD Groningen, The Netherlands\\
  $^{3}$Center for Astrophysics $|$ Harvard \& Smithsonian, 60 Garden Street, Cambridge, MA 02138, USA\\
  $^{4}$Astronomy Department, University of California, Berkeley, CA 94720, USA\\
  $^{5}$Department of Astronomy and Steward Observatory, University of Arizona, 933 N Cherry Ave, Tucson, AZ 85719, USA\\
  $^{6}$Observatorio Astron\'{o}mico Nacional, Universidad Nacional de Colombia, Carrera 45 No 26-85, Bogot\'{a} D.C., Colombia\\
}

\date{Accepted 2020 September 11. Received 2020 September 4; in original form 2020 April 17}

\pubyear{2020}

\begin{document}
\label{firstpage}
\pagerange{\pageref{firstpage}--\pageref{lastpage}}
\maketitle

\begin{abstract}

Emission from active galactic nuclei (AGNs) is known to play an important role in the evolution of many galaxies including luminous and ultraluminous systems (U/LIRGs), as well as merging systems. However, the extent, duration, and exact effects of its influence are still imperfectly understood. To assess the impact of AGNs on interacting systems, we present a Spectral Energy Distribution (SED) analysis of a sample of 189 nearby galaxies. We gather and systematically re-reduce archival broad-band imaging mosaics from the ultraviolet to the far-infrared using data from  {\sl GALEX}, SDSS, 2MASS, {\sl IRAS}, {\sl WISE}, {\sl Spitzer} and {\sl Herschel}. We use spectroscopy from {\sl Spitzer}/IRS to obtain fluxes from fine-structure lines that trace star formation and AGN activity. Utilizing the SED modelling and fitting tool \textsc{CIGALE}, we derive the physical conditions of the ISM, both in star-forming regions and in nuclear regions dominated by the AGN in these galaxies. We investigate how the star formation rates (SFRs) and the fractional AGN contributions ($f_{\rm{AGN}}$) depend on stellar mass, galaxy type, and merger stage. We find that luminous galaxies more massive than about $10^{10} \rm{M}_{*}$ are likely to deviate significantly from the conventional galaxy main-sequence relation. Interestingly, infrared AGN luminosity and stellar mass in this set of objects are much tighter than SFR and stellar mass. We find that buried AGNs may occupy a locus between bright starbursts and pure AGNs in the $f_{\rm{AGN}}$-[\ion{Ne}{v}]/[\ion{Ne}{ii}] plane. We identify a modest correlation between $f_{\rm{AGN}}$ and mergers in their later stages.

\end{abstract}


%%%%%%%%%%%%%%%%%%%%%%%%%%%%%%%%%%%%%%%%%%%%%%%%%%

%%%%%%%%%%%%%%%%% BODY OF PAPER %%%%%%%%%%%%%%%%%%

\section{Introduction}

Over the past decade, a significant body of evidence has accumulated that supports the existence of a so-called main sequence (MS) of star-forming galaxies \citep[e.g.][]{2011A&A...533A.119E,2014ApJS..214...15S}, a tight correlation between galaxy stellar mass and the star formation rate (SFR). This scaling relation is claim to be independent of redshift and luminosity \citep{2011A&A...533A.119E}, but its normalisation does evolve with redshift \citep{2014ApJS..214...15S}. Outliers above the MS are often interpreted as merger-driven starbursts with enhanced SFRs \citep{2015ApJ...801L..29R,2016ApJ...817...76M,2019A&A...631A..51P}. The relatively tight correlation suggests that the bulk of the stars in star-forming galaxies form via secular processes rather than in violent events, such as mergers \citep[and references therein]{2015A&A...576A..10C}. However, this correlation depends in part on the assumptions used to calculate SFRs, star formation histories (SFHs), halo properties, and the degree to which galaxy interactions enhance star formation \citep[e.g.][]{2014MNRAS.445.1598H,2019MNRAS.484..915M}. 

Interacting systems are therefore crucial to our understanding of galaxy assembly over cosmic time, and of the mechanisms that shape the observed scaling relations.  In the local Universe, the most luminous infrared galaxies are almost exclusively systems undergoing significant mergers \citep{2013ApJS..206....1S}.  In these systems, star formation is significantly enhanced by the funnelling of gas and dust into the nuclear region, and the thermal emission from obscured star-forming regions outshines the UV and optical radiation from massive young stars. Systems with luminosities greater than $10^{11}$ L$_{\sun}$ (so-called Luminous InfraRed Galaxies, or LIRGs) are typically found in interacting systems, which results in a strong correlation between enhanced SFR and galaxy interaction \citep{1996ARA&A..34..749S,2013ApJ...778...10S}.  However, this simple description does not capture the full range of observed behaviour. For example, \citet{2013ApJ...768...90L} found no correlation between specific star formation rate (sSFR) and galaxy mergers \citep[see also][]{2018ApJ...868...46S}.

Nuclear starbursts may exist in galaxies that are not undergoing a merger, with about 20\% of all spiral galaxies displaying starburst activity in nuclear rings \citep{2012A&A...543A..61B}. In many of these systems, the active galactic nucleus (AGN) contribution to the luminosity from activity around the supermassive black hole appears to be negligible. These \emph{pure} starbursts are the opposite extreme of systems that are almost entirely dominated by the infrared emission from a dusty torus surrounding an AGN, such as Seyfert galaxies and more distant quasars. To put those two extremes in context, a thorough understanding of the energetics of systems with intermediate AGN contributions is needed.  Although star formation dominates the bolometric luminosity of nearby systems during most of the merger, during the later stages an AGN is thought to become active \citep[and references therein]{1996ARA&A..34..749S,2015ApJS..218....6B}. Presumably, AGNs are fed by the same infalling material that feeds star formation, and the mid-infrared thermal emission from the dusty torus around AGNs can be comparable to that of the dusty star-forming regions \citep{1998ApJ...498..579G}.  

There exists strong theoretical evidence from simulations of mergers \citep{2014ApJ...785...39L,2014MNRAS.445.1598H,2018MNRAS.480.3562D} that AGNs dominate bolometric luminosity during coalescence, and are responsible for quenching star formation in the post-coalescence stages \citep{2011ApJ...740...99D}. This process underlies their transition from the star-forming ``blue cloud'', through the so-called ``green valley'', and onto the passively evolving ``red sequence'' \citep{2015A&A...576A..10C}. There is widespread support of this evolutionary path moving from star-forming galaxies to AGN-dominated galaxies \citep[see][and references therein]{2002A&A...393..821S,2009ApJS..182..628V,2010ApJ...709.1257T,2011ApJS..195...17W}, and it is also supported by simulations showing that AGN activity is strongly correlated with the merger stage \citep{2006ApJS..163....1H}.  Merging galaxies at different interaction stages, ranging from first encounter to post-coalescence, are a natural choice to study 
AGN evolution and star formation of composite galaxies that combine both starburst and AGN processes.

Uncertainties regarding the energy budget in composite starburst-AGN systems, and about 
how the two energy generation processes impact one another and evolve, are among the most 
pressing open questions in astrophysics.  For example, buried AGN have been discovered in systems previously catalogued as pure starbursts \citep[e.g.][]{2009ASPC..408...72H,2011ApJ...740...99D}, and physical models have been proposed  to describe the interplay between the two \citep{2016MNRAS.463.1291I}.  Discriminating between the two processes based on spectral energy distribution (SED) studies is relatively straightforward and reliable when just one dominates the emission, especially at mid-IR wavelengths. Unfortunately, disentangling them becomes much more difficult when their IR luminosities are comparable \citep{2008ApJ...678..686A}. Optical and infrared spectroscopy can potentially separate the two if they cover specific fine-structure lines which are
prominent in the vicinity of AGNs and weak or non-existing in star forming regions.  The best-known example is the BPT diagram \citep{1998ApJ...498..579G,2006MNRAS.366..767F} which separates AGNs from starbursts according to their [O III] $\lambda$5007/H${\beta}$ and [N II] $\lambda$6584/H${\alpha}$ line intensity ratios, among others. 
The BPT diagrams are not always reliable, however, because high dust opacities toward AGNs can significantly attenuate emission lines at optical wavelengths. For this reason, the absolute strengths of specific mid-infrared emission lines have also been used to estimate AGN contributions \citep[e.g.][]{1998ApJ...498..579G}. Others have used the silicate attenuation in the SED, or other SED features \citep{2008ApJS..176..438G,2015A&A...576A..10C}.  But  these techniques, however useful for signaling the presence of AGNs, aren't capable of straightforwardly disentangling the 
relative importance of star formation and AGNs in composite systems.

To understand the physical mechanisms underlying scaling relations such as the MS, it is of crucial importance to account for the AGN contribution to the total luminosity of merging systems and estimate the SFRs and sSFRs at different interaction stages.  The picture at present is somewhat confused.  For example, \citet{2013ApJ...768...90L} and \citet{2018ApJ...868...46S} find no significant change in sSFR with interaction stage, but \citet{2014ApJ...785...39L} do find that sSFR increases during the relatively short times around nuclear coalescence because the SFR increases but the total mass of stars do not change. 
Furthermore, combining simulations and multi-wavelength observations, \citet{2016ApJ...817...76M} 
find that the SEDs of interacting galaxies do change with interaction stage, due to changes in stellar 
mass and SFR, and that these changes affect the location of galaxies within the MS.

Using SED modeling, \citet{2015A&A...576A..10C} showed that the AGN emission could modify 
the MS slope. Overestimations of the SFR due to the presence of a buried AGN are plausible especially at later stages, and the AGN emission can contribute to the observed MS scatter. \citet{2015A&A...576A..10C} verified that these effects can be reduced through broadband SED fitting methods such as \textsc{CIGALE} \citep{2005MNRAS.360.1413B,2009A&A...507.1793N,2011ApJ...740...22S} by taking into account the continuum emission from the AGN to obtain a better interpretation of the star-forming galaxies.

In this work, we apply those SED modeling techniques to four galaxy samples, estimate 
the fractional contributions of AGNs to their output, and elucidate how that depends on interaction stage.  
Our approach includes photometry from the UV to the far-infrared to account for multiple emission 
processes that blend with the AGN emission: UV emission from young stars, optical and near-infrared  stellar photospheric emission, mid-infrared emission from warm dust heated by star formation and evolved AGB stars, and cold dust emission.  
We incorporate photometry from dozens of instruments and surveys.

This paper is organized as follows.  In Sec.~\ref{Sec:Obs}, we present the sample selection and in Sec.~\ref{sec:SEDconstruction} describe the data reduction. Section~\ref{sec:analysis} describes how we use \textsc{CIGALE} to analyse our photometric data and the MIR emission lines.  We present the derived galaxy parameters in Sec.~\ref{sec:results} and discuss their implications in Sec.~\ref{sec:discussion}.  We present our conclusions in Sec.~\ref{sec:conclusions}.  Our photometry and spectroscopy, as well as the derived parameters for all the galaxies, are presented in the Appendix (available online). 
 Throughout this paper we adopt H$_{0}$ = 67.7 km s$^{-1}$ Mpc$^{-1}$ \citep{2016A&A...594A..13P}.
 
\section{The Four Study Samples}
\label{Sec:Obs}

AGN activity ranges from nonexistent to dominant in any particular galaxy.  
During a galaxy merger, AGN activity can increase over time, so that immediately after coalescence, it is -- 
at least briefly -- the dominant contributor to the luminosity \citep{2010MNRAS.407.1701N,2018MNRAS.478.3056B}. Star formation activity is also influenced by mergers, reaching high star-formation intensity in many well-known cases \citep[among others]{2009ApJS..182..628V,2013ApJS..206....1S}.  But not all AGNs arise in mergers, and not all starburst galaxies host 
detectable AGNs.  Here our approach is to address this ambiguity in a statistical sense by comparing
samples of galaxies selected in different ways.  Specifically, we attempt to understand how galaxy 
interactions influence AGN activity by analyzing systems that span wide ranges of 
1) interaction stage, from isolated galaxies to coalescing systems, and 
2) activity, from AGN-dominated to star-formation-dominated.  

We analyze four galaxy samples in the present work. First, we consider a sample of nearby 
systems selected to span a wide range of interaction stages from isolated systems to strongly 
interacting systems, the Spitzer Interacting Galaxies Sample \citep[SIGS,][hereafter \citetalias{2015ApJS..218....6B}]{2015ApJS..218....6B}. Our work in 100 SIGS galaxies (see Sec.\,\ref{ssec:SIGS}) builds on \citet{2013ApJ...768...90L,2014ApJ...785...39L} and \citetalias{2015ApJS..218....6B}, but is based on a more complete sample, and includes spectroscopic diagnostics.  
 The second sample is selected on the basis of {\sl Spitzer}/IRS emission line 
 ratios to be dominated by star formation (the SB sample, 21 galaxies; Sec.\,\ref{ssec:SB}).  
 The third sample is comprised of 29 AGN-dominated galaxies drawn broadly from
 the literature (Sec.\,\ref{ssec:AGN}).  
 Finally, the fourth sample is a set of 49 late-stage merging systems chosen  
 to be in or approaching final coalescence (the Late-Stage Merger or LSM sample; Sec.\,\ref{ssec:LSM}).  We include the LSM galaxies specifically to 
 address a gap in SIGS, which lacks late-stage mergers.   
  
 Thus our work includes not only systems with {\sl a priori} known dominant activity (AGN or star
 formation) selected
 without regard to interaction stage, but also systems with {\sl a priori} known interaction stage
 selected without regard to activity. We add that none of the galaxies in our four samples are radio-loud based on the identification criteria of \citet{2001ApJ...554..803Y} that $\text{L}_{\text{1.4GHz}} \geq 10^{25}$ W Hz$^{-1}$. A summary of the four samples is presented in Table~\ref{tab:4Samples} and are described in detail below.

\begin{table*}
\centering
\caption{Basic data for the four study samples.}
\label{tab:4Samples}
\begin{tabular}{cccl}
\hline
Sample & \# Galaxies& References & Description\\
\hline
SIGS&100& 1,2,3 & Nearby interacting galaxies presented by \citet{2015ApJS..218....6B}.\\
SB&21& 4,5 & Galaxies dominated by star-formation. \\
AGN&29& 5,6,7,8,9,10,11,12,13,14 & Galaxies dominated by AGN.\\
LSM&49$^{\rm a}$& 15,16,17 & Galaxies close to coalescence with a numerical interaction strength of 4 and 5 \citep{2002ApJS..143...47D}. \\
\hline
Total&199$^{\rm b}$&&All galaxies.\\
\hline
\end{tabular}
\begin{flushleft}
\textbf{Note:}  Sample, number of galaxies, references and brief description for the four samples used in work. \\
$^{\rm a}$ We present in this work 38 of the galaxies as 11 of the LSM sample galaxies are presented by \citet{2018MNRAS.480.3562D}. We re-introduce physical parameters of NGC\,2623 as part of the SB sample. \\
$^{\rm b}$ We found reliable SEDs in 189 galaxies (see Sect.~\ref{sec:results}), including the 11 galaxies presented by \citet{2018MNRAS.480.3562D}.\\
References: (1) \citet{1985AJ.....90..708K}, (2) \citet{2013ApJ...768...90L,2014ApJ...785...39L}, (3) \citet{2015ApJS..218....6B}, (4) \citet{2006ApJ...653.1129B}, (5) \citet{2013RMxAA..49..301H}, (6) \citet{2013ApJS..206....1S}, (7) \citet{2009ApJ...690.1105K}, (8) \citet{2010ApJ...709.1257T}, (9) \citet{2010ApJ...716.1151W}, (10) \citet{2010ApJ...725.2270P}, (11) \citet{2011ApJS..195...17W}, (12) \citet{2011ApJ...740...94D}, (13) \citet{2011ApJS..195...17W}, (14) \citet{2012ApJ...747...95G}, (15) \citet{2014MNRAS.442.2739W}, (16) \citet{2008MNRAS.389.1179L,2011MNRAS.410..166L} and (17) \citet{2018MNRAS.480.3562D}.
\end{flushleft}
\end{table*}

\subsection{The Spitzer Interacting Galaxies Sample}
\label{ssec:SIGS}

Our first sample is drawn from SIGS (\citetalias{2015ApJS..218....6B}, 
see Table~\ref{tab:SIGSsample}). The SIGS galaxies are relatively bright, nearby systems compiled by
\citet{1985AJ.....90..708K} in a manner designed to construct a sample free from morphological 
bias.  Specifically, \citet{1985AJ.....90..708K} identified systems containing a spiral galaxy with a 
companion seen in close projection, subject to area and magnitude restrictions. 
 These systems comprise the so-called ``Complete sample''.  To augment the Complete sample with 
 more strongly interacting systems, \citet{1985AJ.....90..708K} also compiled a sample of 
 close pairs with
 pronounced morphological signs of interaction (i.e., tidal tails and asymmetries).  This second sample 
 is known as the ``Arp sample."   The basic properties of all SIGS galaxies are given in 
 Table~\ref{tab:SIGSsample}, in which the Complete and Arp galaxies are indicated with C and A, 
 respectively.   We adopted the distances 
 given in \citetalias{2015ApJS..218....6B} for all SIGS galaxies.

The SIGS galaxies' merger stages were classified by \citetalias{2015ApJS..218....6B}, who assigned 
a numerical interaction strength to each system following \citet{2002ApJS..143...47D}.  
The classification is based on the degree of morphological disturbance, as follows.  
\begin{itemize}
    \item Stage 1 galaxies are 
isolated systems without discernible companions and are therefore, by construction, not present 
in SIGS.
\item Stage 2 galaxies are weakly interacting systems, inferred on the basis of their very mild
 or absent morphological distortions.
 \item Stage 3 galaxies are moderately interacting, have apparent 
 tidal features, and display moderate morphological distortions.
 \item Stage 4 galaxies are strongly 
 interacting, with prominent tidal features, but have two separate nuclei that can still be resolved. 
 \item Stage 5 mergers are at the point of coalescence or are merger remnants, and have only 
 one apparent nucleus (the progenitor nuclei cannot be distinguished). 
\end{itemize}
 As described in \citetalias{2015ApJS..218....6B}, the stage of each system was put to the vote 
 among the authors of that paper, using Digital Sky Survey (DSS) images, and the stage receiving 
 the most votes for each system was assigned.  
 The merger stages classified by \citetalias{2015ApJS..218....6B} are noted in column ``Interaction Stage" of Table~\ref{tab:SIGSsample}.
Ultimately, the original SIGS sample was found to consist of 35 Stage 2 galaxies, 34 Stage 3 galaxies, 33 Stage 4 galaxies, and just 1 Stage 5 galaxy.  The SIGS objects treated here are predominantly early-to-intermediate mergers, with 
 just a few late-stage mergers.  Thus SIGS is most useful as a means of quantifying AGN activity 
 in mergers {\sl before coalescence}. 

SIGS groups 40 and 41 (galaxy pairs NGC\,5544/NGC\,5545 and NGC\,5614/NGC\,5615, respectively) 
overlap too closely to be reliably photometered separately in the {\sl Herschel}/PACS and SPIRE bands.  
We therefore photometered and subsequently modeled these systems as if they were single objects.
NGC\,5846 and NGC\,5846A from SIGS group 42 were similarly entangled, and we treated them the 
same way, although we photometered and modeled the other group 42 galaxy, NGC\,5850, separately.
Thus the apertures given in Table~\ref{tab:SIGSsample} for NGC\,5544, NGC\,5614, and NGC\,5846 
encompass merging pairs instead of individual galaxies.  Taking these considerations into account, 
the SIGS sample is effectively comprised of 100 galaxies.

\subsection{The Starburst Sample}
\label{ssec:SB}

Our second study sample consists of galaxies dominated by star formation. This sample, 
which we refer throughout this work as the SB sample, consists of 21 relatively bright, nearby 
galaxies known from existing high-quality {\sl Spitzer}/IRS \citep{2004ApJS..154...18H} spectra 
taken in Short-High (SH) mode to be dominated by star formation.  This requirement for IRS spectra
was imposed to facilitate interpretation of the energetics and support the modeling effort, as diagnostic
lines of the energetics (e.g. [\ion{Ne}{v}] or [\ion{Ne}{ii}]) fall in the SH bandpass.

The SB sample is a heterogeneous group comprised of two subsamples.  First, it includes 
16 {\sl Spitzer}-selected ``classical'' starbursts galaxies from \citet{2006ApJ...653.1129B}, selected from its enhanced nuclear star-formation.   To these objects 
we added a selection of bright well-known starburst galaxies also having 
SH IRS spectra, some of them also form \citet{2006ApJ...653.1129B}, including NGC\,23, NGC\,253, NGC\,660, NGC\,1797, NGC\,3256, 
NGC\,4088, and NGC\,4945.  A few of the galaxies do have weak AGN signatures as, for example, NGC\,253 is known to host a weak AGN \citep{2010ApJ...716.1166M,2013RMxAA..49..301H}. 
In addition, the following systems are reported to be undergoing interactions: NGC\,660, NGC\,1222, NGC\,1614, NGC\,2623 (see Sect.~\ref{ssec:LSM}), NGC\,4194, NGC\,4676, and NGC\,7252. By using this heterogeneous SB sample, we can compare the other samples and check evolutionary connections between them, from the different levels of intensity of star-formation and AGN \citep{2002A&A...393..821S,2009ApJS..182..628V,2010ApJ...709.1257T,2011ApJS..195...17W}.  

For the SB sample (as well as for the AGN and LSM samples described in detail below), we adopted the redshifts given in NED. 

\subsection{The AGN Sample}
\label{ssec:AGN}

Our third sample consisted of 29 strongly AGN-dominated galaxies.  We created our AGN sample by selecting galaxies with both strong neon emission lines indicative of high ionising flux (i.e., integrated line intensity ratios [\ion{Ne}{v}]/[\ion{Ne}{ii}] > 0.6; see Sec.~\ref{ssec:neon}), and available archival {\sl Herschel}/PACS and/or SPIRE photometry, as described below.

Our AGN sample includes three galaxies from the Great Observatories All-Sky LIRG Survey \citep[GOALS,][a collection of Ultra-Luminous Infrared Galaxies (ULIRGs) with available {\sl Spitzer}/IRS spectra]{2013ApJS..206....1S}, that meet our selection criteria: NGC\,1068, NGC\,7674, and MCG-03-34-63.  We also include NGC\,4151, a composite AGN/starburst galaxy \citep{2013RMxAA..49..301H} in which the AGN is the dominant contributor.  

To these we added galaxies from a batch SIMBAD query \citep{2000A&AS..143....9W} for suitable targets.  Specifically, we retrieved the brightest 20000
galaxies classified by SIMBAD as nearby ($cz<=29999$\,km\,s$^{-1}$) and as AGNs, which also had 
available photometry from {\sl Herschel}/PACS and/or SPIRE.  Of the 20000 galaxies satisfying the
proximity, classification, and data availability constraints, we then searched for suitable neon line
ratios.  We required detections of both [\ion{Ne}{v}] and [\ion{Ne}{ii}], and set a lower limit 
on the measured ratio [\ion{Ne}{v}]/[\ion{Ne}{ii}] > 0.6.   

Estimates of the neon line ratio [\ion{Ne}{v}]/[\ion{Ne}{ii}] for some galaxies appear in different works. 
 In total, we obtained 54 measurements for 26 different galaxies from 
\citet{2009ApJ...690.1105K}, \citet{2010ApJ...709.1257T}, \citet{2010ApJ...716.1151W},
\citet{2010ApJ...725.2270P}, \citet{2011ApJS..195...17W}, \citet{2011ApJ...740...94D}, 
\citet{2011ApJS..195...17W}, and \citet{2012ApJ...747...95G}.  The 26 objects satisfying all 
our selection criteria are classified primarily as Seyferts, including some with hidden broad-line 
regions.   For example, \citet{2008ApJ...676..836T} classify MCG-03-34-63 as a non-Seyfert galaxy, 
but \citet{2010ApJ...709.1257T} and \citet{2010ApJ...716.1151W} discuss a hidden broad-line 
region in this galaxy.  In some cases the estimated line ratios were discrepant.  When multiple measurements were available, we used the most recent, to make use of the best available 
calibration and pipeline for the data in question.

In summary, our sample of 29 AGN-dominated galaxies consists of one object drawn from \citet{2013RMxAA..49..301H} (NGC\,4151), one object from GOALS (MCG-03-34-064), two objects appearing in both GOALS and \citet{2013RMxAA..49..301H}, and 25 objects drawn from our SIMBAD search. The Fundamental properties of the AGN sample galaxies are given in Table~\ref{tab:AGN}.

\subsection{The Late-Stage Merger Sample}
\label{ssec:LSM}

The LSM sample is an extension of SIGS emphasizing mergers whose morphology is consistent with the system being close to coalescence.  Although SIGS 
was designed to span the full range of galaxy interaction parameters by selecting strictly 
on the basis of interaction probability rather than morphology, activity, luminosity, or other 
derivative indicators, SIGS has relatively few systems at stages 4 and 5.  In order to more
thoroughly explore the full range of galaxy interactions, we assembled 
the LSM sample by filtering two catalogs.  The first of these, the Revised \textit{IRAS}-FSC 
Redshift Catalog \citep[RIFSCz;][]{2014MNRAS.442.2739W}, consists of 60\,303 galaxies selected 
from the \textit{IRAS} Faint Source Catalog (FSC) that contains accurate redshifts and 
positions as well as some photometry for the galaxies therein.  The second catalog, 
the Galaxy Zoo Data Release 1 
\citep[GZ1;][]{2008MNRAS.389.1179L,2011MNRAS.410..166L}, consists of almost 900,000 galaxies 
selected from the Sloan Digital Sky Survey \citep[\textit{SDSS};][]
{1998AJ....116.3040G,2006AJ....131.2332G,2000AJ....120.1579Y,2010AJ....139.1628D}. 
GZ1 galaxies were classified by the public 
into different categories including mergers.  Our selection required that galaxies be at redshifts below
 $z=0.06$, and that the fraction of the public votes that the galaxy was a merger 
was greater than 0.33.  These criteria produced the 453 interacting systems that make up the LSM
parent sample.  

The authors then inspected composite SDSS images of all 453 LSM systems and estimated the 
merger stages using the same criteria applied earlier to the SIGS systems, as defined in 
Sec.~\ref{ssec:SIGS}.   In the full LSM sample, only a minority of 24.9\% of the sources 
were classified as being in merger stages earlier than 3, i.e., our selection criteria 
successfully prioritized advanced mergers 
marked by obvious morphological distortions that signify a merger near coalescence.  
In the present work, we analyze all LSM galaxies having available {\sl Herschel}/SPIRE imaging available in the archive.  We excluded galaxies that were truncated by the edges of the SPIRE
 mosaics.  We identified a total of 49 LSM objects with suitable {\sl Herschel}/SPIRE imaging for
 the present work.  The basic properties of 38 of them are given in Table~\ref{tab:LSM}; those 
 for the remaining 12 LSM objects appear in table 1 of \citet[hereafter \citetalias{2018MNRAS.480.3562D}]{2018MNRAS.480.3562D}. NGC\,2623 is a special case of those remaining 12 objects, we re-introduce their physical parameters as part of the SB sample.  
 
In this work we present new SEDs for 188 galaxies.  Adding the 11 galaxies from \citetalias{2018MNRAS.480.3562D} brings the total sample size to 199 galaxies. For reasons fully described in Sec.~\ref{sec:results}, Ten of those 199 galaxies lack SEDs suitable for reliable inferences about the AGN contributions, so we subsequently analyse the implications of the SED fitting for only the 189 remaining galaxies. 
 
\section{SED assembly}\label{sec:SEDconstruction}

In Secs.~\ref{ssec:data_reduction}--\ref{ssec:val} we describe in detail how the SEDs were constructed.  In Sec~\ref{ssec:spectra} we 
also describe additional analysis carried out to retrieve mid-infrared emission line strengths for galaxies in the SB sample.

\subsection{Image Sources}
\label{ssec:data_reduction}

To ensure well-constructed SEDs, our approach was first, to assemble all available archival imaging 
spanning the widest possible wavelength range in the thermal regime, and second, to photometer all galaxies in all images 
within matching apertures.  Thus our resulting SEDs fully reflect 
all the relevant thermal emission mechanisms because they capture the totality 
of the galaxies' output at all thermal wavelengths, 
and they also have reliable colors, allowing us to accurately model the 
separate galaxy components that together comprise the SEDs.  

We drew upon imaging data from the following space- and ground-based missions:

\begin{itemize}
\item[-]{\sl GALEX} \citep[the Galaxy Evolution Explorer]{2005ApJ...619L...1M} for photometry in two ultraviolet bands, the far-ultraviolet  (FUV) band centered at 0.152 $\micron$, and the near-ultraviolet (NUV) band at 0.227 $\micron$.
\item[-]SDSS DR12 \citep[the Sloan Digital Sky Survey]{1998AJ....116.3040G} covering the $u, g, r, i,$ and $z$ bands, at 0.354, 0.477, 0.623, 0.762 and 0.913 $\micron$, respectively.
\item[-]2MASS \citep[Two Micron All-Sky Survey]{2006AJ....131.1163S} covering the $J, H,$ and $K_s$ bands at 1.25, 1.65 and 2.17 $\micron$, respectively.
\item[-]{\sl Spitzer}/IRAC \citep[the Infrared Array Camera]{2004ApJS..154...10F} providing mid-infrared coverage in up to four bands 3.6, 4.5, 5.8, and 8 $\micron$.
\item[-]{\sl Spitzer}/MIPS \citep[the Multiband Imaging Photometer]{2004ApJS..154...25R} covering up to three far-infrared bands at 24, 70, and 160 $\micron$.
\item[-]{\sl WISE} \citep[the Wide-Field Infrared Survey Explorer]{2010AJ....140.1868W} which covered the full sky in four IR bands centred at 3.4, 4.6, 12, and 22 $\micron$.
\item[-]{\sl IRAS} \citep[the Infrared Astronomical Satellite]{1984ApJ...278L...1N}, another all-sky survey mission that provides photometry in four broad bands at 12, 24, 60 and 100 $\micron$.  The {\sl IRAS} photometry used in this work is treated differently in that it was drawn from the Revised IRAS-FSC Redshift Catalogue \citep[RIFSCz,][]{2014MNRAS.442.2739W}, under the assumption that the IRAS data therein are mature and well-characterised, and the photometry is reliable for total galaxy measurements.  We likewise adopted the photometric uncertainties corresponding to the catalogued quality flags for the IRAS bands.  We did not use catalogued upper limits.
\item[-]{\sl Herschel}/PACS \citep[Photoconductor Array Camera and Spectrometer]{2010A&A...518L...2P} covering up to three far-infrared bands at 70, 100 and 160 $\micron$.
\item[-]{\sl Herschel}/SPIRE \citep[Spectral and Photometric Imaging Receiver]{2010A&A...518L...3G} providing far-infrared imaging at 250, 350, and 500 $\micron$. 
\end{itemize}

For {\sl GALEX}, {\sl Spitzer}, {\sl WISE}, and {\sl Herschel} we relied on archived, publicly available mosaics.  We verified the suitability of the available imaging for each galaxy and each band 
by inspection.  Mosaics in which the galaxies were truncated by mosaic edges, and mosaics 
in which the galaxies were saturated, were not considered valid and were not used.  Some archival 
IRAC mosaics for 20 of our galaxies were not suitable for photometry because of saturation of the galaxy nuclei.  Where possible, for these objects we generated our own IRAC mosaics by combining 
only the short exposures (typically 0.6\,sec) from archived IRAC high-dynamic range observations.  These short-exposures mosaics were not, generally speaking, saturated, and were in most instances suitable for the photometric analysis described below. 

For SDSS and 2MASS, we constructed our own mosaics centered at the positions of the sources 
listed in Tables~\ref{tab:SIGSsample}-~\ref{tab:LSM}, ensuring that they were sufficiently large that the 
source-free celestial backgrounds could be reliably estimated.

\subsection{Background Estimation}
\label{ssec:background}

Accurate background subtraction is crucial for accurate photometry.   In the present work, 
background calculation began with masking of mosaic pixels containing unphysical values, e.g., 
unexposed pixels not suitable for photometry.  We also created a mask for
potential contaminating foreground sources (Milky Way stars) by flagging all pixels with
a SNR higher than 3.0 for point sources.  This step is crucial for accurate background estimation.

We tested two background estimation techniques on our masked science mosaics, 
both within the Python package \textit{photutils}\footnote{https://github.com/astropy/photutils} 
\citep{larry_bradley_2018_1340699},
 an affiliated package of Astropy \citep{2013A&A...558A..33A}.\footnote{Further documentation is at \url{https://photutils.readthedocs.io/en/stable}.} 
 The first technique was \textit{Local Background Subtraction}, where we used an external 
 elliptical annulus of width equal to 10\% of the elliptical aperture radius to estimate the 
 background level around the galaxy.  
 The second technique was \textit{Global Background Subtraction}, where the image 
 was analyzed using sigma-clipped statistics, and an overall background estimate of the image was obtained. 
 We compared the global and local background calculations to those from our own custom 
 calculation -- also based on masked mosaics -- within square regions far from the target
 galaxies.  We found that \textit{Global Background Subtraction} was significantly more accurate
 than the local technique, so we adopted it subsequently for all our photometry.  This choice was
 validated when we found that our resulting {\sl Herschel}/PACS+SPIRE photometry agreed with
 published values, within the uncertainties, for sources having published photometry.  We speculate
 that the annuli used for the local background estimation were contaminated by low-level emission from the target galaxies at large radii.    

\subsection{Apertures, Inclinations, and Flux Densities}
\label{ssec:apertures}

We used elliptical apertures to estimate total fluxes for all galaxies considered here.  
Specifically, for a given galaxy, the same aperture was used in every photometric band, 
to ensure accurate colors and thus reliable SEDs.  Each aperture was sized to encompass
the maximum apparent extent of each galaxy, as measured either in the {\sl GALEX}/NUV or
3.6\,$\micron$ IRAC mosaic (or, if the latter was unavailable, the 3.4\,$\micron$ {\sl WISE} mosaic).
We inspected all mosaics of all galaxies with the apertures overlaid to ensure that no flux fell 
outside them.  Based on those inspections, in some instances it was necessary to 
enlarge or shift the apertures and re-measure the photometry.  Ultimately, all apertures were appropriately 
sized and located to enclose all of a galaxy's flux in all available bands.

We applied appropriate {\sl Herschel}/PACS aperture and colour corrections to account 
for missing flux due to incomplete sampling of the point spread function (PSF) in each of the 
PACS bands.  

The pixel values within apertures were summed and converted to flux densities using the flux 
calibrations in the instrument handbooks. 
We accounted for absolute calibration error by adding appropriate instrument-dependent uncertainties
in quadrature to the measurement uncertainties calculated in the standard way.  These were as follows: 
10\% for {\sl GALEX} \citep{2007ApJS..173..682M}, 
2\% for SDSS \citep{2010AJ....139.1628D}, 
2\% for 2MASS  \citep{ 2003AJ....126.1090C}, 
3\% for IRAC \citep{2003AJ....125.2645C},  
4\% for MIPS \citep{2007PASP..119..994E}, 
6\% for {\sl WISE} \citep{2010AJ....140.1868W}, 
10\% for PACS \citep{2010A&A...518L...2P},
and 7\% for SPIRE \citep{2010A&A...518L...4S}.   Typically, the calibration errors were much larger 
than the measurement errors for these relatively bright objects.
No additional uncertainties were added to those 
already adopted from the RIFSCz for the flux densities measured in the {\sl IRAS} bands. 

We present our {\sl GALEX} and SDSS photometry in 
Table~\ref{tab:all_phot_galex_sdss},
our 2MASS and {\sl Spitzer}/IRAC photometry in 
Table~\ref{tab:all_phot_2mass_irac},
our {\sl WISE} and {\sl Spitzer}/MIPS photometry in
Table~\ref{tab:all_phot_wise_mips}, and finally 
our {\sl Herschel}/PACS+SPIRE photometry in
Table~\ref{tab:all_phot_herschel}.
When the photometry was consistent with zero flux density (i.e. the estimated uncertainty 
was greater than the estimated flux density) we chose not to include it in our SED models.   

\subsection{Photometry Validation}
\label{ssec:val}

We verified that our approach yields high-quality photometry by comparing our measurements to previously published photometry.  Specifically, we compared our {\sl Spitzer}/IRAC+MIPS 24\,$\mu$m photometry to that published previously by \citetalias{2015ApJS..218....6B}.   Overall we found good agreement.  In the following we describe the comparison in detail.

\begin{figure}
\begin{center}
\includegraphics[width=0.98\columnwidth]{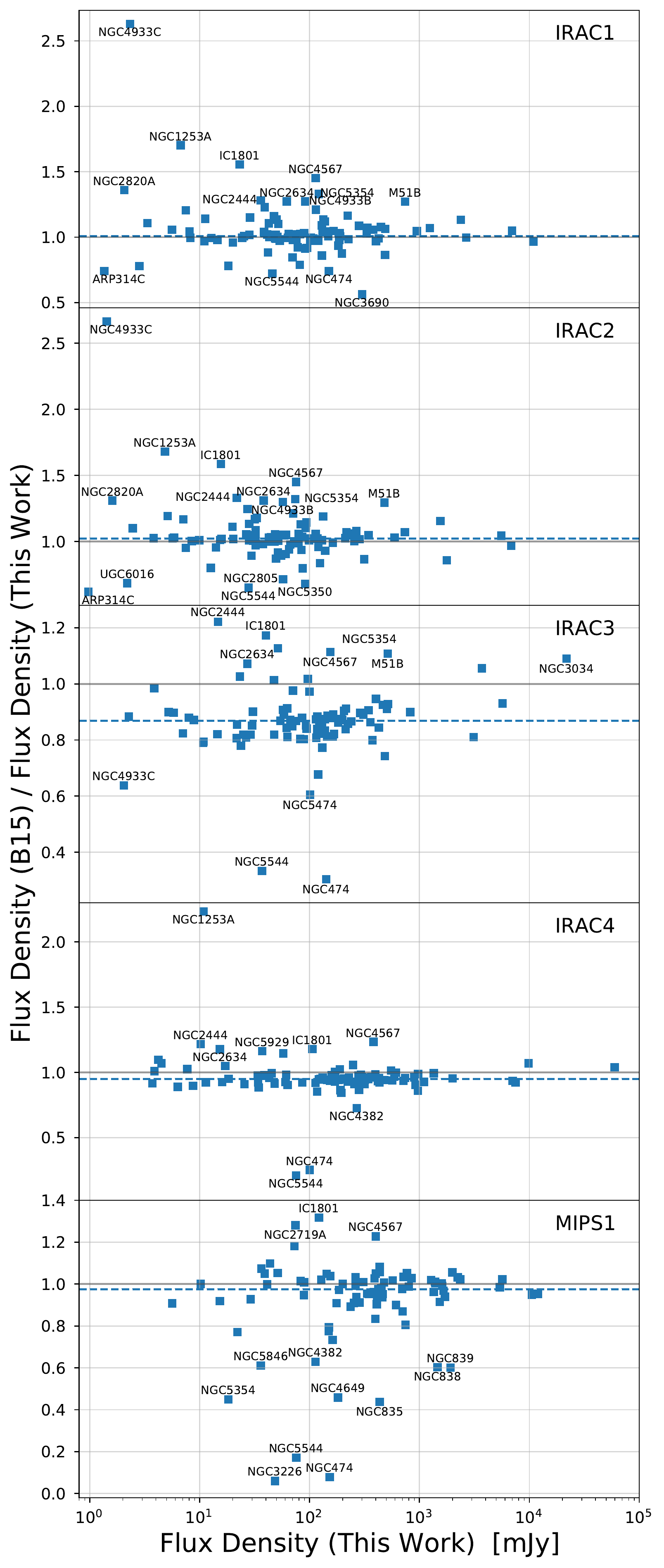}
\end{center}
\caption{A comparison of our {\sl Spitzer}/IRAC+MIPS global photometry to that of \citetalias{2015ApJS..218....6B} for the SIGS galaxies.  The photometry is consistent on average, but differences for individual galaxies differences are apparent.  Most of the discrepancies are traced 
to different apertures,  as described in Sec.~\ref{ssec:val}.}
\label{fig:Phot}
\end{figure}

Figure~\ref{fig:Phot} compares our IRAC and MIPS 24\,$\micron$ photometry with that of\ \citetalias{2015ApJS..218....6B} for all systems common to both studies. 

Outliers are apparent, however, and some are significant.  To understand
the causes of the discrepancies, we obtained from \citetalias{2015ApJS..218....6B} their Source
Extractor (SE; \citet{1996A&AS..117..393B}) output files and examined them in light of our 
own output from \textit{photutils}.  Our findings are listed below in order of significance.

{\sl NGC\,4933}: For two galaxies in this system we used significantly smaller apertures than \citetalias{2015ApJS..218....6B} 
(the \citetalias{2015ApJS..218....6B} aperture diameters were factors of roughly 15 and 4 times  
those we used for NGC\,4933C and B, 
respectively), which allowed us to avoid the nearby potentially contaminating IR-bright source SSTSL2 J130402.66$-$112854.1.  We also shifted our aperture center for NGC\,4933A by 11\arcsec\ relative to 
\citetalias{2015ApJS..218....6B} to avoid potential contamination from NGC\,4933B.  

{\sl NGC\,1253A:} The \citetalias{2015ApJS..218....6B} aperture is roughly 10.4 times the size of ours. It is a faint source compared to its companion NGC\,1253; a nearby bright star (TYC\,4711-231-1) lies within the \citetalias{2015ApJS..218....6B} aperture. 	

{\sl IC\,1801:} The \citetalias{2015ApJS..218....6B} aperture is roughly 4.6 times the size of ours, and contains part of the core of NGC\,935 and a nearby source (2MASS\,J02281028+1934207).  

{\sl NGC\,4567:} The \citetalias{2015ApJS..218....6B} aperture is roughly 2.7 times the size of ours, and overlaps with the core of NGC\,4568.   We shifted our aperture center by 23\arcsec\ to avoid potential contamination from NGC\,4568. 

{\sl NGC\,2820A:} The \citetalias{2015ApJS..218....6B} aperture is roughly 5.5 times the size of ours, and therefore includes 
a nearby star (2MASS\,J09212802+6413442) that likely contaminates their 3.6 and 4.5\,$\mu$m photometry.
    
{\sl NGC\,5354:} The \citetalias{2015ApJS..218....6B} aperture is roughly 6 times the size of ours, and covers the core of 
NGC\,5353.  The \citetalias{2015ApJS..218....6B} MIPS 24\,$\mu$m photometry is only marginally lower ($<2$\,mJy) 
than our {\sl WISE} band 4 photometry.  Our photometry is however 
consistent with \citet{2016ApJ...818..182V}.   Due to contamination from the nearby
NGC\,5353, there appears to be considerable variation in the tabulated photometry of 
NGC\,5354 in the literature \citep{2016ApJ...821..113Z, 2018A&A...609A..37C}.
    
{\sl NGC\,2444:} The \citetalias{2015ApJS..218....6B} aperture is roughly 2.1 times the size of ours.  We offset our aperture center by 10\arcsec\ relative to the galaxy center to avoid potential contamination from the nearby
galaxy NGC\,2445.   NGC\,2445 is faint at 24\,$\mu$m so the contamination in the MIPS 24\,$\mu$m
band is not significant.
    
{\sl IC\,694 and NGC\,3690:} It appears that the \citetalias{2015ApJS..218....6B} aperture attributed to IC\,694 actually 
corresponds to a portion of NGC\,3690, and that the \citetalias{2015ApJS..218....6B} aperture for the latter is undersized.

{\sl NGC\,2634:} The \citetalias{2015ApJS..218....6B} aperture is roughly 4.2 times the size of ours, potentially admitting contaminating flux from several nearby sources.  
    
{\sl M51B:}  The \citetalias{2015ApJS..218....6B} aperture is roughly 2.7 times the size of ours, and encompasses part of one arm of M51A. This is significant only for the 3.6 and 4.5\,$\mu$m bands because the relevant portion of that arm of M51A is relatively faint at longer wavelengths.

{\sl NGC\,5544:} In this work we treat NGC\,5544 and 5545 as a single system because they are inseparable 
at {\sl Herschel} spatial resolution, whereas \citetalias{2015ApJS..218....6B} photometered them separately.

{\sl NGC\,3034:} Our aperture is roughly 2.4 times the size of that in \citetalias{2015ApJS..218....6B}, explaining the differences in IRAC3.

{\sl NGC\,474:} Our aperture is roughly five times the size of that in \citetalias{2015ApJS..218....6B}.   

{\sl NGC\,5474:} The centroid of our aperture is offset from that of \citetalias{2015ApJS..218....6B} by 32\arcsec\ for this diffuse galaxy.  Our 5.8\,$\mu$m photometry is similar to that in \citet{2005ApJ...633..857D}. 

{\sl Arp\,314C:} Our aperture is roughly 1.4 times the size of that in \citetalias{2015ApJS..218....6B}.  This is a faint galaxy, and
is likely strongly affected by stars lying within the aperture, especially in the IRAC 3.6 and 4.5\,$\mu$m bands. 

{\sl UGC\,6016:}  Our aperture is 1.3 times the size of that in \citetalias{2015ApJS..218....6B}, and is offset by 10\arcsec\ to avoid potential contamination of this relatively faint galaxy from nearby bright stars in the 3.6 and 4.5\,$\mu$m bands.   

{\sl NGC\,5929:} The \citetalias{2015ApJS..218....6B} aperture is roughly 1.5 times the size of ours, and our aperture is shifted
relative to the galaxy center by 11\arcsec\ to avoid potential contamination from the nearby NGC\,5930.

{\sl NGC\,4382:}  Our aperture is roughly 0.77 times the size of that in \citetalias{2015ApJS..218....6B}.  Our MIPS 24\,$\mu$m and {\sl WISE} 22\,$\mu$m photometry is consistent with \citet{2014A&A...570A..69B}.  Our
IRAC 8\,$\mu$m photometry is consistent with \citet{2014ApJ...783..135A}.

{\sl NGC2719A:} \citetalias{2015ApJS..218....6B} aperture is roughly twice the size of ours.

{\sl NGC3226:} Our aperture is less than half the size of that in \citetalias{2015ApJS..218....6B}.  Our MIPS 24\,$\mu$m and 
{\sl WISE} 22\,$\mu$m photometry is consistent with the {\sl WISE} photometry reported in 
\citet{2016ApJ...818..182V} and \citet{2014A&A...565A.128C}.

{\sl NGC\,4649:}  Our MIPS 24\,$\mu$m and {\sl WISE} 22\,$\mu$m photometry is consistent with the {\sl WISE} photometry reported in \citet{2016ApJ...818..182V} and \citet{2014A&A...565A.128C}.

{\sl NGC\,835, NGC\,838 and NGC\,839:}  We obtain higher mid-infrared flux densities than \citetalias{2015ApJS..218....6B} for these galaxies.  Our MIPS 24\,$\mu$m photometry is consistent with \citet{2007AJ....134.1522J} and \citet{2011A&A...533A.142B}.  
Our {\sl WISE} 22\,$\mu$m photometry is consistent with \citet{2016ApJ...821..113Z}.

Having reached down to discrepancies of order 20\% (specifically, 23, 23, 13, 19, and 22\% 
in the IRAC 3.6, 4.5, 5.8, and 8.0\,$\mu$m bands and the MIPS 24\,$\mu$m band, respectively) 
 relative to \citetalias{2015ApJS..218....6B} without
finding any serious faults with our photometry, we carried the comparison no further.  

\begin{figure}
\begin{center}
\includegraphics[width=\columnwidth]{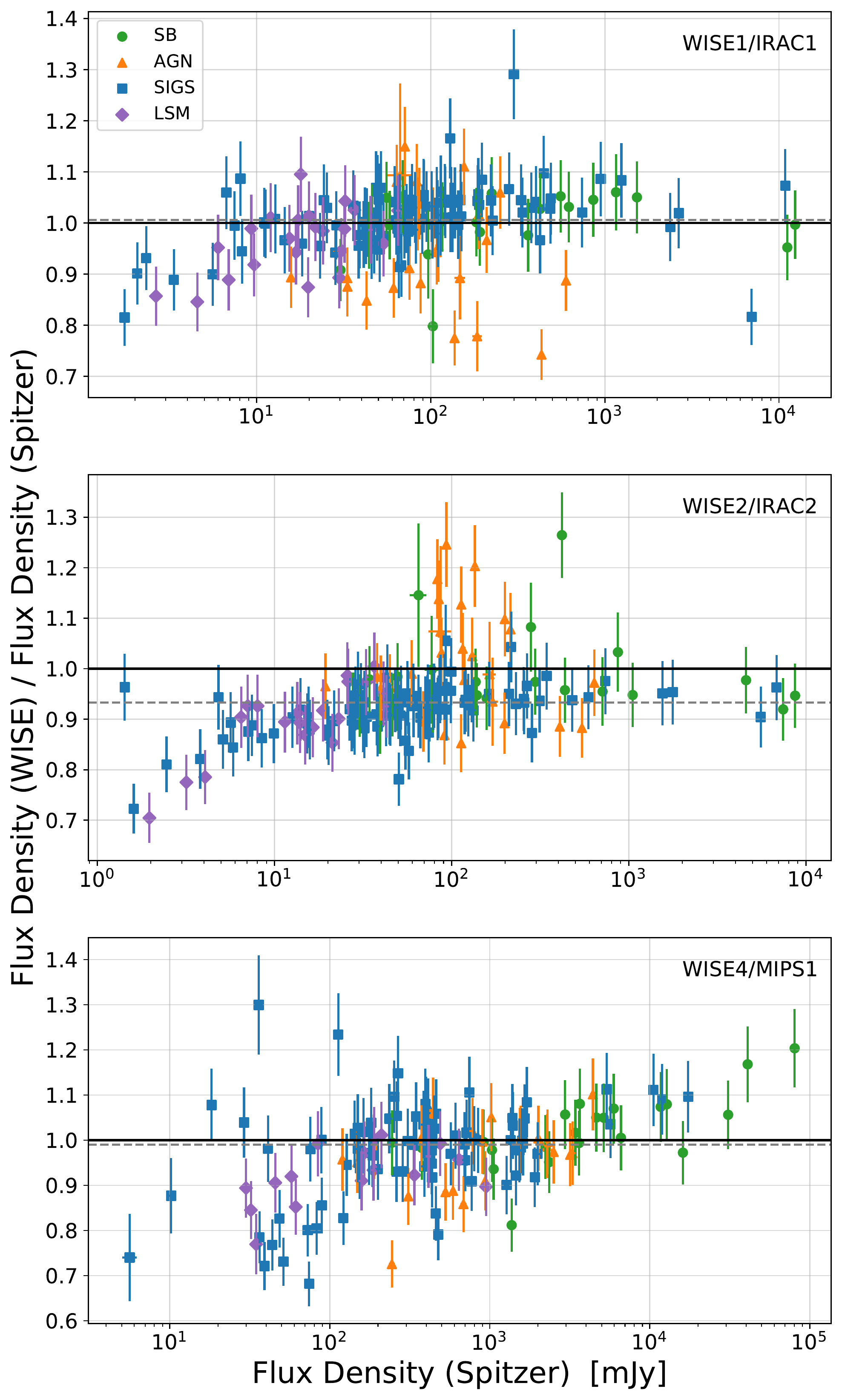}
\end{center}
\caption{A comparison of our {\sl Spitzer}/IRAC 3.6, 4.5, and MIPS 24\,$\mu$m global photometry 
to that measured in the {\sl WISE} 3.4, 4.6, and 22\,$\mu$m bands for all galaxies in the SIGS, SB, AGN, 
and LSM samples. The gray dashed line shows the median value of the {\sl WISE}-to-{\sl Spitzer} ratio for all galaxies per panel.}
\label{fig:compare}
\end{figure}

We also compared the {\sl photoutils} photometry for the IRAC 3.6 and 4.5\,$\mu$m bands to
what we measured with the same aperture in the very similar {\sl WISE} bands 1 and 2.  
In addition, we compared our
MIPS 24\,$\mu$m photometry to that obtained in the similar {\sl WISE} band 4 at 22\,$\mu$m.
The results are shown in Fig.~\ref{fig:compare}.  In general, the agreement is excellent. A small systematic flux underestimation is present in {\sl WISE} for low IRAC fluxes showing that the background level is overestimated, but this only affects a few galaxies. We were able to
resolve most of the discrepant cases with small shifts in aperture centers or diameters, 
or (in a few cases) by correcting an erroneous background estimate.  When we were unable 
to understand and resolve a pair of discrepant bands, we chose not to use either of them in the subsequent analysis.  

On the basis 
of these two comparisons -- of our photometry measured in similar bands and measured by \citetalias{2015ApJS..218....6B} -- and 
the fact that we visually inspected every mosaic for every galaxy 
with our \textit{photutils} aperture overlaid, we are confident that our photometry is sound and that
 suitable for the SED modeling described in Sec.~\ref{ssec:SED}.
 
\subsection{Mid-Infrared Spectroscopy of Galaxy Nuclei}
\label{ssec:spectra}

Mid-infrared spectroscopy provides useful constraints on galaxy energetics because 
emission lines in the mid-infrared regime reveal the excitation conditions in the ISM nearly 
free of the usual complications from dust attenuation.  For this reason we made use 
 of {\sl Spitzer}/IRS spectroscopy to help quantify the AGN contributions to our sample
galaxies. Specifically, we used IRS short-high (SH) spectra  of our SB sample 
galaxies to better understand their energetics via their neon and PAH features.  In this Section we describe how we reduced and 
analysed those spectra.  IRS spectra were taken for all our AGN sample galaxies as well, 
but we did not reduce them ourselves, we took the published neon line ratios from the literature to consistently fulfil the AGN sample selection criteria.

For each galaxy in the SB sample, we began with the SH basic calibrated data (BCD) produced by the IRS pipeline, covering wavelengths from 10 to 20\,$\mu$m.  
We reduced the data in the standard way, first using \textsc{IRSCLEAN} to mask 
cosmic rays and bad pixels.   We set the aggressive keyword to 0.5, so that
a pixel which exceeds the sigma threshold could only be flagged as bad if it 
had no neighbours that also satisfied this criterion.  
We then used the CUbe Builder for IRS Spectra Maps \citep[\textsc{CUBISM}]{2007PASP..119.1133S} to combine the spatial and spectral information of the datasets, perform background subtraction, and generate a 
one-dimensional spectrum for each galaxy.  We then used \textsc{PAHFIT} 
\citep{2007ApJ...656..770S} to estimate the strengths of the emission features in our spectra.

In general, this procedure worked well, although there were some exceptions. As was also
found by \citet{2006ApJ...653.1129B}, the nuclei of NGC\,520 and Mrk\,52 were observed slightly 
off-center.  A more severe mis-pointing was revealed for NGC\,3310.  Thus for these three
sources, our spectra do not represent all the emission from their nuclei. 
Results from these emission lines are bias to nuclear regions, so comparing with other galaxies not observed only in the nuclei can lead to different estimations. We assume that varying the physical scale of the systems will give similar results in terms of line ratios. The results of our IRS 
spectroscopy are described in Sec.~\ref{ssec:neon} and the emission line strengths are tabulated in 
Table~\ref{tab:SHlines}.

\section{Analysis}\label{sec:analysis}

This Section details how the SEDs compiled in Sec.~\ref{ssec:data_reduction} were modeled 
 to estimate the contributions from young and old stellar populations, thermal emission from dust, 
 and AGNs to the overall emission of each galaxy in our four study samples.
 
\subsection{SED Modeling with \textsc{CIGALE}}
\label{ssec:SED}

This work relies primarily on \textsc{\textsc{CIGALE}}\footnote{\url{http://cigale.lam.fr/}, version 0.12.1} as the
means of interpreting galaxy SEDs. \textsc{CIGALE} is a widely used fitting code, based on an energy balance principle, that 
attempts to model galaxy SEDs in terms of a combination of a small number of separate 
components that overlap in wavelength.  A detailed description of the mechanics
of \textsc{CIGALE} are available from \citet{2015A&A...576A..10C} and 
\citet{2019A&A...622A.103B}; here we summarize only the main points relevant to our analysis.

\textsc{CIGALE} works by first populating a 
high-dimensional parameter grid of SED models consisting of all combinations of user-specified
components that contribute to the emission, and then computes the 
goodness of fit for each model. \textsc{CIGALE} identifies the best-fit SED model
by minimising the $\chi^2$ statistic, and produces probability distribution functions for the model 
grid parameters by assuming Gaussian measurement errors \citep{2005MNRAS.360.1413B,2009A&A...507.1793N,2011ApJ...740...22S}.   Most relevant to the present work is the fact that \textsc{CIGALE} implements
convenient templates for emission from an obscured AGN, based on the models described in \citet{2006MNRAS.366..767F}. 

\begin{table*}
\centering
\caption{\textsc{CIGALE} grid parameter values adopted for the modeling described in Section~\ref{ssec:SED}}
\label{tab:Par_CIG}
\begin{tabular}{p{0.15\textwidth}p{0.25\textwidth}p{0.45\textwidth}}
\hline
\hline
Parameter & Values & Description \\
\hline
\multicolumn{3}{c}{Star formation history (SFH): Delayed}\\
$\tau_{\textnormal{main}}$&50, 500, 1000, 2500, 5000, 7500& e-folding time of the main stellar population model (Myr).\\
Age &500, 1000, 2000, 3000, 4000, 5000, 6000&Age of the oldest stars in the galaxy (Myr).\\
\hline
\multicolumn{3}{c}{Single-age stellar population (SSP): \citet{2003MNRAS.344.1000B}}\\
Separation Age& 10& Age of the separation (to differentiate) between the young and the old star populations (Myr).\\
\hline
\multicolumn{3}{c}{Dust attenuation: \citet{2000ApJ...533..682C}}\\
E$(B-V)_{\textnormal{young}}$ &0.1, 0.25, 0.4, 0.55, 0.7& Color excess of the stellar continuum light for the young population.\\
E$(B-V)_{\textnormal{old factor}}$ &0.22, 0.44, 0.66, 0.88 & Reduction factor for the  E$(B-V)$ of the old population compared to the young one.\\
Power-law slope ($\delta$) & 0.0, 0.25, 0.5& Slope delta of the power law modifying the attenuation curve.\\
\hline
\multicolumn{3}{c}{Dust emission: \citet{2014ApJ...784...83D}}\\
$\alpha$ & 1.0, 1.25, 1.5, 1.75, 2.0, 2.25, 2.5, 2.75, 3.0 & Alpha from the power-law distribution in eq.~\ref{eq:alpha}. \\
\hline
\multicolumn{3}{c}{AGN model: \citet{2006MNRAS.366..767F}}\\
R$_{\textnormal{max}}$/R$_{\textnormal{min}}$ & 10.0, 30.0, 60.0, 100.0, 150.0 & Ratio of the maximum to minimum radii of the dust torus. \\
$\tau$ & 0.1, 0.6, 1.0, 6.0, 10.0 & Optical depth at 9.7 $\micron$. \\
$\beta$ & $-$1.00, $-$0.75, $-$0.50, $-$0.25, 0.00 & Beta from the power-law density distribution for the radial component of the dust torus (eq. 3 of Fritz 2006).\\
$\gamma$ & 0.0, 2.0 & Gamma from the power-law density distribution for the polar component of the dust torus (eq. 3 of Fritz 2006).\\
Opening Angle ($\theta$) & 60.0, 100.0, 140.0 & Full opening angle of the dust torus (Fig 1 of Fritz 2006). \\
$\psi$ & 30.1$^{\rm a}$ & Angle between equatorial axis and line of sight. \\
$f_{\rm{AGN}}$ & 0.1 -- 0.9 in steps of 0.05 & Fraction of AGN torus contribution to the IR luminosity (fracAGN in Equ. 1 of Ciesla 2015)\\
\hline
\end{tabular}

\begin{flushleft}
$^{\rm a}$ The apparent precision was adopted to accommodate an idiosyncrasy in \textsc{CIGALE}'s mode of operation, fractional degree precision is not implied.
\end{flushleft}
\end{table*}

We used the parameters and values given in Table~\ref{tab:Par_CIG} to define the \textsc{CIGALE} grid 
of model galaxy SEDs.  Except as noted below, the parameter settings were identical to
those of \citetalias{2018MNRAS.480.3562D}.  
For all parameters not shown in Table~\ref{tab:Par_CIG}, we adopted the
\textsc{CIGALE} default settings.  All fits were performed assuming the distances given in Tables~\ref{tab:SIGSsample}-\ref{tab:LSM}.

We treated the galaxies' star formation histories (SFH) with a delayed SFH model, taking that
as a reasonable approximation for the SF history during the last $\sim 10$\,Myr.  This approach 
assumes a single past starburst event \citep{2015A&A...576A..10C}.  The parameters that
control the delayed SFH model are the age of the oldest stars in the galaxy, and the 
folding time ($\tau_{\textnormal{main}}$) of the exponential decay in star formation 
after the starburst occurs. Depending in the combination of these two parameters, we can simulate ongoing or recent starburst events.

The stellar emission was modeled with the standard  \citet{2003MNRAS.344.1000B} population synthesis libraries, weighted by the SFH.  We used the default \textsc{CIGALE} nebular emission module. The module controlling UV attenuation followed  \citet{2000ApJ...533..682C} and \citet{2002ApJS..140..303L}. This module is parameterized by 
the young population color excess E$(B-V)_{\rm young}$ of stellar continuum light, 
the reduction factor of the color excess for the old population E$(B-V)_{\rm old\ factor}$ 
as compared with the young population, the UV bump central wavelength, FWHM, and amplitude 
(the \textsc{CIGALE} default values for these parameters of 2\,175\AA, 350\AA, and 0 were used), 
and the power-law slope ($\delta$) which modifies the attenuation curve. 

The dust emission was modeled following \citet{2014ApJ...784...83D}, implementing a 
modified blackbody spectrum with a power-law distribution of dust mass at each temperature,
\begin{eqnarray}\label{eq:alpha}
dM \propto U^{-\alpha} dU
\end{eqnarray}

where $U$ is the local heating intensity. 

We adopted the same overall AGN model as \citetalias{2018MNRAS.480.3562D} to estimate
the AGN fraction $f_{\rm{AGN}}$ in our sample galaxies, i.e., the \citet{2006MNRAS.366..767F}
model.   
Because one of our primary goals is to investigate the emission fraction coming from 
the obscured AGNs in our sample galaxies, we sampled the AGN fraction parameter $f_{\rm{AGN}}$
somewhat more finely than 
\citetalias{2018MNRAS.480.3562D},  in steps of 0.05 between 0.1 and 0.9, as well as at 0.0 (i.e., zero AGN contribution).  
We adopted a single value for the viewing angle into the AGN ($\psi = 30.1$), as intermediate between type 1 and 2 AGNs. We tested the effect of varying the viewing angle in the samples of this work. We run a similar grid as in Table~\ref{tab:Par_CIG} with half of the steps for $\alpha$ and $\beta$, and adding $\psi = 70.1$. In general, changing the angle does not usually make a significant difference in the output parameters. However, we also find that \textsc{CIGALE} can identify Type 1 AGNs: their output parameters, especially stellar mass, suddenly become sensitive with $\psi = 70.1$ and a lower $\chi^2$ compared with $\psi = 30.1$. We detected six AGN galaxies that fall into this category, all of them are already known to be Type 1 Seyfert AGN. We present the derived parameters for those galaxies with $\psi = 70.1$ in Table~\ref{tab:CIGALE1Psi}. We use these lower $\chi^2$  values in all the Figures of this work. A new version of \textsc{CIGALE}, ``\textsc{X-CIGALE}'' \citep{2020MNRAS.491..740Y}, has been recently released that is specifically designed to be more attentive to the angle and to the high-energy contributions to the SED. A study focusing on Type 1 AGN will benefit from both this new version and from a more detailed angle analysis, but is beyond the current work.

We also sampled $\alpha$ in increments of 0.25 between 1.0 and 3.0, and extended the values for the slope delta power-law modifying the dust attenuation curve (0.25 and 0.5 in addition to 0), the optical depth at 9.7 $\micron$ (including 0.1) and the density radial exponent of the torus (adding the values $-$0.5, $-$0.25, and 0).  
Our tests indicate that by choosing a compact grid of values for $\alpha$ and $f_{\rm{AGN}}$, and a single value for the viewing angle into the AGN $\psi = 30.1$, we can obtain well-behaved 
PDFs for these grid-parameters (i.e., they are well resolved probability distributions).  We might therefore 
expect an improvement relative to the measurements of \citetalias{2018MNRAS.480.3562D} because
of the more finely sampled parameter space.  

Our strategy was to run two different families of \textsc{CIGALE} models.  The first family included
AGNs parameterized according to Table~\ref{tab:Par_CIG}, while the otherwise identical second 
family was run without. We adopted this approach because \citetalias{2018MNRAS.480.3562D} found that $f_{\rm{AGN}}$ was typically uncertain 
by $\pm 10$\%.  Thus, cases when $f_{\rm{AGN}}< 0.20$ are not inconsistent with 
$f_{\rm{AGN}}=0$, i.e., no AGN being present.  We therefore chose to treat cases for which $f_{\rm{AGN}} < 0.20$ as if they had no AGN.  We thereby avoid the pitfall noted by \citet{2015A&A...576A..10C}
i.e., that the AGN contribution can be overestimated, 
an effect often seen when deriving low-valued parameters with truncated PDF analysis \citep{2009A&A...507.1793N}.

We present some examples of the best SED fitting for each of the four samples presented in this work in Figures~\ref{fig:SED_SB}-\ref{fig:SED_LSM}.

\subsection{Neon emission lines}
\label{ssec:neon}
The MIR provides spectral features that are excited by the intense UV radiation from massive young stars. Among the most prominent infrared emission features are the PAHs bands that arise in the photon-dominated regions (PDR) around HII regions. Also, the forbidden nebular lines emitted by ionised atomic gas play an essential role in the characterisation of the gas physics. 

Strong radiation fields such as those around AGNs are necessary to reach the ionization potential (IP) of the [\ion{Ne}{v}] emission line at 14.3 $\micron$ (97.1 eV). Such radiation strength is unlikely to be produced by star formation
\citep{2002A&A...393..821S,2006ApJ...653.1129B}. This line is therefore used as a tracer of AGN activity.

An additional advantage of using the [\ion{Ne}{v}] emission line relates to the fact that dust extinction at 15 $\micron$ is small and typically independent of the orientation \citep{2011ApJS..195...17W}. \citet{2009MNRAS.398.1165G} show that optical spectroscopic surveys, in contrast, can miss approximately half of the AGN population due to extinction through the host galaxy. \citet{1998ApJ...498..579G} found a correlation between the strength of emission lines, higher stages of ionisation, and the level of AGN activity. Therefore, [\ion{Ne}{v}] can be used to quantify AGN activity. The forbidden [\ion{O}{iv}] at 25.9~$\micron$ is also used for similar reasons. This line has a lower IP (54.9 eV) and is detected in a more significant fraction of AGNs, but can also be produced in star-forming galaxies, mainly in the presence of WR stars and ionising shocks \citep{2008ApJ...678..686A,2010ApJ...725.2270P,2010ApJ...716.1151W,2012ApJ...744....2A}. Although [\ion{O}{iv}] has proved to be useful as an AGN tracer by some authors \citep{2009ApJS..182..628V,2016MNRAS.458.4297G}, here we have decided to use the [\ion{Ne}{v}] emission only, in order to avoid any contamination from star formation.

The [\ion{Ne}{ii}] low ionisation line at 12.8 $\micron$ (IP = 21.6 eV) traces the thermal stellar emission in star-forming galaxies \citep{2002A&A...393..821S}. Therefore, comparing its strength to that of the [NeV] line provides a straightforward measurement of the relative contribution from star formation and AGN to the overall energy budget. The proximity of the two neon lines in wavelength implies that both of them are subject to similar extinction \citep{2010ApJ...709.1257T}. A caveat is that the [\ion{Ne}{ii}] line blends with the  PAH feature at 12.7 $\micron$ and the [\ion{Ne}{v}] line blends with the [\ion{Cl}{ii}] line. We work under the assumption that the effect of this blending in the estimation of the lone strengths is not very significant. As noted in \citet{2009MNRAS.398.1165G}, it is safe to make this assumption when high  signal-to-noise data is available, as is our case. In the present work the uncertainty in line strengths due to blending is smaller than the uncertainty due to instrumental and detection effects. 

The [\ion{Ne}{v}]/[\ion{Ne}{ii}] line ratio has been used to calibrate the relative AGN contribution to the total infrared luminosity of galaxies \citep[among others]{1998ApJ...498..579G, 2002A&A...393..821S,2008ApJ...676..836T,2009ApJS..182..628V,2010ApJ...709.1257T,2011ApJS..195...17W,2011ApJ...740...99D,2017ApJ...846...32D}. For example, \citet{2011ApJS..195...17W} estimate a 100\% AGN contribution corresponding to a [\ion{Ne}{v}]/[\ion{Ne}{ii}] ratio of $\approx$ 1.0 and a 0\% AGN contribution corresponding to a [\ion{Ne}{v}]/[\ion{Ne}{ii}] ratio of $\approx$ 0.01. As pointed out by \citet{2011ApJ...730...28P}, discrepancies in the measured contribution of the AGN to the bolometric luminosity can be due to different calibrations of the line ratio. For instance, \citet{2010ApJ...725.2270P} have argued that for pure AGN emission, the line ratio should be closer to [\ion{Ne}{v}]/[\ion{Ne}{ii}] $\approx$ 2-3. 
%%% Howard thinks that the last part of the paragraph can be dropped because is weak

The fluxes of the [\ion{Ne}{v}] lines for 19 of the 23 galaxies in the SB sample, are presented in Table~\ref{tab:SHlines} along with other useful lines in the Spitzer SH mode\footnote{With a slit size of 4.7\arcsec $\times$ 11.3\arcsec}: the  [\ion{Ne}{iii}], [\ion{S}{iii}], [\ion{S}{iv}], [\ion{Fe}{ii}] and H$_{2}$ S(2) and H$_{2}$ S(1). For Mrk\,52, NGC\,23, NGC\,253 and NGC\,7714 we took upper limits from the literature \citep{2009ApJS..184..230B,2010ApJ...725.2270P} and we use those values for comparison. 

Although the results for the SB sample fall in a region dominated by upper limits in the detection of the [\ion{Ne}{v}]/[\ion{Ne}{ii}], some of the results could be affected by the adopted emission-line detection procedure and signal-to-noise ratio threshold used in weak cases, as noted by \citet{2009MNRAS.398.1165G}.
Most of the spectroscopic data come from the nuclear region of the galaxies, so when we compare with the global values of the galaxies in the SED for the given apertures, we are comparing global characteristics with a measure of the central emission (most predominant) region of the samples. 

\begin{figure}
\includegraphics[width=\columnwidth]{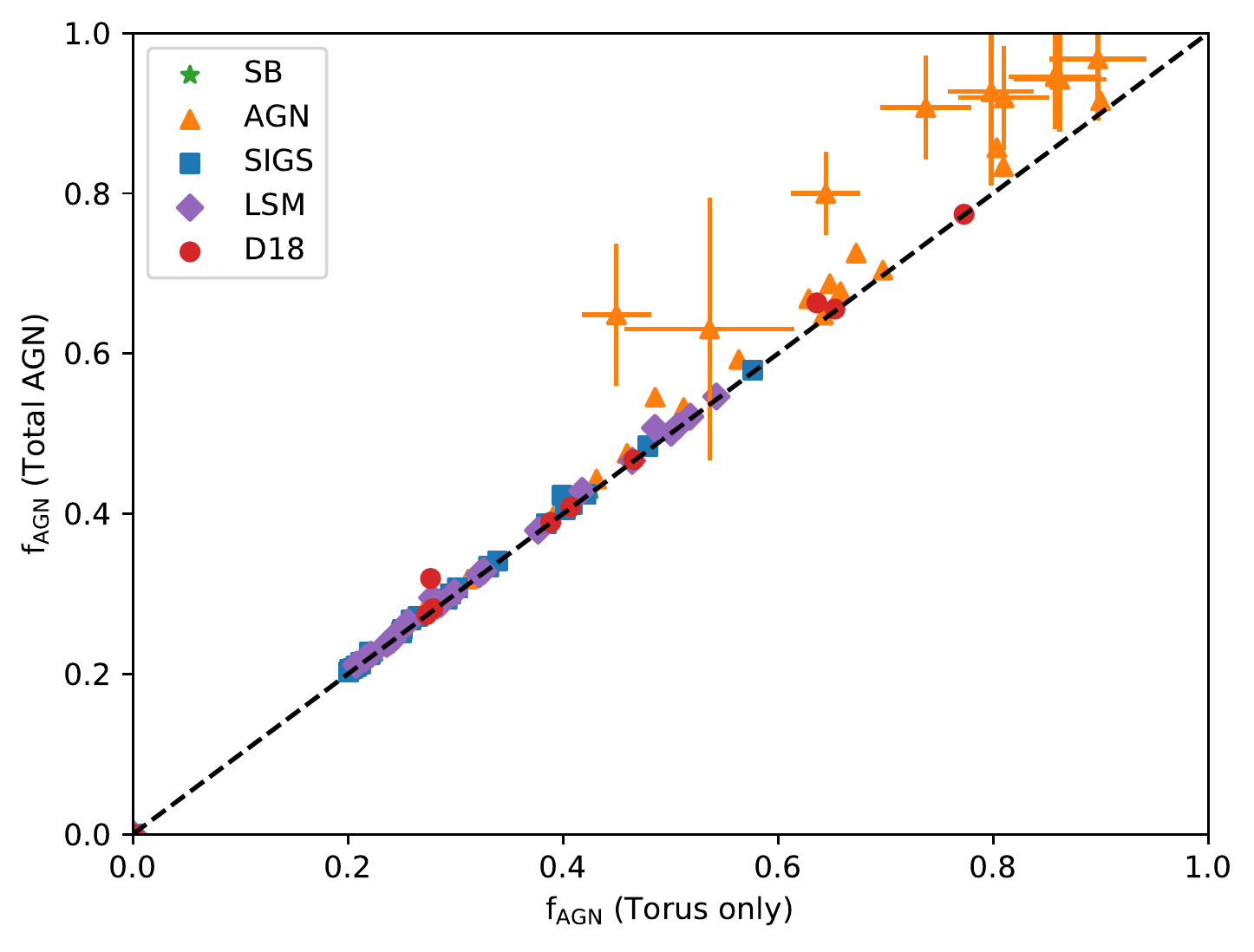}
\caption{
The total AGN contribution to the infrared luminosity, $f_{\textnormal{AGN}}$, compared to the fractional contribution from the molecular torus alone in \textsc{CIGALE}'s implementation of the Fritz AGN model.   For clarity, error bars are only shown for galaxies that deviate
significantly from the line of equality.  The uncertainties for points without drawn
error bars are similar.   Symbols indicate which of the four subsamples the objects 
belong to (Sec.~\ref{Sec:Obs}); the D18 galaxies are also shown.
Here and in the main text we use $f_{\textnormal{AGN}}$ to 
indicate the total AGN fraction, which differs from the popular convention that considers emission only from the torus 
(see Sec.~\ref{sec:results}).  
The distinction matters for a significant number of objects in the AGN sample.
 }
\label{fig:AGNdefs}
\end{figure}

\section{Distributions of Derived Galaxy Properties}
\label{sec:results}

This section describes the \textsc{CIGALE}-based SED fitting results for the 199
objects in the SIGS, SB, AGN, and LSM subsamples described in Sect~\ref{Sec:Obs}. The large overall sample size and the 
well-defined subsets facilitate some useful statistical comparisons. A total of 94 objects have $f_{\rm{AGN}}\ge0.2$; for these objects
we present the \textsc{CIGALE} parameters as computed.  For the remaining objects
we present the \textsc{CIGALE} parameters as computed with their AGN 
contribution set to zero.  Ten galaxies (marked with an $^{\rm b}$ in Tables~\ref{tab:SIGSsample}-\ref{tab:LSM}) have sparse photometric coverage and consequently
lack reliable SED fits; we omit these objects from further analysis. 

Out of the $\sim60$ parameters that \textsc{CIGALE} estimates,
we focus on those most relevant to star formation and AGN activity; the parameters 
 we emphasize  are listed in Tables~\ref{tab:CIGALE1AGN}--\ref{tab:CIGALELSM}.  

\textsc{CIGALE} treats the AGN as a composite object consisting of contributions from three
elements in the context of the Fritz model, namely: 
1) the primarily mid-IR emission arising from the molecular torus, 
2) the emission from the accretion disk in the optical and near-IR, and
3) light scattered from the torus. 
The \textsc{CIGALE} parameter $f_{\textnormal{AGN}}$ is typically used to
mean the ratio of the mid-IR emission from the torus only to the total infrared luminosity (see for example \citetalias{2018MNRAS.480.3562D}). However, our investigations show that the emission from the torus does not accurately account for the total AGN output for some of our galaxies (Fig.~\ref{fig:AGNdefs}). In most cases this  
 makes very little difference, as demonstrated by the near one-to-one correlation between 
 $f_{\textnormal{AGN}}$ (TOTAL)
 hand the fractional contribution arising only from the torus ($f_{\textnormal{AGN}}$ (Torus Only), in  Fig.~\ref{fig:AGNdefs}). Therefore, throughout this work we define $f_{\textnormal{AGN}}$ as the contribution coming from the torus.
We find that nine objects (NGC\,3516, NGC\,5548, ESO\,141-55, Mrk\,771, Mrk\,841, Mrk\,1383, Mrk\,1513,
Mrk\,335, and 2XMM\,J141348.3+440014) are significant outliers of this correlation.
All of them are characterized by an accretion-disk luminosity that exceeds that of the torus, including both the thermal and the scattering components, as calculated by \textsc{CIGALE} using the Fritz model.  They all have good wavelength coverage in their SEDs and
reliable \textsc{CIGALE} fits, with reduced-$\chi^2$ between 1 and 3.  No other \textsc{CIGALE} parameters single out these objects as having high accretion or identify them as unusual in other ways.  
The SFRs in this set, for example, vary from about 30.5\,M$_\odot$\,yr$^{-1}$ (Mrk\,1513) to 
0.04\,M$_\odot$\,yr$^{-1}$ (NGC\,3516).  The most extreme outlier, Mrk\,771, has an accretion luminosity almost five times larger than its torus emission. This object is noted for having soft X-ray excess emission of $0.15\times10^{-11}$\,erg\,cm$^{-2}$\,sec$^{-1}$ \citep{2016A&A...588A..70B}. The excess in the soft band and the high accretion luminosity favors an interpretation in which UV photons from the accretion disk are comptonized by the electrons in the hot plasma (comptonization) as the cause for excess soft emission\citep{2016A&A...588A..70B}.

\begin{figure*}
\includegraphics[width=\textwidth]{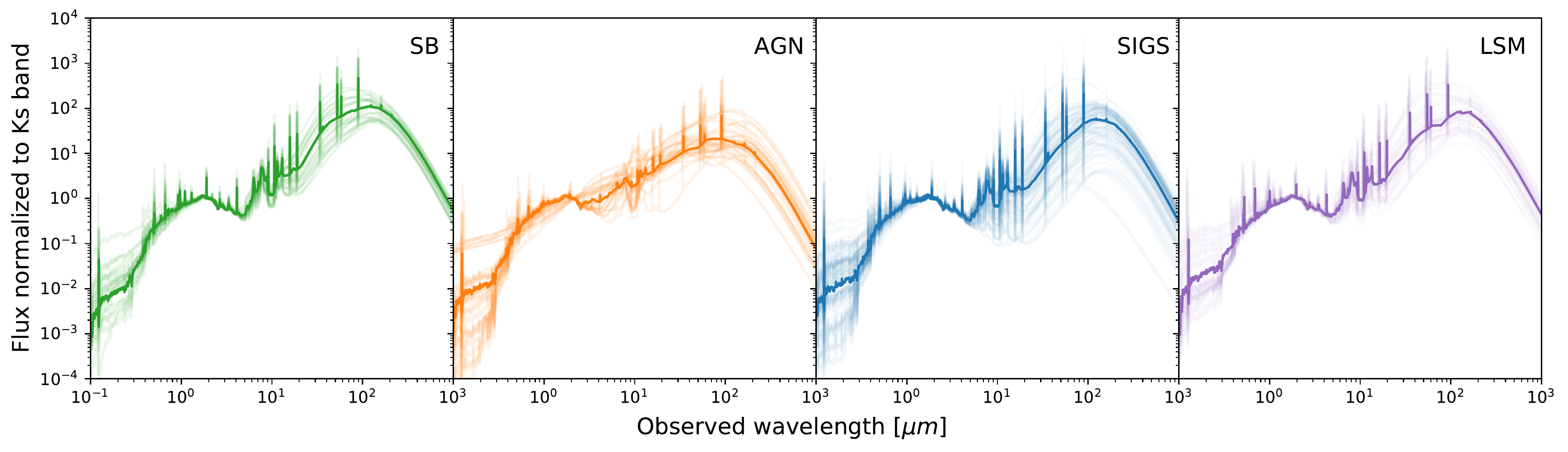}
\caption{Model spectra normalized in the $K_s$ band for all galaxies in this work, separated by sample.  From left to right: the SB sample (in green), the AGN sample (in orange), the SIGS sample (in blue), and finally the LSM sample (in violet).   Faint spectra are those of individual galaxies.  The bold lines indicate the median-averaged spectra for the full subsamples.  This comparison highlights the differences seen among the subsamples
on average, especially those of the AGN sample in the mid-infrared relative to the other samples.}
\label{fig:AVGSED}
\end{figure*}

For purposes of qualitative illustration, we collect the \textsc{CIGALE} model SEDs in Fig.~\ref{fig:AVGSED}.  
We indicate the median-averaged SEDs with bold lines, normalized to their $K_s$ flux densities, 
for each of our four subsamples, together with the most likely fitted SEDs.  
Some aspects of the SEDs are immediately evident.  For example, the SB and LSM subsamples 
show qualitatively similar overall behavior, which suggests that star formation dominates the SEDs of the LSM galaxies.  There is however a weak bump in the median SB and LSM SED at about
50\,$\mu$m of uncertain origin; it may reflect the presence of warm dust.
The AGN sample has a higher median ratio
of NIR to FIR flux than do any of the other samples.  It also, unsurprisingly, has a higher median ratio of MIR to FIR flux, 
reflecting the presence of the hot dust associated with the
AGN component.  Finally, the SIGS sample shows the greatest variety in individual galaxy SEDs.  The latter can be understood as reflecting the much larger variety of star formation activity present throughout the merger sequence as compared with our  other samples.

We calculated IR luminosities ($\rm{L}_{\rm{IR}}$) by integrating the best fitted 
SEDs from 5 to 1000\,$\micron$.  We chose this definition to conform to that 
in \citet{2006MNRAS.366..767F}, to account for PAH features between 5 and 8\,$\micron$,
and to avoid the near-infrared stellar emission that enters into the 1-5\,$\micron$ window. When the contribution of the AGN to the SED drops below about 20\% it becomes increasingly difficult to use the SED to reliability determine the  Fritz parameter values.  

\begin{figure*}
\includegraphics[width=\textwidth]{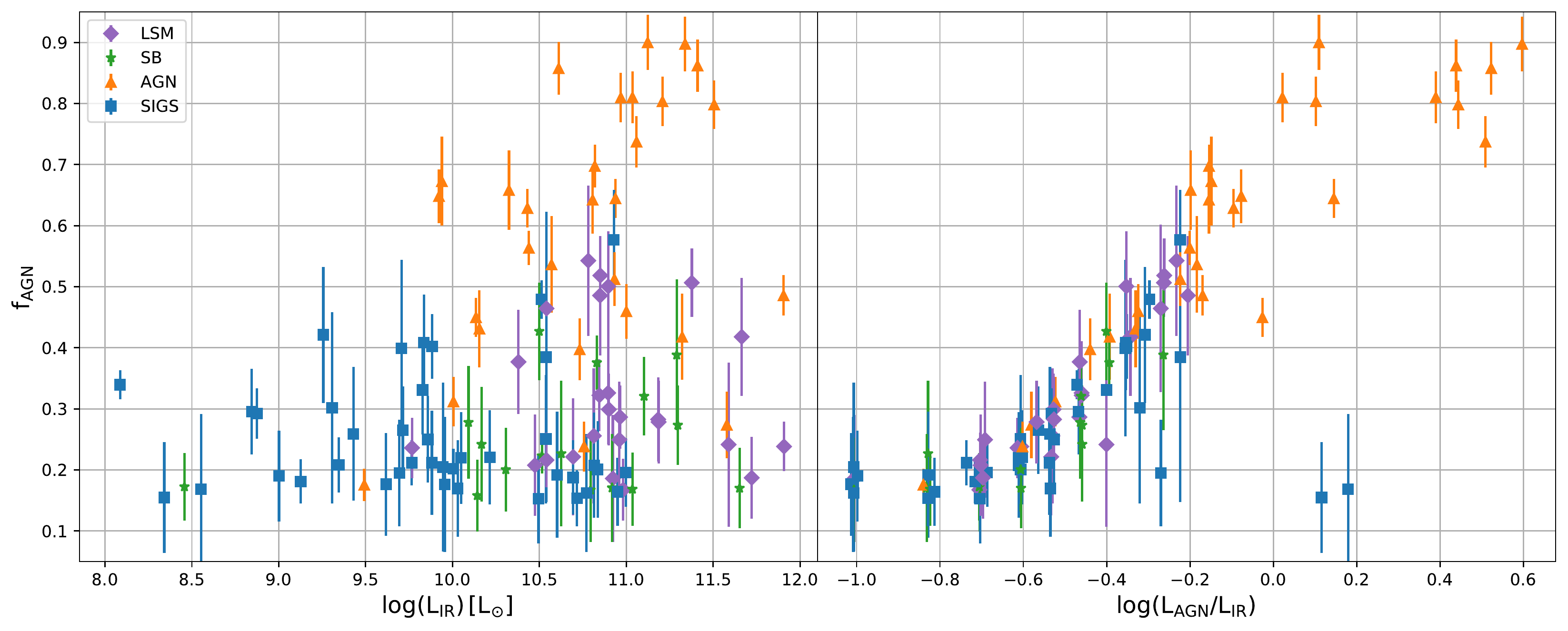}
\caption{\textsc{CIGALE}-estimated total  fractional AGN contribution to the IR luminosity versus \textsc{CIGALE}-estimated total luminosities for the SIGS (blue squares), SB (green stars), AGN (orange triangles) and LSM (violet squares) galaxies.  {\sl Left panel}: AGN fraction as a function of $\rm{L}_{\rm{IR}}$. Unsurprisingly, the AGN sample behavior is distinct from that of the other three subsamples.  {\sl Right panel:} AGN fraction as a function of the ratio of AGN luminosity and $\rm{L}_{\rm{IR}}$.   AGN emission at UV wavelengths can in some cases drive the ratio
of AGN luminosity to IR luminosity to values greater than unity.}
\label{fig:Lbol}
\end{figure*}

To illustrate the effect of low AGN fractions in the accuracy of our results, in Figure~\ref{fig:Lbol} we include all fits with $f_{\rm{AGN}}$ fraction values larger than 15\%, i.e., we include objects that are below the 20\% threshold of what we consider reliable AGN fractions. The uncertainties of these points in the plots are not any larger than those of higher AGN fraction, but a closer look at the SED fits and their reduced-$\chi^2$ in all samples prompt us to use the 20\% cutoff in the remaining figures so that genuine effects can be highlighted.  (A galaxy whose AGN contribution is less than 20\% is then reanalyzed with \textsc{CIGALE} with the AGN parameters set equal to zero, and the other resultant parameter values are the ones listed in the Tables.)
The left panel of Fig.~\ref{fig:Lbol} shows no clear relation between
$f_{\rm{AGN}}$  and $\rm{L}_{\rm{IR}}$ for any of the subsamples considered here, but it does
clearly reveal the tendency of the AGN sample galaxies to host strong and in some cases dominant AGNs. Measurable AGN contributions are only present at IR luminosities above $\sim10^{9.5} L_{\sun}$ .

$f_{\rm{AGN}}$ is plotted as a function of $\rm{L}_{\rm{AGN}}/\rm{L}_{\rm{IR}}$ ($\rm{L}_{\rm{AGN}}$ from \textsc{CIGALE} output) for all modeled galaxies in the right panel of Figure~\ref{fig:Lbol}.  The expected relationship is apparent, and at the smallest values of $f_{\rm{AGN}}$ the flattening confirms our decision to limit further analyses to values exceeding 20\%.  
Toward large ratios, the trend in $f_{\rm{AGN}}$ flattens and becomes more scattered. The flattening is the result of a larger fraction of the AGN luminosity being emitted at UV wavelengths for the brightest AGNs. The scattering, on the other hand, is explained by the different levels of obscuration in each case, related to geometrical (i.e., inclination) effects, and the contribution of the $f_{\rm{AGN}}$ coming only from the torus (Fig.~\ref{fig:Lbol}).

\subsection{Galaxy Properties by Merger Stage}
\label{ssec:properties}

We segregated the LSM+SIGS galaxies 
by merger stage and compared them to the SB and AGN subsets using a Kolmogorov-Smirnov (KS) test to determine how statistically different the derived parameters are between samples. The KS statistic is a measure of how likely it is that two distributions are consistent with being two realizations of the same underlying distribution. The higher the KS probability between two parameter distributions, the more likely it is that they are coming from the same parent distribution.

\citet{2013ApJ...768...90L} reported KS tests on a smaller sample of mergers in the original SIGS program (before the availability of Herschel data) and tentatively did not find statistically significant correlations between SED shape, merger stage, and star-forming properties. With our enlarged study sample, analyzed with an SED code that does take into account the AGN contribution, we are more successful at finding meaningful statistical differences.

\begin{figure*}
\includegraphics[width=0.49\textwidth]{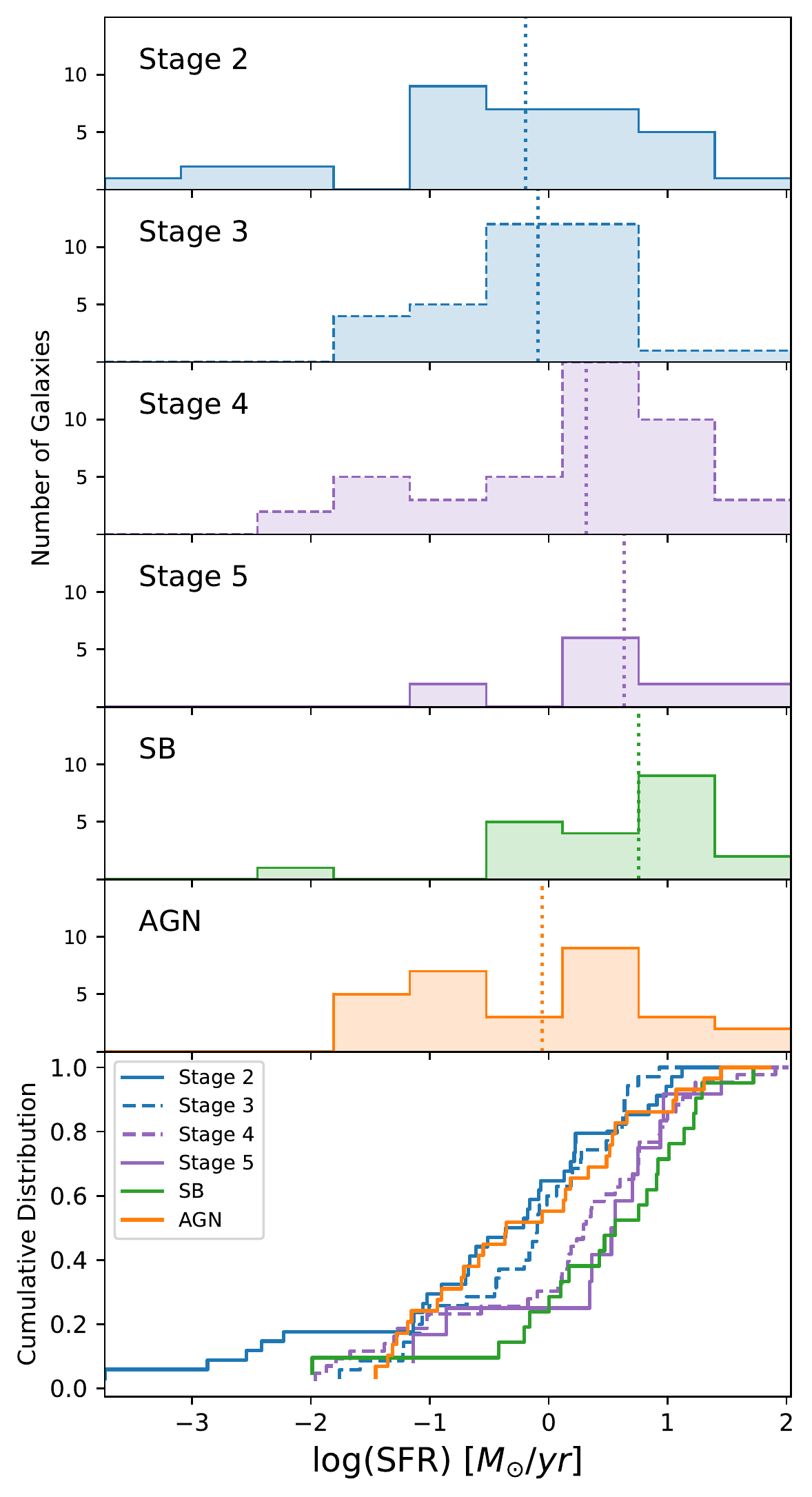}
\includegraphics[width=0.49\textwidth]{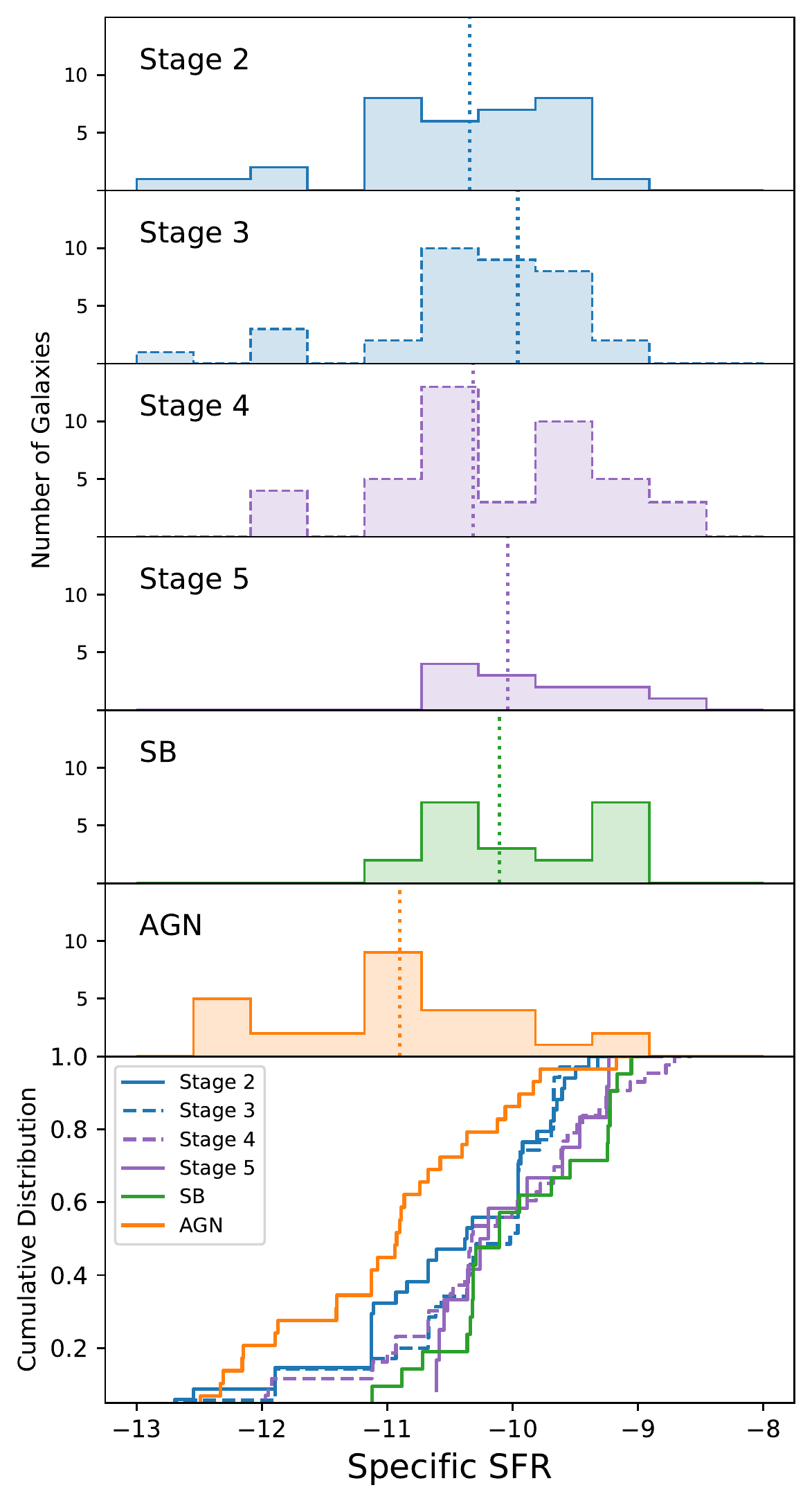}
\caption{Histograms (top) for the stages of the combined SIGS+LSM sample and the normalized cumulative distributions (bottom) of SFR ({\sl Left panel}) and sSFR ({\sl Right panel}). In the histograms we present the stage 2 (blue), 3 (dashed blue), 4 (dashed purple) and 5 (purple) from the SIGS+LSM sample, and the SB (green) and  AGN (orange) samples. The respective median value is represented in the histograms by dotted vertical lines. In the cumulative distributions we present all the previous samples.}
\label{fig:Hist}
\end{figure*}

We summarize the results of this comparison for selected parameters in Table~\ref{tab:KS} and    
present the distributions for SF and SFR in Fig.~\ref{fig:Hist}.  Analogous results for the other derived parameters are presented in Fig.~\ref{fig:Hist2}.

\begin{table*}
\centering
\caption{Results of KS tests comparing fitted parameters by subsample.}
\label{tab:KS}
\begin{tabular}{cccccccccccc}
\hline
Samples & $f_{\rm{AGN}}$ & L(AGN) & E$(B-V)$ & E$(B-V)$ & $\alpha$ & L(dust) & SFR & age(stars) & Mgas & M$_*$ & sSFR \\
Compared & & & (Old Stars) & (Young Stars)& (Dust) \\
\hline
2-3 & 0.44 & 0.42 & 0.95 & 0.96 & 0.24 & 0.88 & 0.91 & 0.62 & 0.59 & 0.92 & 0.24 \\
2-4 & 0.3 & 0.05 & 0.01 & 0.01 & 0.38 & <0.01 & 0.01 & 0.07 & 0.42 & 0.26 & 0.5 \\
2-5 & 0.12 & 0.01 & 0.03 & 0.03 & 0.34 & 0.02 & 0.05 & 0.75 & 0.49 & 0.49 & 0.07 \\
2-AGN & <0.01 & <0.01 & <0.01 & 0.66 & <0.01 & 0.02 & 0.84 & 0.19 & <0.01 & <0.01 & 0.09 \\
3-4 & 0.43 & 0.18 & <0.01 & 0.01 & <0.01 & 0.01 & 0.02 & 0.26 & 0.63 & 0.49 & 0.1 \\
3-5 & 0.21 & 0.02 & 0.03 & 0.08 & 0.06 & 0.05 & 0.17 & 0.8 & 0.22 & 0.22 & 0.14 \\
3-AGN & <0.01 & <0.01 & <0.01 & 0.86 & <0.01 & <0.01 & 0.76 & 0.03 & <0.01 & <0.01 & <0.01 \\
4-5 & 0.61 & 0.42 & 0.51 & 0.97 & 0.84 & 0.83 & 0.98 & 0.92 & 0.51 & 0.51 & 0.38 \\
4-AGN & <0.01 & <0.01 & 0.31 & 0.06 & 0.08 & 0.43 & 0.10 & 0.01 & 0.01 & 0.04 & <0.01 \\
5-AGN & <0.01 & 0.02 & 0.56 & 0.15 & 0.38 & 0.56 & 0.15 & 0.17 & 0.07 & 0.07 & <0.01 \\
SB-2 & 0.31 & 0.03 & 0.01 & 0.01 & 0.11 & <0.01 & 0.01 & 0.2 & 0.43 & 0.41 & 0.05 \\
SB-3 & 0.61 & 0.05 & <0.01 & 0.01 & <0.01 & <0.01 & 0.02 & 0.44 & 0.35 & 0.35 & 0.1 \\
SB-4 & >0.99 & 0.85 & 0.22 & 0.43 & 0.42 & 0.43 & 0.43 & 0.76 & 0.98 & 0.99 & 0.18 \\
SB-5 & 0.53 & 0.38 & 0.08 & 0.64 & 0.93 & 0.26 & 0.48 & 0.85 & 0.58 & 0.64 & 0.96 \\
SB-AGN & <0.01 & <0.01 & 0.07 & 0.01 & 0.60 & 0.07 & 0.04 & 0.03 & 0.04 & 0.09 & <0.01 \\
noAGN-AGN & <0.01 & 0.01 & 0.18 & 0.03 & <0.01 & 0.28 & 0.04 & 0.01 & 0.02 & 0.02 & <0.01 \\
\hline
\end{tabular}	

\begin{flushleft}
{\bf Note.} KS-derived probabilities indicating the likelihood that the \textsc{CIGALE}-derived 
parameter distributions (column headings) for two galaxy subsamples were drawn from the same parent sample.  The left-hand column indicates the two samples tested.
For this test, the SIGS and LSM samples were combined and then divided into merger-stage-based
 subsamples.  The numbers in the left column indicate the 
merger stage used.  The 9th column [age(stars)] refers to the mean age of the stellar
population.  Because the subsamples were relatively small (e.g., 29 AGN galaxies), 
we made no attempt to refine probabilities below 1\% ($>99$\% confidence that the two
samples differ).
\end{flushleft}
\end{table*}

The KS tests reveal a number of trends:
\begin{itemize}
    \item First, the parameter distributions of advanced merger stages (4,5), especially $f_{\rm{AGN}}$, $M_*$, and $\alpha_{\rm{dust}}$, are most similar to those of the starburst sample, and are less statistically correlated with the parameters of the AGN sample.
    \item Second, KS scores for consecutive stages (2-3, 3-4, 5-6) are higher than for non-consecutive stages, with the smallest correlation occurring for stages that are farther apart along the merger evolution (e.g. 2-5), which is expected if the properties of the system evolve gradually as the merger progresses.
    \item Third, apart from their dust luminosities and $E(B-V)$ values, the AGN sample parameters have a very small statistical correlation with the parameters of any other samples, but the KS scores are slightly larger between the AGN galaxies and the advanced merger stages than between AGN and the early merger stages.
    \item Fourth, lowest statistical correlation occurs between the SB and the AGN samples, even for parameters that tend to be correlated between all the other samples.
    \item Finally, the parameters that show more dispersion between samples are $f_{\rm{AGN}}$ and $\alpha_{\rm{dust}}$, which implies that they are the most useful parameters to discriminate between galaxy types.
\end{itemize}

The picture that emerges from these results (and from the overall SED shapes in Fig.~\ref{fig:AVGSED}) is in agreement with a classical interpretation: in the local Universe, mergers trigger starburst activity in galaxies \citep{2006ApJS..163....1H,2019A&A...631A..51P}. They also trigger AGN activity, but to a lesser extent \citep{2006ApJS..163....1H,2017ApJ...848..126S,2018MNRAS.478.3056B,2018MNRAS.480.3562D}. 

\subsection{Dust spectral slope ({$\alpha$}) and the star formation main sequence}
\label{ssec:alpha}

\begin{figure*}
\includegraphics[width=\textwidth]{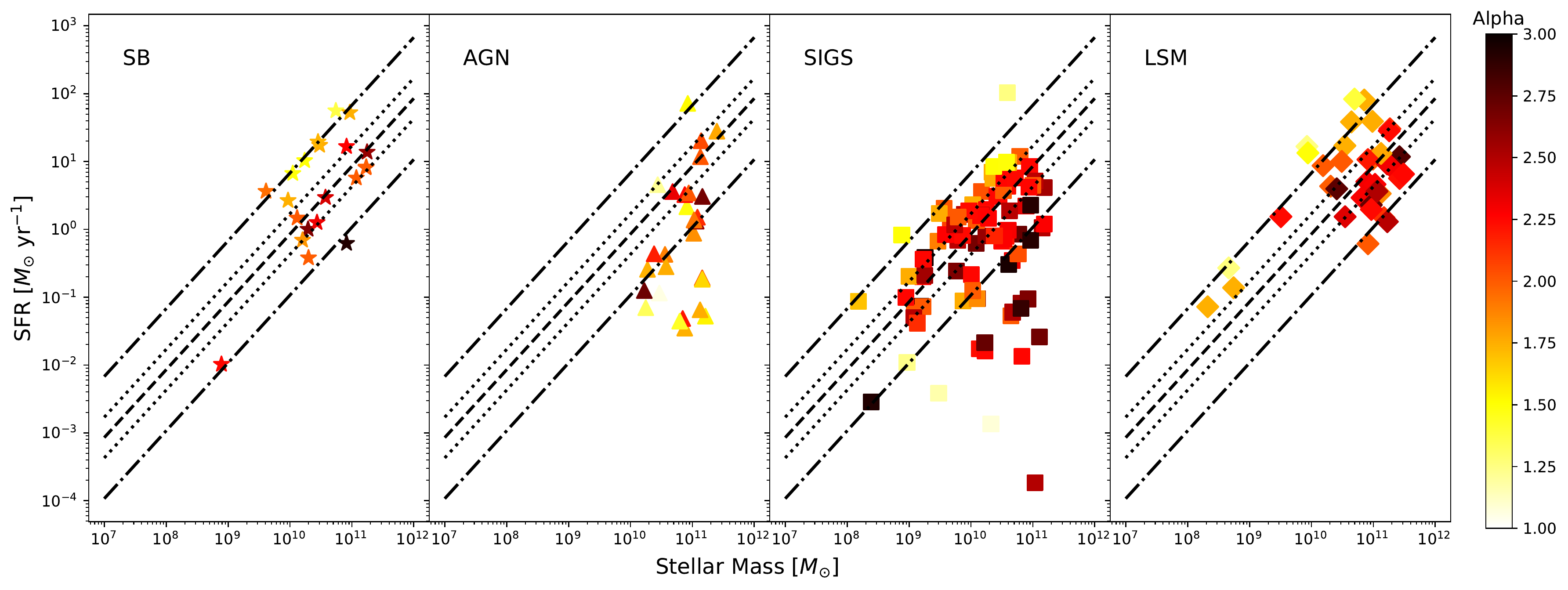}
\caption{SFR as a function of stellar mass with selected galaxies color-coded by the derived parameter $\alpha$ \citep{2014ApJ...784...83D} obtained from \textsc{CIGALE}. The dashed, dotted, and dashed-dotted lines  
in every panel indicate respectively the $z = 0$ MS and the 0.3 and 0.9\,dex limits above and below it.  We compare the SB ({\sl left panel}), AGN ({\sl middle-left panel}), SIGS ({\sl middle-right panel}) and LSM samples ({\sl right panel}).  All the SB galaxies follow the MS and the gradient of $\alpha$ seems to agree with \citet{2016ApJ...817...76M} with the parameter $C$, as we discuss in Section~\ref{ssec:alpha}.  As expected, the AGN sample galaxies do not follow the MS, and no trend with $\alpha$ is evident.  The SIGS sample shows a large scatter across the MS and some SIGS galaxies appear in the region where SF is quenched.}
\label{fig:MS1}
\end{figure*}

Fig.~\ref{fig:MS1} shows how galaxies from the four samples populate the $M_*-\rm{SFR}$ plane, using $M_*$ and $\rm{SFR}$ results from \textsc{CIGALE}. The so-called star formation main sequence (MS) is usually defined in terms of a positive correlation followed by star-forming galaxies between  star formation rate and stellar mass. 
Both Figs.~\ref{fig:Hist} and~\ref{fig:MS1} show that the SB and stage 5 galaxies indeed lie in a narrow and relatively high range of SFRs. 
We have color coded the symbols according to their \textsc{CIGALE}-derived estimates for $\alpha$, the exponent of the power law defined in Eq.~\ref{eq:alpha}, which parametrizes the average dust temperature.  
For the same range of masses, the AGN and SIGS samples extend to lower ($<10^{-1}M_\odot$\,yr$^{-1}$) SFRs compared to the other two samples. 

The bulk of the SIGS galaxies follows the MS over three orders of magnitude in stellar mass, and over four orders of magnitude in SFR. There are some outliers at low SFR, consistent with these being quiescent galaxies. The LSM and SB galaxies also seem to follow the MS, but they are more massive, than the SIGS sample and consequently show higher values of SFR. The AGNs in our sample have masses limited to a narrow range between $10^{10}~M_{\odot}$ and $4\times 10^{11}~M_{\odot}$ and a broad range of SFRs. 

The behavior of the AGN galaxies is notable especially when we consider dust temperatures as parametrized by $\alpha$. For all the other samples, dust temperature positively correlates with sSFR, that is, for a given stellar mass, dust becomes hotter (alpha decreases) as SFR increases. Using simulations and observations of SIGS galaxies, \citet{2016ApJ...817...76M} have shown that this evolution of the dust temperature as galaxies depart the MS is linked to the interaction stage: initially (at early interaction stages) galaxies have low SFRs and relatively cool dust temperatures, but SFR and dust temperature 
both increase as the systems approach coalescence.  This is related to an increase in the \textit{compactness} of the ISM, i.e, the average distance of the dust to the heating sources, normalized by the luminosity of the source. 

We observe a similar evolution of dust temperature with distance from the MS for the SB and LSM samples. The AGN sample, however, is different.  For AGN sample galaxies the $\alpha$ parameter is completely uncorrelated with the location of the system relative to the MS, and moreover, SEDs compatible with hot dust are observed at very low SFRs.  
The average dust temperature is therefore not controlled by star formation in AGN-dominated galaxies, and the concept of compactness should be interpreted from a different perspective for these systems.  

Perhaps more relevant here is that by disentangling AGN and SF activity
we can obtain more reliable SFR estimates for these systems unbiased by the thermal emission from the AGN. 
We also corroborate that mergers can be an important factor in contributing to the scatter of the MS, since galaxies move away from the MS as they evolve into later phases of the merger. One additional note has to do with quenching. Although the SIGS galaxies lying below the MS 
(as indicated by the dot-dashed lines in Fig.~\ref{fig:MS1}) are most likely quenched systems with small or negligible gas reservoirs, some of the galaxies that we would infer to be actively forming stars might actually be recently quenched systems where the stars formed right before the quenching are still dust-enshrouded, as demonstrated in \citet{2014MNRAS.445.1598H}. Additionally, the fact that we find AGN systems below the MS suggests  AGN activity persists after the quenching, even at very low levels of SF. 

\begin{figure}
\includegraphics[width=\columnwidth]{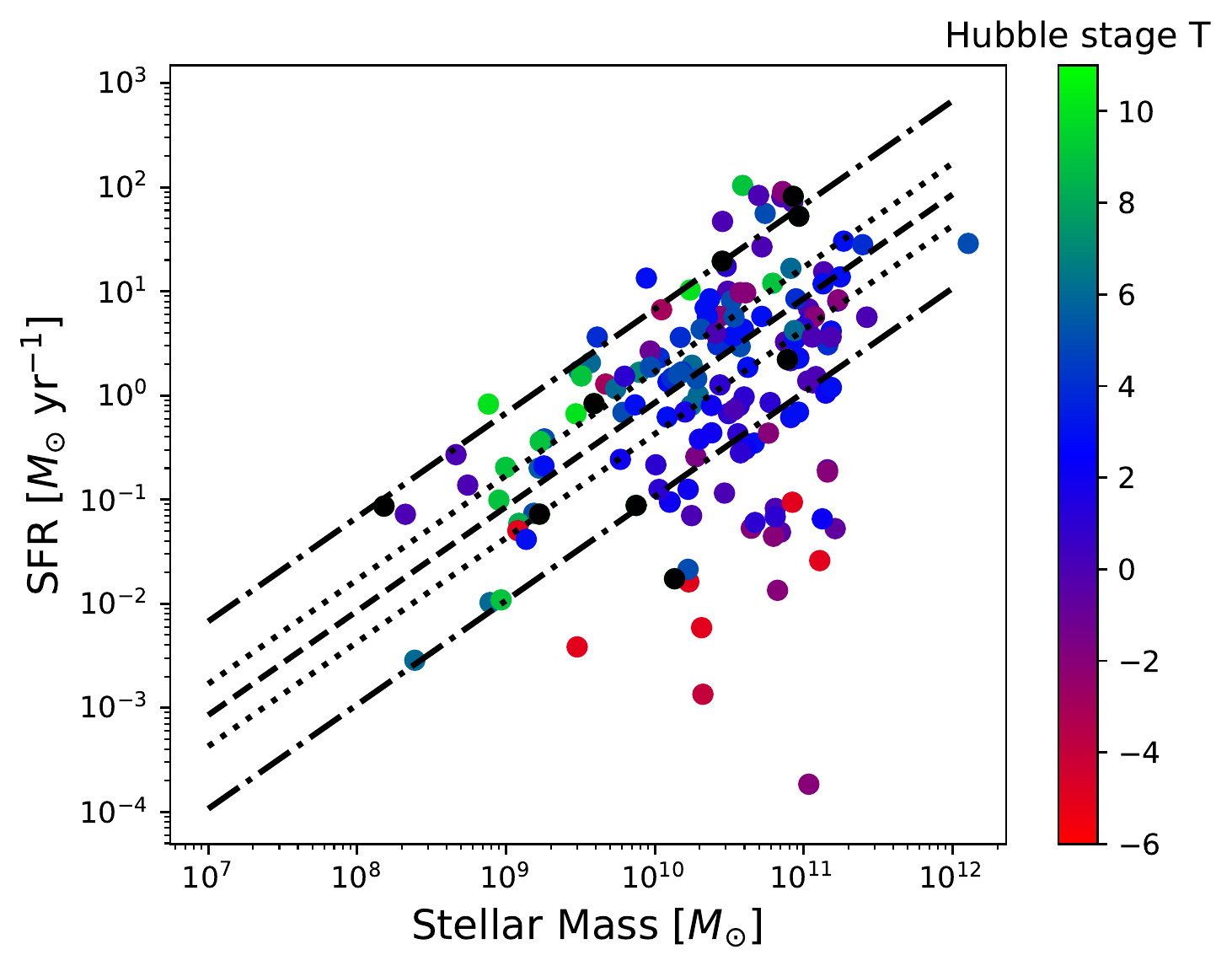}
\caption{Main sequence of star-forming galaxies showing the galaxies in this work, but color-coded by Hubble stage (T). Peculiar galaxies outside of range of colors shown for T in the color bar are plotted in black. The dashed lines are 
identical to those in Figure~\ref{fig:MS1}.  Early types (elliptical and lenticular galaxies; T$<0$) tend to have low  SFRs while the spirals follow the MS.  Irregular galaxies  scatter over in several order of magnitude in SFR for a given stellar mass.}
\label{fig:MS3}
\end{figure}

Figure~\ref{fig:MS3} shows how galaxy morphology affects location within the star-forming main sequence by using the method of  \citet{1991rc3..book.....D}, which assigns numerical values to the Hubble stages \citep[usually called T values,][]{1977egsp.conf...43D}. 
We were unable to classify four galaxies (2XMM\,J141348.3+440014, 4U\,0557-385, LEDA\,68751, Mrk\,1383), which was unsurprising given that these are among the most distant galaxies in this work.  
For three galaxies with morphologies not available in
\citet{1991rc3..book.....D} we use the NASA/IPAC Extragalactic Database (NED). The three galaxies classified using NED are ESO\,033-02 (a SB0; T$=-3$), Mrk\,841 as (E; T$=-5$), and MCG-03-34-064  (S0/a and S0+ are both given; we 
assigned it T$=-0.5$).  
Not all galaxies have T values in the range from $-6$ to 11. Six are classified as Non-Magellanic irregulars (NGC\,2820A, NGC\,2968, NGC\,3034, NGC\,3077, NGC\,3448, M51B) and three as peculiar irregulars (NGC\,2623, NGC\,3256, NGC\,520). 
For close mergers (NGC\,4038/4039 and NGC\,5614/5615) we only use the information of the most prominent galaxy.  
We were unable to determine morphologies for seven galaxies with NED or a literature search; these object were ESO\,141-55, ESO\,383-35, Mrk\,1502, Mrk\,1513, Mrk\,335, Mrk\,771, and IC\,694).  
They do not appear in Figure~\ref{fig:MS3}.

Unsurprisingly, most of early type galaxies lie in the quiescent region (below the lowest diagonal line) of the MS diagram, while most of the spiral galaxies follow the MS.
No obvious trend is apparent for the irregular galaxies, which scatter widely in SFR for a given stellar mass.
This is in harmony with the demographics of 
disk-dominated (``blue cloud'') and spheroid-dominated (``red sequence'') systems \citep{2003MNRAS.341...54K,2004MNRAS.351.1151B,2015ARA&A..53...51S}. 

The emission from warm dust is an essential contributor to the SED in most stages of mergers. \citet{2016ApJ...817...76M} found that the  compactness parameter $C$ that relates the distribution of dust temperatures with the geometry of the environment is correlated with the sSFR. The position of our galaxies in the MS, (Figure~\ref{fig:MS1} and~\ref{fig:MS3}) supports a picture in which the dust within star-forming
galaxies evolves as those galaxies evolve and transform their morphologies. 

\subsection{Comparisons with other AGN indicators}

In this Section we compare our fitted $f_{\textnormal{AGN}}$ and other \textsc{CIGALE}-derived
AGN-related parameters to widely used AGN indicators.

\subsubsection{IRAS 60/25 $\micron$ and Neon emission-line ratios} \label{sec:IRASNeon}

\begin{figure}
\includegraphics[width=\columnwidth]{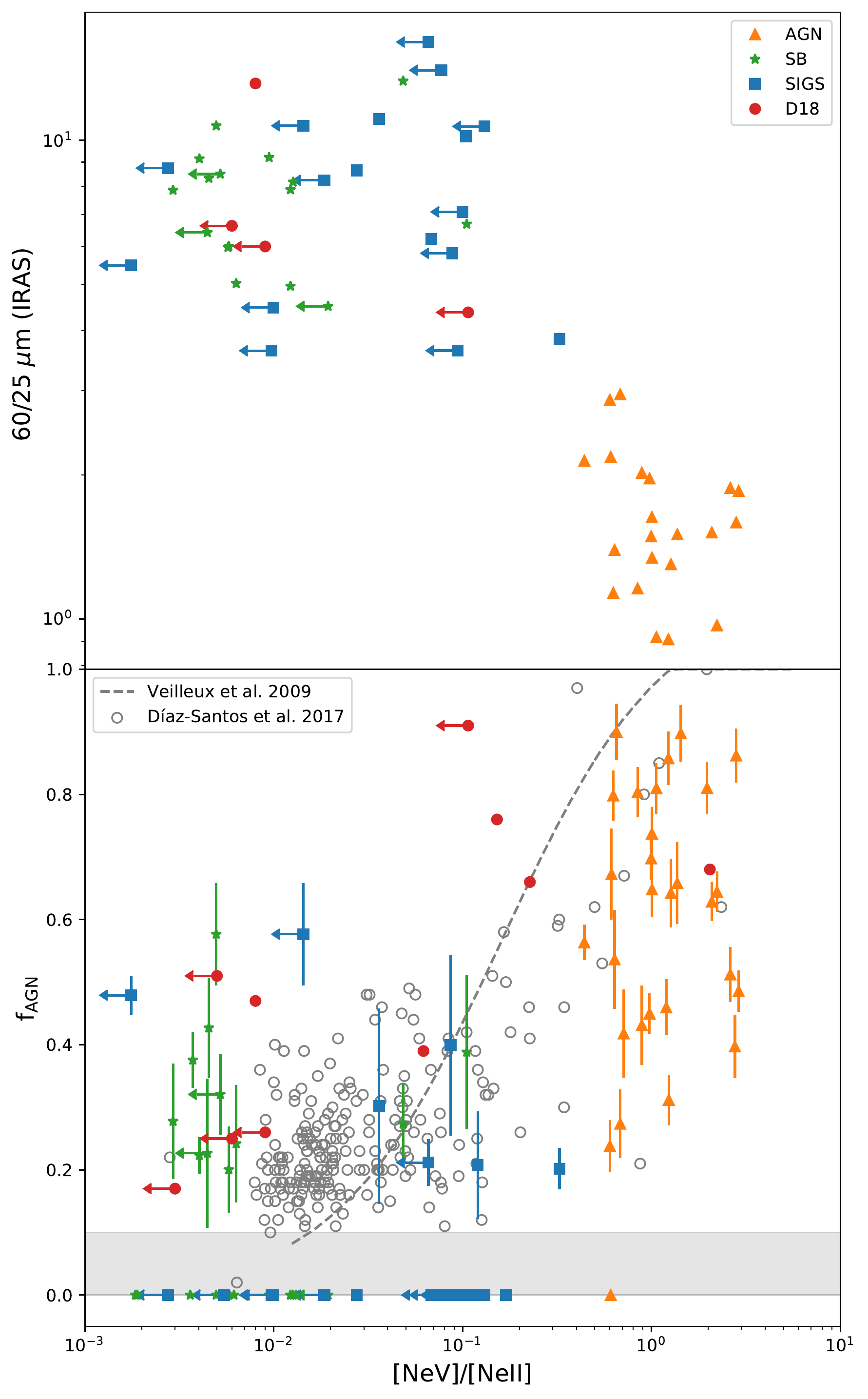}
\caption{{\sl Top panel:} The ratio of integrated IRAS flux densities at 60 and 25 $\micron$ $f_{60}/f_{25}$ as a function of the [\ion{Ne}{v}]/[\ion{Ne}{ii}] integrated line intensity ratio. The symbols are the same as in Figure~\ref{fig:Lbol} with the addition of values from \citetalias{2018MNRAS.480.3562D}.  Upper limits are indicated with arrows.   
{\sl Bottom panel:} Total AGN fraction as a function of the [\ion{Ne}{v}]/[\ion{Ne}{ii}] integrated line intensity ratios. The markers are the same as in the top panel, with the errorbars on $f_{\rm AGN}$ indicating
the uncertainty estimates taken from \textsc{CIGALE}. We added the estimations from \citet{2009ApJS..182..628V} (dashed line) and \citet{2017ApJ...846...32D} (empty gray circles) to compare with literature results. The grey region indicates where the $f_{\rm AGN}$ estimates are not believed to be reliable; we treat objects in this region as if
they have $f_{\rm AGN}=0$.  In both panels the SB and AGN samples are well-separated. The SIGS galaxies tend not to have significant AGN contributions.}
\label{fig:FAGN}
\end{figure}

The IRAS 60~$\micron$ to 25~$\micron$ flux ratio $f_{60}/f_{25}$ is an indicator of hot dust content \citep{2011ApJS..195...17W}, and thus suggestive of the strength of the AGN relative to ongoing 
star formation because dust in the AGN torus is on average hotter than in star formation regions. 
In combination with mid-infrared emission-line ratios, $f_{60}/f_{25}$ can be quite effective at separating AGNs from starbursts \citep{2008ApJ...676..836T,2009ApJS..182..628V,2010ApJ...709.1257T}.  In the top panel of Fig.~\ref{fig:FAGN}  the $f_{60}/f_{25}$ ratio is plotted as a function of the [\ion{Ne}{v}]/[\ion{Ne}{ii}] integrated intensity ratio for our galaxies.   
There is a clear separation between star formation-dominated galaxies ([\ion{Ne}{v}]/[\ion{Ne}{ii}]$~\lesssim 0.7$, $f_{60}/f_{25} \gtrsim 3$) and AGNs, in agreement with \citet{2010ApJ...709.1257T} and \citet{2013RMxAA..49..301H}. The result, together with Fig.~\ref{fig:AVGSED}, supports the common assumption that the shape of the mid-IR continuum of galaxies with significant emission from AGNs can be approximated by a power-law and thus that the continuum in this region is a good discriminator between galaxies with and without strong AGN emission \citep{2006ApJ...653.1129B,2008ApJ...676..836T,2011ApJ...740...99D}. \citet[figures 24-26]{2009ApJS..182..628V} found a similar relationship with the analogous [O IV]/[\ion{Ne}{ii}] ratio and less dramatically with the [\ion{Ne}{v}]/[\ion{Ne}{ii}] ratio. They find a progression of these line ratios from low to high starting with star-forming galaxies, followed by Seyferts 2, Seyferts 1, ULIRGs, and finally QSOs.

The SIGS sample unfortunately only possesses a few published measurements or upper limits for these lines. Most of them have ([\ion{Ne}{v}]/[\ion{Ne}{ii}]$~\lesssim 0.1$ and $f_{\rm AGN} < 0.2$.  We would expect the interacting systems in the SIGS sample to have a different fraction of AGN emission as they move from early on in the interaction towards the coalescence phase, to fall in between the two regimes of [\ion{Ne}{v}]/[\ion{Ne}{ii}] ratio presented here.

We observe that those SIGS galaxies for which we were able to collect line emission and that are not upper limits, do fall in between the SB and the AGN galaxies, with intermediate cases. The AGN galaxies with the highest {\sl IRAS} ratio (cooler dust) are NGC\,3281, ESO428-14, NGC\,4941, NGC\,4388 and NGC\,7674, and the SB with the most elevated [\ion{Ne}{v}]/[\ion{Ne}{ii}] are NGC\,2623, NGC\,1365, NGC\,4088, NGC\,4194 and NGC\,4676. There is one upper limit published for a LSM galaxy (2MASX\,J10591815+2432343, with [\ion{Ne}{v}]/[\ion{Ne}{ii}]$<0.02$), so we do not include that galaxy in Fig.~\ref{fig:FAGN}.

\subsubsection{Comparing galaxy parameters with their emission line ratios}

The fine structure Ne-lines can help us discriminate between the SB and AGN samples as well as with the SIGS galaxies because they flag the presence of an AGN even in galaxies otherwise classified as a star-forming \citep{2008ApJ...678..686A}. One of the most useful outputs of \textsc{CIGALE} is the fraction of AGN derived from the SED (Sect.~\ref{ssec:SED}). The bottom panel of Figure~\ref{fig:FAGN} offers confirmation of the  \textsc{CIGALE}  estimated AGN contributions, compared with other estimations not using SED from \citet[table 12]{2009ApJS..182..628V} and \citet[table 2, column 7]{2017ApJ...846...32D}. We can see that the AGN sample and SB sample separate very well in Figure 9. However, the estimated AGN fraction can in some cases have a value near to 40\% in the SB sample galaxies (NGC\,1365 and NGC\,660). In the same way, three AGN galaxies have a AGN fraction below 0.3 (NGC\,4941, NGC\,7674 and NGC\,4388).

Most of the SIGS sample galaxies have weak or no AGN contributions \citep{2013ApJ...768...90L}. A particular outlier of this behaviour is NGC\,3034, with the highest $f_{\textnormal{AGN}}=0.48 \pm 0.03$ in the sample but yet a very low [\ion{Ne}{v}]/[\ion{Ne}{ii}]. \citet{2013ApJ...768...90L} show that this galaxy was very difficult to fit with MAGPHYS \citep{2008MNRAS.388.1595D}, possibly due to the high obscuration, the presence of an outflow, or some other unaccounted for activity. We were also unable to obtain a good fit with \textsc{CIGALE} except when we include the AGN component model; in that case our  reduced-$\chi^2$ was 3.29, which is low enough to be considered a reliable fit. Therefore, even when strong star formation is present, $f_{\textnormal{AGN}}$ estimates with \textsc{CIGALE} could reveal a hidden AGN, that is invisible in optical wavelengths. This estimate can only be checked by the emission spectra of high IP lines like [\ion{Ne}{v}].

\subsubsection{Infrared color diagnostics}

\begin{figure*}
\includegraphics[width=\columnwidth]{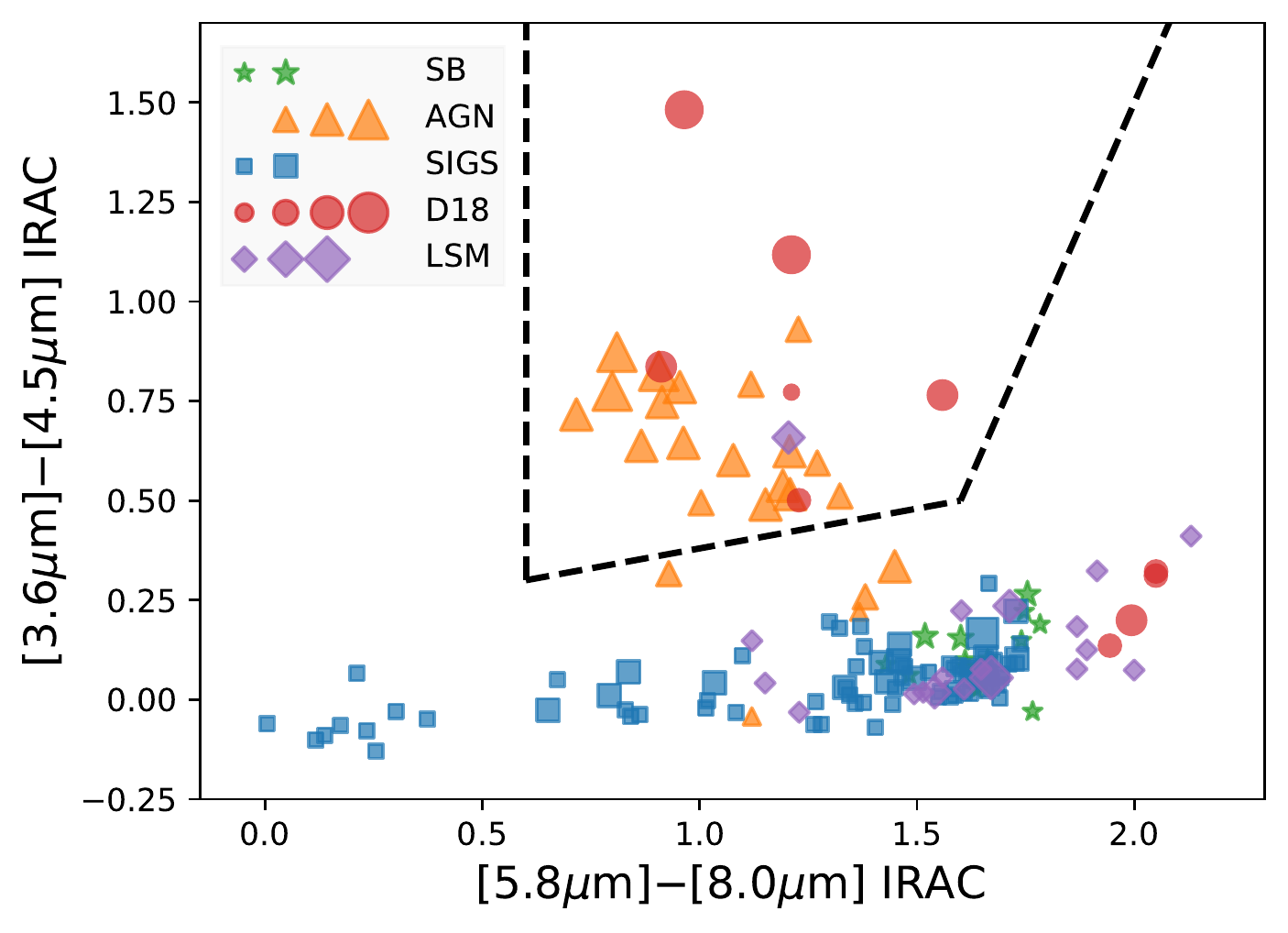}
\includegraphics[width=\columnwidth]{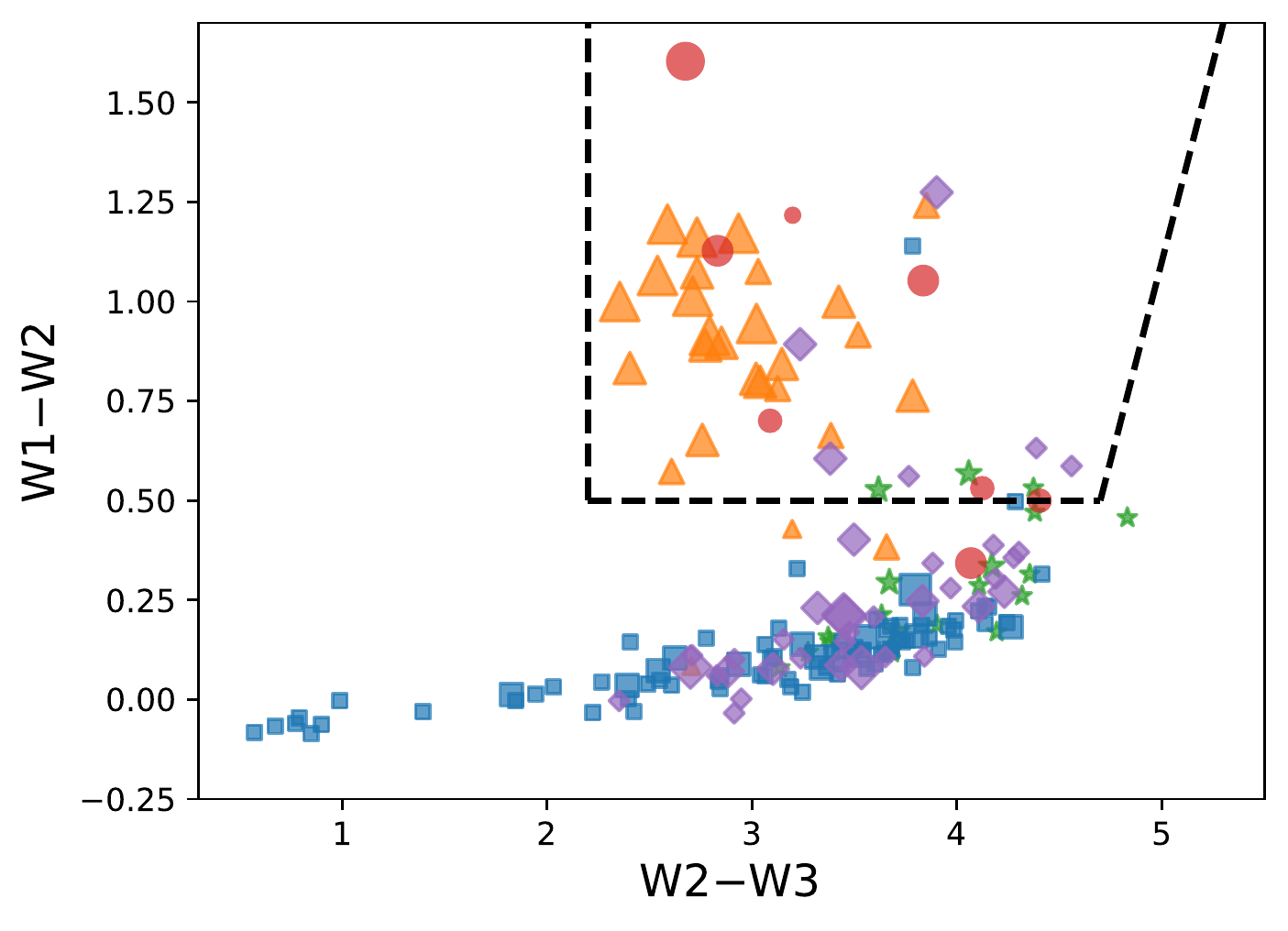}
\caption{ Mid-infrared color-color classification diagrams showing the SIGS (blue squares), SB (green stars), LSM (purple diamonds) and AGN (orange triangles) galaxies together with late-stage mergers (red circles) from \citetalias{2018MNRAS.480.3562D}. The markers sizes are divided in four ranges for the $f_{\rm AGN}$ $<$0.25,0.25-0.5,0.5-0.75 and $>$0.75, where not all the samples cover the four ranges. {\sl Left panel:} $[5.8]-[8.0]$ vs. $[3.6]-[4.5]$ color-color diagram following \citet{2005ApJ...631..163S} for all sample galaxies with available IRAC photometry. Sources in the region enclosed by the dashed lines (commonly referred to as the Stern wedge) are dominated by AGN emission in the mid-infrared, while sources below this region are dominated by star-formation. {\sl Right panel:} $W2-W3$ vs. $W1-W2$ color-color diagram, with an AGN wedge (dashed lines) per \citet{2018MNRAS.478.3056B}. Most late-stage mergers and AGN fall in this wedge with some galaxies from the SIGS and SB sample. The separation of these wedges with the $f_{\rm AGN}$ is in agreement with the expected behaviour of AGN activity onset.}
\label{fig:stern}
\end{figure*}

Infrared colors are well-known diagnostics of the energy sources powering infrared-luminous galaxies; two salient examples are the color-color diagrams developed by 
\citet{2005ApJ...631..163S} and \citet{2004ApJS..154..166L} 
to discriminate between galaxies dominated by star formation and AGN emission. More strict criteria can be applied \citep[e.g.][]{2012ApJ...748..142D}, but they depend on other factors, as luminosity. We can use our more precise AGN and SFR measurements to test the reliability of these diagrams.
Fig.~\ref{fig:stern} shows {\sl Spitzer}/IRAC and {\sl WISE} color-color diagrams for our galaxies.  Galaxies that lie within the wedge enclosed by the dotted lines in left panel 
are expected to be AGN-dominated. Only five of the 29 galaxies from our AGN sample (ESO428-14, NGC\,3516, NGC\,4941, NGC\,4388, IC5063)  lie outside the wedge.  This is not unexpected: \citet{2011ApJ...730...28P} find that faint AGNs with measurable PAH 6.2\,$\micron$ EW fall outside the wedge.  

AGN- and star formation-dominated galaxies are also efficiently segregated in a complementary
 {\sl WISE} color-color diagram (Fig.~\ref{fig:stern}, right panel).  All but three of our
AGN sample galaxies lie in the AGN selection region described by 
 \citet{2018MNRAS.478.3056B};  two of the outliers are located 
close to the boundary.  However, some SB populate the AGN wedge in this plot.   
We include the late-stage mergers from \citepalias{2018MNRAS.480.3562D} in Fig.~\ref{fig:stern}.
Many of the \citepalias{2018MNRAS.480.3562D} late-stage mergers have weak or undetected AGNs; they appear to populate both the AGN- and star formation-dominated regions of both
panels of Fig.~\ref{fig:stern}.  

\citet{2012ApJ...753...30S} proposed that  {\sl WISE}  $W1-W2\ge0.7$ colour is a robust indicator of AGN emission. The majority (22) of our 29 AGN sample galaxies meet this criterion.
A less conservative $W1-W2\geq0.5$ colour cut \citep{2009ApJ...701..428A}
is similar to the lower boundary of the wedge in the left panel of Fig.~\ref{fig:stern}, and is identical to that shown in the right panel. The two AGN galaxies which 
fail to satisfy the less stringent criterion are ESO\,428-14 and NGC\,4388. Our most-likely
\textsc{CIGALE} models for them yield total AGN fractions of  
$f_{\rm{AGN}}=0.31\pm0.04$ and $0.24\pm0.04$ respectively, which are significant although not large enough to make them AGN-dominated, and help indicate the reliability limits of these diagrams. 

NGC\,4941 is the only AGN having a blue $W1-W2$ colour comparable to those of the SIGS galaxies, with $f_{\rm{AGN}}$<20\% yet also possessing 
a high [\ion{Ne}{v}]/[\ion{Ne}{ii}] ratio (Sec.~\ref{sec:IRASNeon}). Its blue mid-IR color and high Ne line ratio is consistent with NGC\,4941 being 
a heavily absorbed low-luminosity AGN \citep{2013ApJ...770..157K} and illustrates 
how the [\ion{Ne}{v}] emission can help identify the AGN contribution in highly obscured cases. 
Overall, we confirm that mid-IR color diagnostics in general do identify AGNs, 
and with $f_{\rm{AGN}}$ we quantify their contribution to the total galaxy output. 

Some of the SB and SIGS galaxies lie close to the AGN wedge in the right-hand panel of 
Fig.~\ref{fig:stern}. We decided to test whether their location in Fig.~\ref{fig:stern} could be interpreted straightforwardly
to mean that they are composite objects hosting significant AGN and star formation but not
necessarily dominated by either.  We examined galaxies having intermediate {\sl WISE} colours $W1 - W2 > 0.3$.   
This included five SIGS galaxies (NGC\,838, NGC\,839, NGC\,3034, NGC\,3227 and 
NGC\,3690) 
and eight SB galaxies (NGC\,660, NGC\,1222, NGC\,1365, NGC\,1614, NGC\,2146, NGC\,2623, NGC\,3256 and NGC\,4194).

We consider the SIGS galaxies first.  For NGC\,838 and 839 the most likely \textsc{CIGALE} fits give $f_{\rm{AGN}}\sim 0.0$ for both galaxies.  \citetalias{2015ApJS..218....6B} classify NGC\,839 as a low-ionization nuclear emission-line region \citep[LINER, e.g.,][]{2006MNRAS.372..961K}. The most likely $f_{\rm{AGN}}$ estimate for NGC\,3690 is 0.0, but \citetalias{2018MNRAS.480.3562D} notice different classification for this galaxy, as LINER, AGN and star-forming. The optical spectroscopic classifications \citep{2010ApJ...725.2270P}
for NGC\,3034 and 3227 are HII and Seyfert 1, these galaxies have an SED that is consistent with an AGN contribution of $f_{\rm{AGN}}= 0.48\pm0.03$ and $0.20\pm0.03$, respectively.

Next we consider the SB galaxies. 
For NGC\,660 we estimated$f_{\rm{AGN}}= 0.43\pm0.08$.
This object is usually classified as star-forming \citep{2011ApJ...730...28P,2011ApJS..195...17W}, and there are also signs of interaction.  
NGC\,1365 has one of the highest [\ion{Ne}{v}]/[\ion{Ne}{ii}] ratios in the SB sample.  Its  calculated $f_{\rm{AGN}}$ estimate is 0.39. \cite{2015ApJ...803..109H} obtained 0.6.  
NGC\,1614  has $f_{\rm{AGN}}$=0.0;  \citet{2015ApJ...803..109H} obtained 0.3.  This object is classified as an ``uncertain AGN'' by \citet{2014MNRAS.439.1648A}.  
NGC\,1222, NGC\,2146 and NGC\,3256 have $f_{\rm{AGN}}\sim0.0$, 0.37 and 0.0, respectively. 
Finally, NGC\,4194 has a high ratio of [\ion{Ne}{v}]/[\ion{Ne}{ii}] and seems to be undergoing 
a merger, but the \textsc{CIGALE} results show a $f_{\rm{AGN}}=0.0$, 
consistent with there being no AGN contribution.  It is nonetheless classified as a Seyfert 2 galaxy \citep{2010ApJ...725.2270P}.

Lastly, for NGC\,2623 the [\ion{Ne}{v}]/[\ion{Ne}{ii}] ratio is among the highest in the SB sample (see Sect.~\ref{ssec:LSM}),  with $f_{\rm{AGN}}=0.27\pm0.06$. Unfortunately, we obtained a relatively poor fit to the SED of this source, so our \textsc{CIGALE} model for it may be unreliable.  It has been classified as starburst \citep[e.g.][]{2011ApJS..195...17W}, composite \citep{2014MNRAS.439.1648A} and AGN \citep[e.g.][]{2015ApJ...803..109H} where a fraction of AGN of 0.44 is obtained, comparable to the AGN fraction of $0.39\pm0.05$ from \citetalias{2018MNRAS.480.3562D}.  These are intermediate cases in which both an AGN presence and also intense star-formation are underway, and color alone is insufficient to categorize the source unambiguously.

The last sample we examine in Figure~\ref{fig:stern} is the LSM. Twelve galaxies are above the cut at $W1 - W2 > 0.3$ and six of them with values of $W1 - W2 > 0.5$. Galaxies 2MASX\,J01221811+0100262 and 2MASX\,J08434495+3549421 have the highest $f_{\rm{AGN}}$ of the LSM sample, $0.42\pm0.10$ and $0.48\pm0.10$ respectively. The former was found to be a tentative dual AGN with mixed signs of star-formation by \citet{2017ApJ...848..126S}. The latter is classified as Seyfert 2 by \citet{2010A&A...518A..10V}. The rest of the galaxies have values for $f_{\rm{AGN}}$ below $\lesssim$0.25.

In summary, these five SIGS and eight SB galaxies that lie close to the AGN wedge in
Fig.~\ref{fig:stern} appear to be composite systems, and the LSM systems inside the wedge are classified as AGN as one would expect: the \textsc{CIGALE}
models indicate varying fractional contributions of AGNs to their emission, but in none of them there is a dominant AGN that was somehow `missed' by the color-color diagram. We can identify systems like the ones presented here, which do have AGN contribution, but not large enough to make the Stern cut, and quantify the amount of the contribution in terms of $f_{\rm{AGN}}$. Likewise,
our SED analysis is consistent with the implications of the canonical
infrared color-color diagnostic diagrams.

\subsubsection{sSFR and stellar mass estimates in light of prior results}

\begin{figure*}
\includegraphics[width=0.49\textwidth]{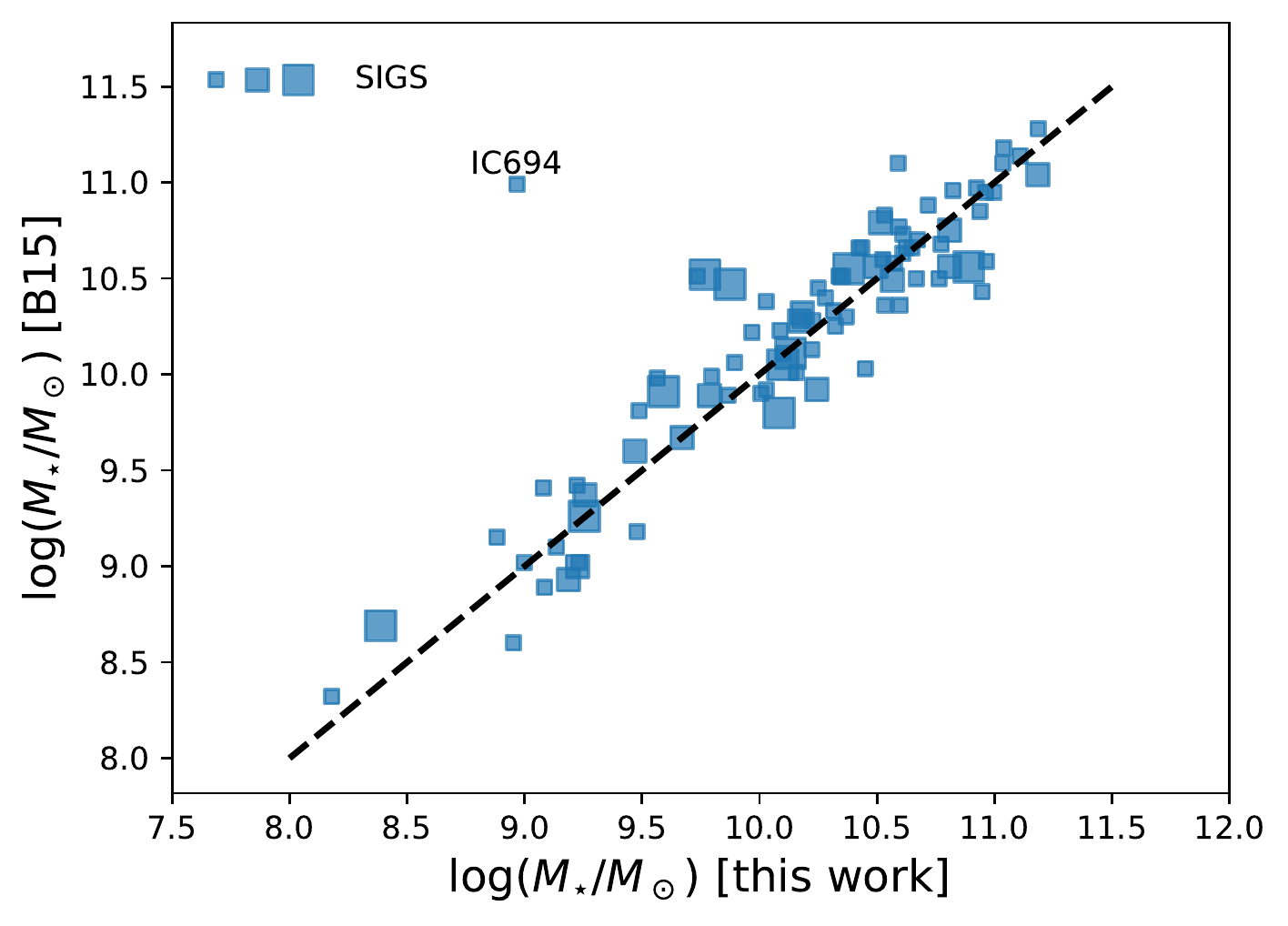}
\includegraphics[width=0.49\textwidth]{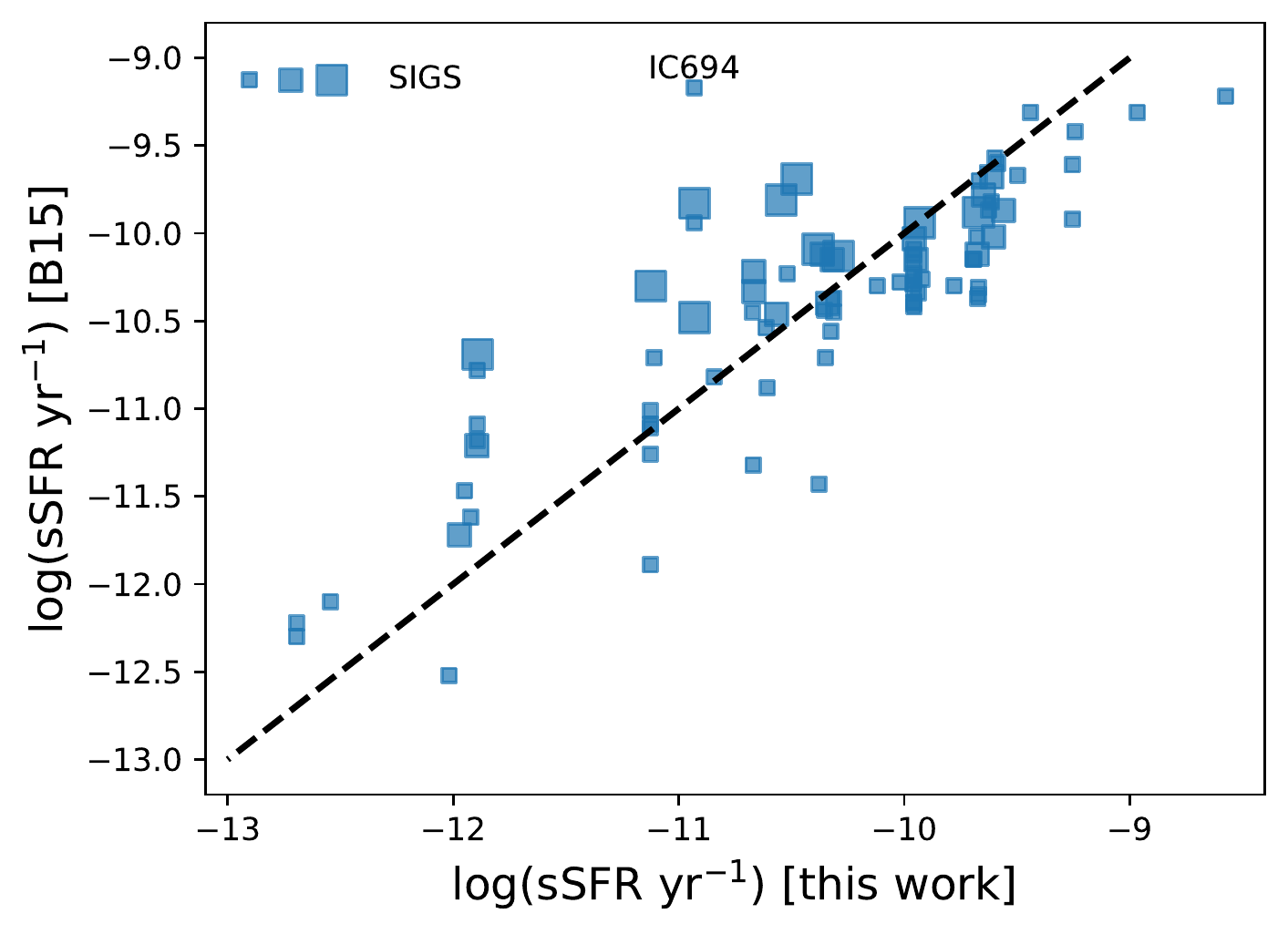}
\caption{Stellar mass ({\sl Left panel}) and specific star formation rate ({\sl Right panel}) estimates derived in this work compared to those obtained with MAGPHYS by \citetalias{2015ApJS..218....6B} for the SIGS galaxies.  Symbol sizes indicate the total AGN contributions $f_{\rm AGN}$: 
the smallest symbols indicate an AGN contribution below 20\%, 
the intermediate symbols indicate an AGN contribution between 20 and 30\%,
and the largest symbols indicate an AGN contribution above 30\%.
The line of equality is indicated with the dashed
line in both panels.  The stellar masses are in good overall agreement, with large scatter. 
The sSSR estimates also show approximate overall agreement, but also with 
significant scatter, especially below log(sSFR)$=-11.5$.
}
\label{fig:Brass}
\end{figure*}

\citetalias{2015ApJS..218....6B} and \citet{2013ApJ...768...90L} use the MAGPHYS SED
 fitting code (a version that did not include an AGN component) plus DECOMPIR \citep{2011MNRAS.414.1082M} to characterize the SIGS galaxies.  
 In Fig.~\ref{fig:Brass} we compare our results with those from \citetalias{2015ApJS..218....6B}, focusing on the stellar mass and specific star formation rate, parameters only indirectly influenced by the presence of an AGN.

The left panel of Fig.~\ref{fig:Brass} shows that our stellar mass estimates agree on average
with those from \citetalias{2015ApJS..218....6B}, but with a large scatter. No obvious dependence
on the total AGN fraction is apparent, which can be related to the selected sample. SED models with AGN and without AGN reproduce similar stellar masses and specific star formation rates, although objects with high AGN fraction show a small trend for being outliers in stellar mass.  IC 694 is a conspicuous outlier because of aperture 
issues (Sec.~\ref{ssec:val}).  Even more scatter is observed in the comparison for derived
sSFR (right panel of Fig.~\ref{fig:Brass}), especially below $\log$(sSFR)$=-11.5$, where the accuracy of SED-based methods in measuring the sSFR significantly decreases \citep{2017MNRAS.465.3125E}. The coarse \textsc{CIGALE} input parameter grid 
below $\log$(sSFR)$=-11.5$ produces the entirely artificial grouping of galaxies along
discrete sSFR values.  NGC\,4933A, NGC\,5353 and NGC\,5481 are not shown because
their estimated sSFRs fall in an extremely low, likely unreliable, sSFR regime ($<10^{-13}$). 

\section{Discussion}
\label{sec:discussion}

In this Section we discuss the implications of the SED modeling for star formation and AGN emission in interacting galaxies.  

\subsection{Interaction Stage}

\citetalias{2015ApJS..218....6B} note that almost all their galaxies with $\log$(sSFR) $<-11.0$ are at early interaction stages. 
Morphologically disturbed systems lie along a broad range of sSFRs, with the range occupied by stage 4 galaxies extending to higher sSFRs than the earlier stages (Fig.~\ref{fig:Hist}).  Our results are equivocal with respect to stage 4 systems and sSFR, however, because the KS tests applied in Sec.~\ref{ssec:properties} do not provide compelling evidence, even with our enlarged sample, that stage 4 systems differ significantly from stage 2 or 3 systems (Table~\ref{tab:KS}).  
However, the outcome is different for the stage 5 systems. They differ significantly in the aggregate from the AGN sample, marginally  from the stage 2 and 3 systems (7\% and 14\% chance of being drawn from the same parent sample, respectively), and there is no evidence that they differ from the starburst systems.  Collectively, the evidence thus favors a picture in which sSFRs are greatest in stage 5 systems, i.e., in or approaching coalescence. This is in agreement with hydrodynamical simulations \citep{2014MNRAS.445.1598H,2016ApJ...817...76M}, which show a steep increase of the SFR very close to coalescence (more for massive systems).

As discussed in Sect.~\ref{ssec:properties}, \citet{2013ApJ...768...90L} did not find statistically significant correlations between SED shape, merger stage, and star-forming properties. In the present work we have enlarged the study sample and the available photometry, and analyzed these data with \textsc{CIGALE}, i.e., a code that explicitly accounts for AGN emission. With these enhancements, we observe that $f_{\textnormal{AGN}}$ does show a correlation with luminosity (see Sect.~\ref{secDis:AGNcrucial}). Our results also point to a weak correlation between $f_{\textnormal{AGN}}$ and interaction stages, with a larger fraction of late stage mergers showing a higher $f_{\textnormal{AGN}}$. SED analysis can therefore be used to infer the physical conditions associated with different stages.

\subsection{The Schmidt-Kennicutt Relation}

Star-forming galaxies form a relatively narrow distribution in the two-dimensional parameter space defined by total stellar mass and SFR, commonly known as the star-forming main sequence (MS). This scaling relation has been widely used to study the relationship between galaxy morphology, star formation, and SED shapes \citep[e.g.,][]{2011A&A...533A.119E}. However, the SFR is not a directly measured quantity, as it is indirectly derived from different observables; total stellar mass estimations, although generally robust, also suffer from being model-dependent. As an example, the vast majority of papers discussing the MS rely on a version of Schmidt-Kennicutt (S-K) relation to infer SFRs from the infrared luminosity of galaxies, that is first converted to a dust mass, and then, via a gas to dust ratio, to a gas mass. In this sub-section we re-examine the reliability of that relation for LIRGs. For the analysis, we rely on parameters derived from our full \textsc{CIGALE} modeling of the SEDs.

In order to estimate the obscured SFR, the 8\,$\mu$m or 24\,$\mu$m luminosities are often used \citep[see][for a more detailed discussion]{2013ApJ...768...90L}, but when possible, it is convenient to use the integrated infrared luminosity between 5--1\,000\,$\mu$m, which is related to the thermal emission from dust heated by star formation (at wavelengths shorter than about 
5\,$\mu$m the SED is dominated by emission from stellar photospheres rather than dust heated by star formation). The Schmidt-Kennicutt relation, often formulated as the relationship between gas surface density and SFR, can also be formulated as a relationship between the total SFR and the infrared luminosity \citep{2013ApJ...768...90L}. This has been a very useful SFR diagnostic ever since the infrared was first made accessible by the {\sl IRAS} and {\sl ISO} satellites. With panchromatic datasets now at hand, incorporating photometry from 
{\sl Spitzer}, {\sl WISE}, {\sl AKARI}, and {\sl Herschel}, there is room for considerable improvement.

\begin{figure*}
\includegraphics[width=0.9\textwidth]{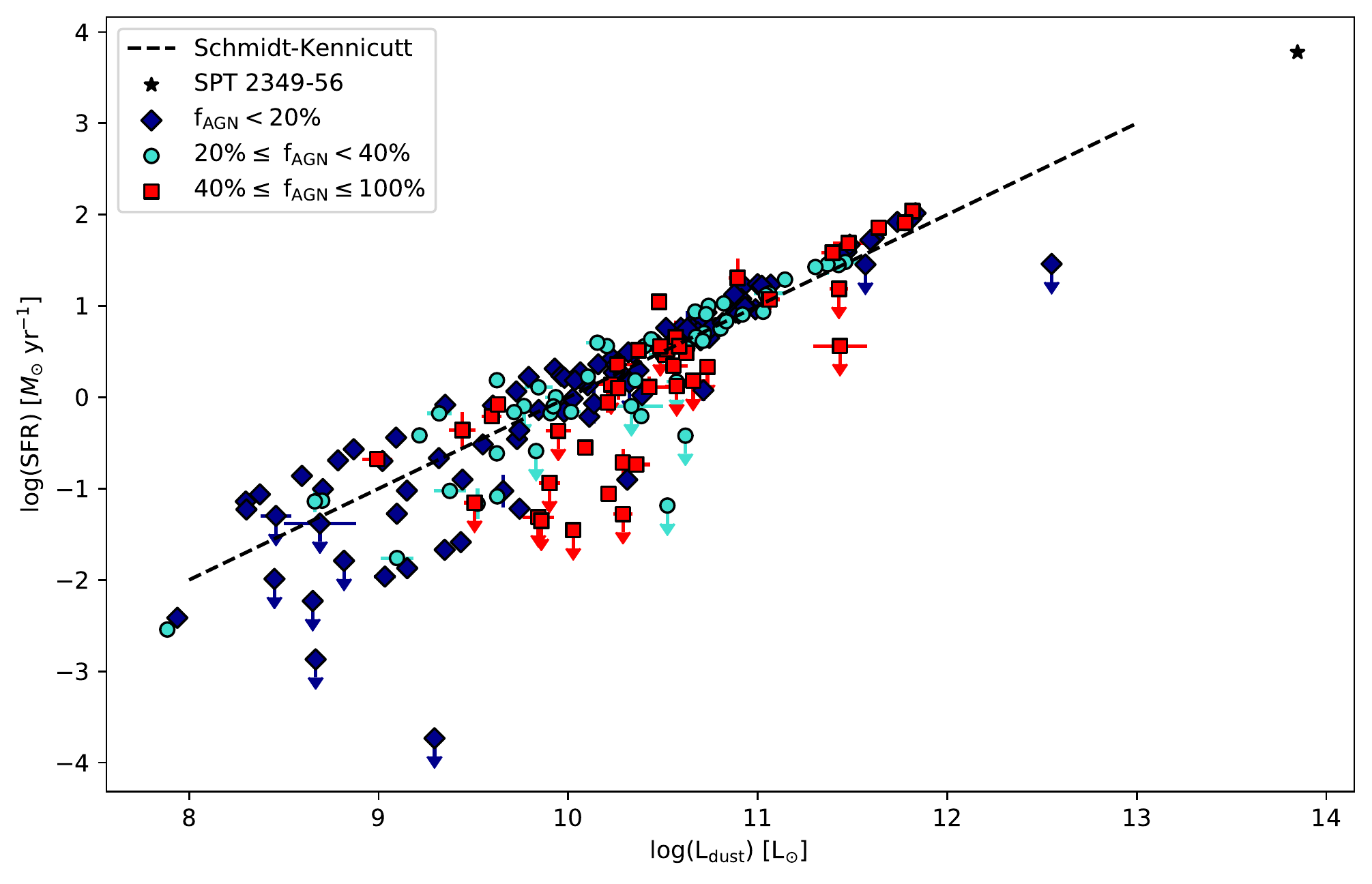}
\caption{
Total derived SFR as a function of total dust luminosity for all galaxies modeled in this work. 
Red squares, cyan circles, and blue diamonds respectively indicate galaxies with 
total AGN fraction $>40$\%, in the range 20--40\%, and below 20\%.  
The dashed line is the S-K relation.  Also plotted for comparison is an IR-luminous 
cluster SPT2349$-$56 at $z=4.6$ \citep[black star,][]{ 2018Natur.556..469M}. Arrows indicate the galaxies for which the SFR are only estimated to within a factor of roughly two (see Table~\ref{tab:CIGALE1AGN}).
}
\label{fig:fracAGN}
\end{figure*}

An important caveat in this conventional approach is that the infrared emission is often interpreted as originating from thermal emission from dust heated only by star formation. However, other heating mechanisms are often in place and need to be accounted for, such as AGN activity and older stellar populations.
\citep{2010MNRAS.402.1693H,2013ApJ...768...90L, 2019ApJ...882....5W}.
Although for most star-forming galaxies this additional contribution is small or negligible, in some cases it can be significant or even dominate the infrared luminosity of the entire galaxy. For the galaxies included in this work, which by construction emphasize infrared-luminous systems and mergers at various stages, we show that the AGN alone can contribute up to $\sim 80\%$ of the infrared luminosity for these systems.

Figure~\ref{fig:fracAGN} plots derived SFR as a function of dust luminosity for all galaxies in Tables~\ref{tab:SIGSsample}--\ref{tab:LSM}, color-coded by the estimated $f_{\rm{AGN}}$. We observe that for dust luminosities above roughly $10^{10.5}\: {\rm L}_{\sun}$, the S-K diagnostic provides a good measure of the amount of star formation taking place. In particular, the fact that at those luminosities there is so little scatter at all levels of $f_{\rm{AGN}}$ indicates that our approximation correctly accounts for other sources of dust heating, allowing a better estimation of the SFR. At lower luminosities we observe a significant scatter in the SFR at a given luminosity. This indicates that the S-K diagnostic might not provide a reliable measure of the SFR at these lower dust luminosities. A plausible explanation for this is that low dust luminosities also implies a larger relative amount of unobscured star formation that is not accounted for by the infrared diagnostics. Less luminous, less morphologically disturbed systems are less optically thick to UV radiation, and therefore a pan-chromatic approach such as the one we have adopted here is more likely to provide reliable estimates.

Figure~\ref{fig:fracAGN} also indicates that galaxies that are less luminous in the infrared have a broader range of AGN contributions skewed towards smaller contributions, i.e., very few galaxies with infrared luminosities below $10^{9}\: {\rm L}_{\sun}$ have $f_{\rm{AGN}} > 20\%$. This supports an scenario in which significant AGN emission occurs preferentially in highly disturbed/obscured systems, and is in agreement with hydrodynamical simulations. Finally, our results are also consistent with a wider dynamical range of SFR at lower luminosities, in absolute terms. At low luminosties, star formation can range from being a negligible to being a significant factor in the galaxy evolution, but at high infrared luminosities, the dust heating from SFR and AGN completely dominates the galactic evolution. This also has to do with the timescales of this evolution. Luminous systems are morphologically disturbed with starburst-like, short-term episodes of star formation and AGN accretion, whereas low luminous systems evolve more secularly, impacting the range of possible fractional SFR contributions.

\begin{figure*}
\includegraphics[width=15cm]{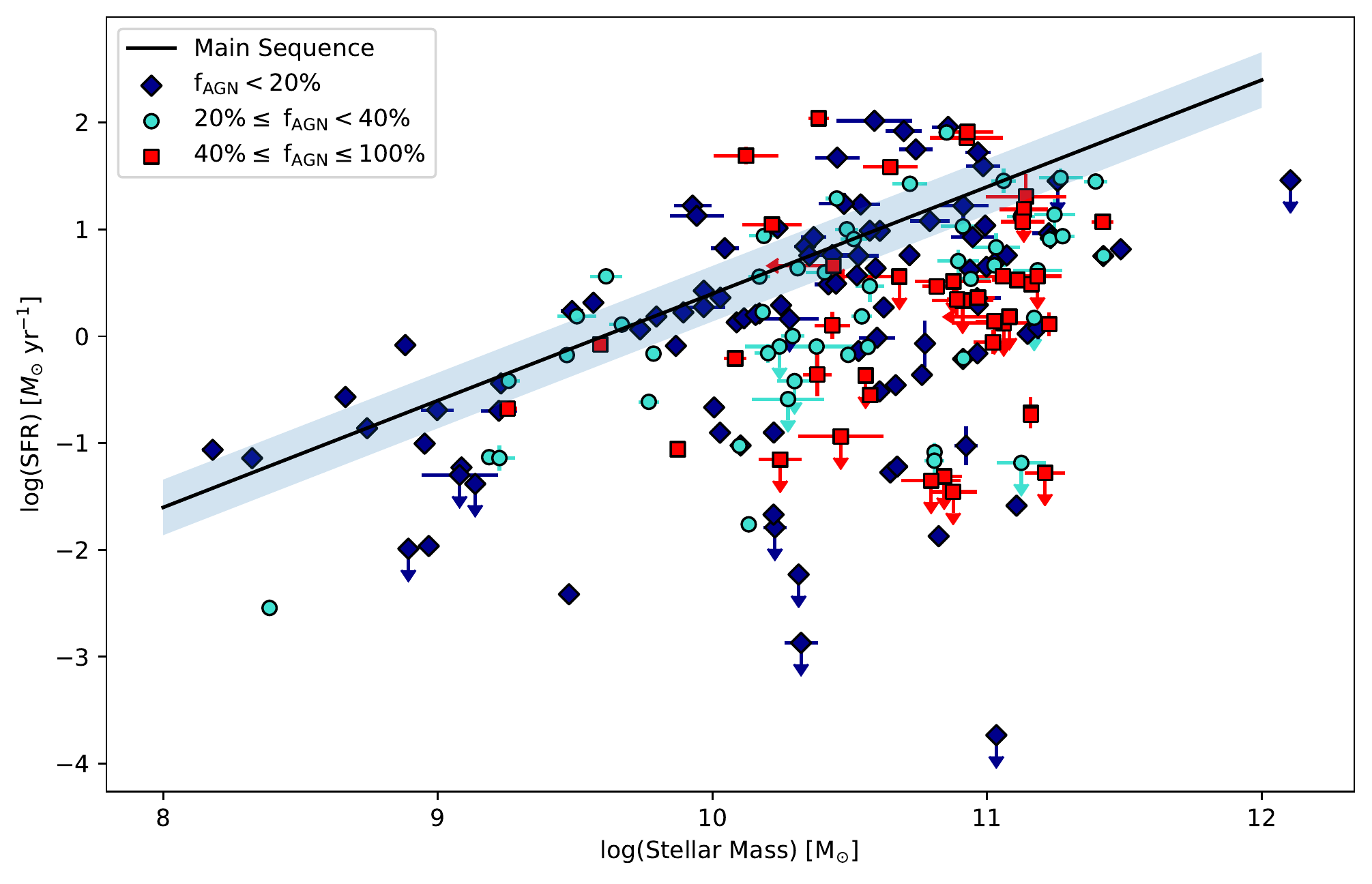}
\caption{Integrated derived SFR as a function of total stellar mass populated with all galaxies modeled 
in this work.  Red squares, cyan circles, and blue diamonds respectively indicate galaxies with 
total AGN fraction $>40$\%, in the range 20--40\%, and below 20\%.
The star forming main sequence per \citet{2011A&A...533A.119E} is indicated with the 
black line; the shaded region
extends to $\pm$0.26 dex about it.  
Values of SFR below $-1$ in the log  are relatively uncertain. Arrows indicate the galaxies for which the SFR or stellar masses are only estimated to within a factor of roughly two (see Table~\ref{tab:CIGALE1AGN})
}
\label{fig:MS}
\end{figure*}

\subsection{The Galaxy Main-Sequence: New Subtleties and Issues}

In Fig.~\ref{fig:MS} we plot the \textsc{CIGALE}-derived stellar masses and SFRs for our galaxies, together with the location of the MS as derived by \citet{2011A&A...533A.119E} for comparison. Several interesting inferences follow from the way our sample galaxies populate this two-dimensional space. First, only a minority of galaxies are located on the nominal MS locus. Some lie above it, in the zone associated by \citet{2011A&A...533A.119E} with starburst-like star formation, whereas a significant amount lie below it, even by a few orders of magnitude. At high stellar masses, the majority of those systems above the MS have higher AGN contributions, which indicates that not only the SFR per unit stellar mass in enhanced, but also the AGN activity. 

What is more puzzling is the significant amount of systems that we observe below the MS. Significant divergences from the MS have been reported both in observations of high redshift galaxies within protoclusters, due to environmental quenching \citep[e.g.][]{2019ApJ...887..183Z}, and in cosmological simulations that relate the growth of galactic halos to that of stellar mass, in which the MS scatter depends on the timescale of star formation variability \citep[e.g.][]{2019ApJ...872..160H}. 

In our case, the large scatter is probably due to the way we assembled our sample. We have selected preferentially galaxies that are luminous and that are morphologically disturbed through mergers. In some of these cases, it is impossible to tell from the morphology alone whether the system has undergone coalescence, and it is likely that in some of those systems star formation has been suppressed due to negative feedback from the AGN, after coalescence. This interpretation is consistent is supported by the results shown in Figure \ref{fig:MS3}, with early type galaxies showing significantly lower SFRs. Our sample is therefore not representative of the secular stages of star formation in galaxies. Instead it represents systems with enhanced star formation through the effect of mergers, and systems where AGN feedback has probably quenched star formation. The fact that both high and low $f_{\rm{AGN}}$ values are similarly represented below the MS indicates that quenching takes place very raplidly after the onset of the AGN. In an upcoming paper (Della Costa et al. in prep.), we discuss this latter conclusion in more detail. 

\begin{figure*}
\includegraphics[width=0.95\textwidth]{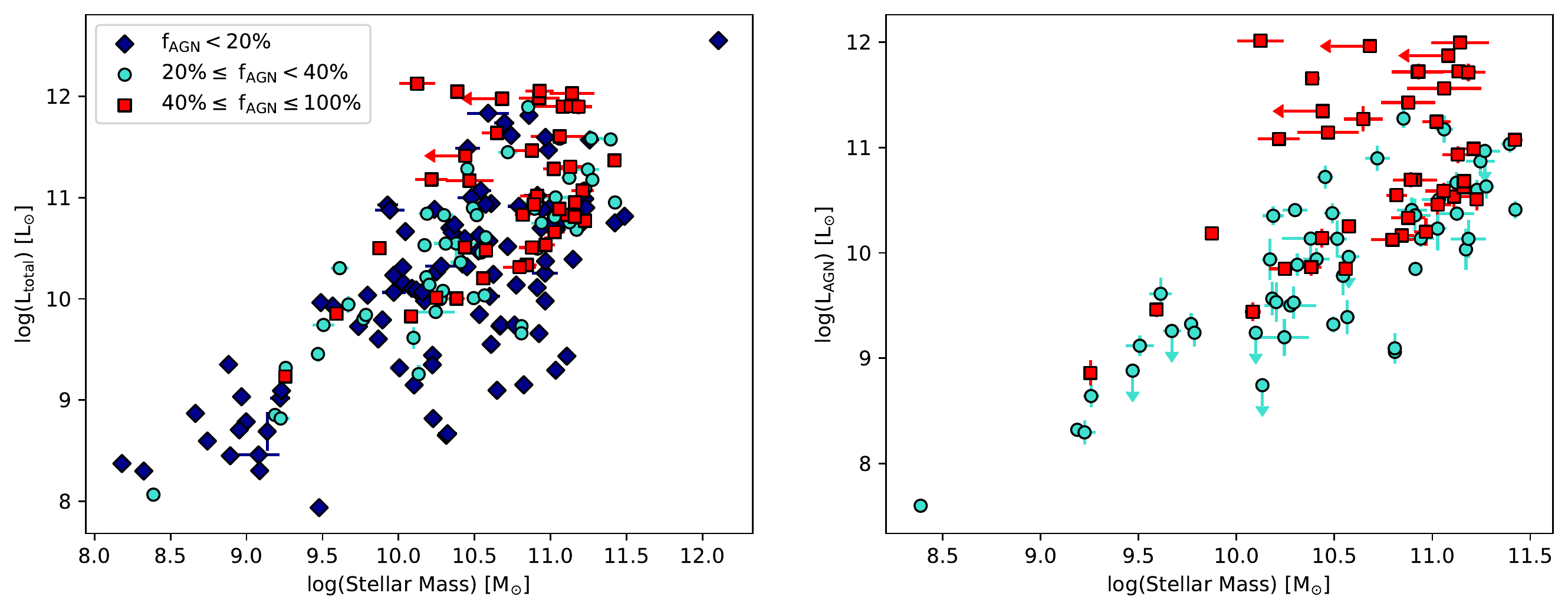}
\caption{
Left panel:  The total dust luminosity as a function of stellar mass. Note that sources with a \textsc{CIGALE}-estimated AGN fraction below 20\% are recalculated and plotted here assuming its AGN contribution is zero.
Right panel: The AGN luminosity only as a function of stellar mass. Arrows indicate the galaxies for which the luminosities or stellar masses are only estimated to within a factor of roughly two (see Table~\ref{tab:CIGALE1AGN})}
\label{fig:HAS-MStar}
\end{figure*}

In Fig.~\ref{fig:HAS-MStar} we compare the stellar mass to the total infrared luminosity from dust heated by stars and AGN in the left panel and to the AGN luminosity only in the right panel. We note that out of a total of 188 galaxies, 42 galaxies have $f_{\textnormal{AGN}} \ge$ 40\% whereas 51 of them have $f_{\textnormal{AGN}} \ge$ 20\%. We observe a mild correlation between stellar mass and luminosity for systems with low contribution from AGN. Presumably, in these systems the infrared luminosity is dominated by star formation, and the correlation confirms that more massive systems tend to have more dust heating, but not always and there are wide variations. A similar correlation is found for systems with a significant contribution form the AGN, but notably the scatter is much smaller. For galaxies with AGN, we also observe a correlation that implies that most luminous AGNs tend to love in the most massive galaxies regardless of the fraction of total luminosity that the AGN contributes, as long as it is above 20\%. The apparent scatter for systems above $10^{10} M_{\odot}$ is most likely due to larger uncertainties in the determination of stellar masses for these systems. In the context of galaxy assembly, this positive correlation supports a joint evolution of super massive black holes and their hosts.

\begin{figure}
\includegraphics[width=8cm]{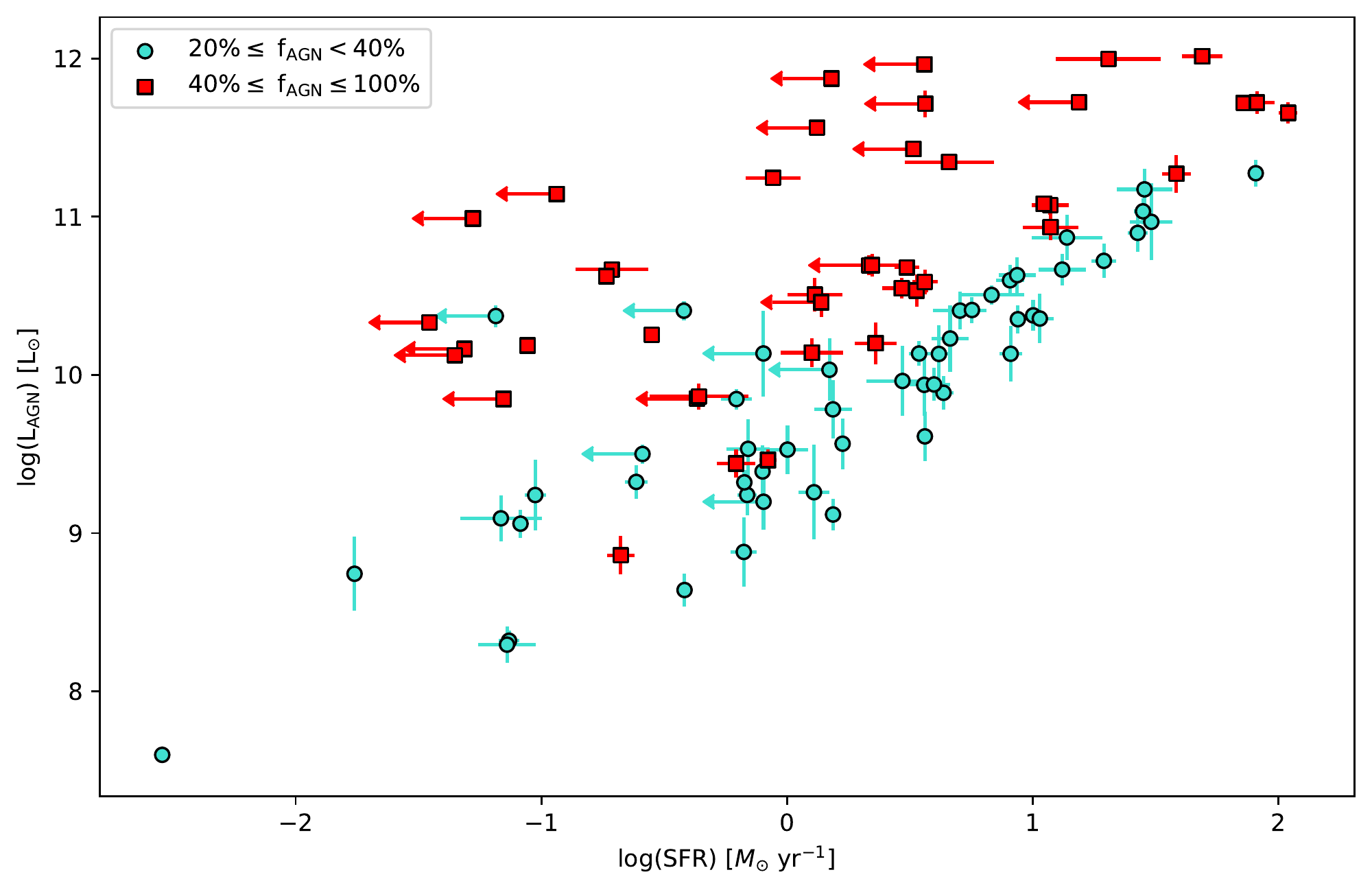}
\caption{
Total AGN luminosity as a function of SFR.  Red squares indicate objects with 
$f_{\textnormal{AGN}}>40$\%, whereas cyan dots indicate those with $f_{\textnormal{AGN}}$ 
in the range of 20--40\%. Arrows indicate the galaxies for which the luminosities or SFR are only estimated to within a factor of roughly two (see Table~\ref{tab:CIGALE1AGN})
 }
\label{fig:HAS-6}
\end{figure}

Fig.~\ref{fig:HAS-6} shows the correlation between SFR and AGN luminosity for those systems with $f_{\textnormal{AGN}}>20$\%. We observe that systems with intermediate AGN contributions (cyan dots) show a tight correlation in this plane over 5 orders of magnitude in SFR. For systems with high AGN contributions (red dots), we observe a similar correlation, but there is a larger scatter, and also, the luminosity of the AGN at a given SFR is higher in comparison with the intermediate systems. Similar correlations between AGN lumonosity and SFR have been found in more uniform samples, such as the COSMOS field \citep{2017A&A...602A.123L}; such correlations support scenarios proposed in recent galaxy evolution models in which black hole accretion and star formation are correlated due to the compression of large amounts of gas in nuclear regions. 

The fact that the correlation is less tight at higher AGN contributions, at evolutionary stages closer to coalescence, can be interpreted in terms of star formation quenching: as hydrodynamical simulations show, the AGN reaches a peak in luminosity right after coalescence, and star formation gets quenched very rapidly. This supports a real effect of AGN feedback on SFR, as opposed to recent studies \citep{ 2019arXiv191201020H} that suggest that AGN activity does not quench galaxy wide AGN. One possible explanation of this discrepancy could be the difference between galaxy-wide star formation and the nuclear, merger-induced star formation that we are measuring in the present work.

\subsection{The AGN as crucial ingredient in galaxy evolution}\label{secDis:AGNcrucial}

We have shown (e.g., Figs.~\ref{fig:MS}-\ref{fig:HAS-6} and related discussions), that accounting for the AGN emission is a necessary step in order to gain a better understanding of the physics and energy budget of infrared-luminous galaxies, specially in the late stage of mergers. In particular, we have provided evidence that dust heating by the AGN can be a dominant factor in the galaxy SED, and that the latter can provide hints as to the specific stage of the merger. We have also shown that the rapid evolution of AGN accretion an SFR right before and right after coalescence creates significant spread in the so-called Main Sequence, partly due to star formation quenching. This conclusion echoes that by earlier works \citep[e.g.,][]{2012ApJ...744....2A,2014MNRAS.445.1598H}. Here we elucidate some implications of these findings.

\citet{2012ApJ...744....2A}  suggest that the prevalence of galaxies with $f_{\rm AGN}>0.05$ correlates with IR luminosity, increasing from $f_{\rm AGN} \sim 0.3$ for $10^{11}\rm{L}_{\sun} <  \rm{L}_{\rm{IR}} < 10^{12}\rm{L}_{\sun}$ to $f_{\rm AGN} \gtrsim 0.5$ in $\rm{L}_{\rm{IR}} > 10^{12} \rm{L}_{\sun}$. 
Our Figure~\ref{fig:Lbol} points to a similar trend. They also found that the AGN contribution is close to 50\% for LIRGs, and that it reaches 80\% in the case of ULIRGs. Our results are consistent with those numbers, with galaxies in the AGN sample reaching the highest values of fractional AGN contribution. We also see an increase in the fractional AGN contribution for bins of increasing luminosity, For bins centered at $10^{9},\,10^{10},\,10^{11},\,10^{12}$ in $\rm{L}_{\rm{IR}}$, the respective means in $f_{\rm AGN}$ values are 5\%, 14\%, 20\% and 36\%. Similar results were found by \citet{2018MNRAS.478.4238D} in a larger sample of U(LIRGS) with a measured AGN contribution. Including the AGN contribution in the SED analysis of luminous infrared galaxies is therefore a required step in any reliable study of their physical properties.

There are a number of diagnostics that can be better interpreted if a reliable determination of the merger or AGN onset stages can be made from SED analysis, as we are doing in this work. For example,  \citet{2010ApJ...709.1257T} notice that if the AGN ionising continuum is switched off, the photoionized narrow line region (NLR) could still be detected due to its large extension and long line recombination time ($\sim 300$yr, depending on the density). They also suspect that this delay might be related to the existence of many different types of AGNs as classified by their line emission properties. Our work provides a sample that can be used as a proxy to study how line emission properties change as a function of AGN fractional contribution and amount of star formation quenching. A combination of $f_{\rm AGN}$ estimations in addition to fine structure lines ratios (as Fig.~\ref{fig:FAGN}) can lead to a reliable analysis of AGN stages and their relation with interacting galaxies, as with the $f_{\rm{AGN}}$-[\ion{Ne}{v}]/[\ion{Ne}{ii}] plane. We have confirmed that line ratios are a more reliable proxy for AGN presence that the SED analysis alone.

Funneling of large amounts of gas into the nuclear regions during a merger can trigger large episodes of star-formation and AGNs activity \citep{2010ApJ...716.1151W,2017A&A...602A.123L}. The resulting obscuration can significantly attenuate the AGN optical emission \citep{2018MNRAS.478.3056B}. Our results support a co-evolution of star formation and AGN activity during a merger, while correcting by obscuration effects by using a pan-chromatic, energy-conserving approach.  

In their study of post-starbursts, \citet{2017ApJ...843....9A} find that, well before coalescence, merging galaxies  are generally located in the ``green valley'' and show bluer $W1-W2$ {\sl WISE} colours, characteristic of AGN activity (see also our Fig.~\ref{fig:stern}). These galaxies thus may contain buried AGNs that emit in the infrared and are better traced by infrared emission lines (see Figure~\ref{fig:FAGN}). They suggest that the AGNs do not radiatively dominate the post-starburst phase, and that a better census of these post-starbursts can be constructed if there are reliable tracers of the AGN activity during the early phases of quenching. Our work provides an example of such search, and the fact that we see a significant number of less luminous AGNs below the main sequence in Fig.~\ref{fig:MS1} agrees with their results.

The need for more extensive samples and better indicators of the interaction stage specially during the obscured and mirphologically disturbed phase is critical for an improved understanding of the evolution of mergers. Our work provides a pilot study of what JWST and the SPace IR Telescope for Cosmology and Astrophysics (SPICA) \citep{2017PASA...34...57S} will be able to do with their improved sensitivity and spectral resolution. 

\section{Conclusions}
\label{sec:conclusions}

We have re-reduced and re-analysed photometric observations from the UV to the FIR on 199 luminous and ultra-luminous galaxies in four different sets of objects including mergers and AGN, analyzed their physical properties using the \textsc{CIGALE} SED modeling code, and presented an analysis of the results. This is the largest systematic, wide, multi-band SED analysis program yet done on an ULIRG sample.  In particular, our approach included galaxies over a broad range of AGN activity as reported in the literature. Our goal was to accurately measure the fractional AGN contribution to the total luminosity in these systems, and to assess how this contribution impacts popularly used SED diagnostics of star formation and ISM properties. We also aimed at examining the evolving effects of AGN activity across the merger sequence. From the original sample, we excluded ten objects that either had limited or uncertain datasets and/or unreliable SEDs. Our primary conclusions apply to the remaining 189 objects. Here are our major findings:

\begin{enumerate}

\item[1.] A reliable measure of the fractional contribution of AGN emission to the total luminosity of galaxies is essential in the understanding of galaxy-wide physics, such as star formation evolution and total energy output. About half of the galaxies in our sample have more that 20\% AGN contribution to their total luminosities, and about a quarter of the systems had contributions over 40\%. This results in warmer dust temperatures that can be wrongly associated to star formation if AGN is not included in the modelling. Overall, we find only a weak correlation between the merger stage and the AGN fractional contribution, in agreement with other studies \citep[e.g., ]{2013ApJ...768...90L}.

\item[2.] AGN activiy can be responsible for a significant displacement of galaxies across the so-called "Main Sequence" of star formation. Outliers of this correlation must therefore be interpreted in terms of their AGN activity, and not only in terms of their star formation properties. We have produced a carefully remeasured SFR-$M_{*}$ plane that shows significant deviations from the MS correlation, both above and below it. These deviations are only partially explained by increased star formation, as increased AGN activity and feedback-driven star formation quenching can have a significant role in the emission properties, specially at merger stages just before and just after coalescence.

\item[3.] As a result of the previous statement, infrared galaxies at intermediate and high redshifts should not have their physical properties, specially those related to star formation, interpreted as if they were local MS infrared galaxies, without first accounting for their merger stage and AGN fractional contribution. Possible diagnostics to do this from their SED and spectra include their location in the $f_{\rm{AGN}}$-[\ion{Ne}{v}]/[\ion{Ne}{ii}] plane.

\item[4.] At high (>40\%) fractional AGN contributions, both the star formation luminosity and the AGN luminosity independently correlate with the total stellar mass of the galaxy. This is in agreement with findings in large uniform surveys such as COSMOS, and supports scenarios in which both black hole accretion and star formation are driven by gas compression in the nuclear regions during the merger. The lack of correlation between the total luminosity and the stellar mass at low AGN fractional contributions calls into question the use of infrared diagnostics alone to estimate SFR in the early stages of mergers.

\item[5.] The SFR-$M_{*}$ plane for our 189 luminous galaxies reveals significant outliers from the Main Sequence, specially among systems larger than about $10^{10} \rm{M}_{*}$.  In particular, many systems in late stages of the merger fall up to a few orders of magnitude below the MS. This suggests that the MS paradigm stops being valid for luminous merging systems near to coalescence, due to the rapid quenching of star formation by the AGN feedback. This is in agreement with recent evidence of quenching in intermediate redshift galaxies affecting the MS, and implies that the MS paradigm needs to be evaluated carefully for samples without a though SED analysis that includes AGN emission. A similar study with a much larger sample of galaxies might be more conclusive in this respect.

\item[6.] Our sample is significantly larger than many other studies and (we argue) the SED method is more accurate; using it we support and refine earlier conclusions that $f_{\rm AGN}$ correlates with $\rm{L}_{\rm{IR}}$, with the average AGN contribution to a galaxy's $\rm{L}_{\rm{IR}}$ increasing from about 5\% to 36\% as $\rm{L}_{\rm{IR}}$ increases from $10^{9}$ to $10^{12} \rm{L}_{\sun}$.

\end{enumerate}

\textsc{CIGALE} was in many cases able to identify Type 1 AGNs by varying the viewing angle Psi and looking for a minimum $\chi^2$ value (and other parameters changes); Type 1's had optimum fits with $\psi \gtrsim 70$. This feature may be of particular value in studies of high-z objects whose morphology is unknown but whose line strengths cannot be properly analyzed without attention to the possible extinction corrections.  A new version of \textsc{CIGALE} has recently been released which we plan to use in a more detailed analysis of viewing angle effects.

In the near future, planned and proposed facilities such as \emph{JWST} and \emph{SPICA}, and ground-based telescopes, will provide better resolution and new insights into the physical processes at work in galaxies and their evolution. In particular, they will begin to piece together the  cosmic history of galaxies in the universe. The method of meticulous SED modelling, as presented in this work, can play an important role in the interpretation of these new datasets.

\section*{Acknowledgements}

The authors gratefully thank Steve Willner, Andreas Zezas, Lingyu Wang, Elizabeth Hora, Madison Hora, John Della Costa III, and Antonio Frigo for their assistance and advice.  They particularly thank Denis Burgarella and Laura Ciesla from the \textsc{CIGALE} team for their expert guidance and assistance. We acknowledge the anonymous referee for a careful reading of the manuscript and very helpful questions and comments. HAS, MA, and JRM-G acknowledge partial support from NASA Grants NNX14AJ61G and NNX15AE56G. 

The SAO REU program is funded by the National Science Foundation REU and Department of Defense ASSURE programs under NSF Grant AST-1659473, and by the Smithsonian Institution.  

This publication makes use of data products from the Two Micron All Sky Survey, which is a joint project of the University of Massachusetts and the Infrared Processing and Analysis Center/California Institute of Technology, funded by the National Aeronautics and Space Administration and the National Science Foundation.

This research is based on observations made with the Galaxy Evolution Explorer, obtained from the MAST data archive at the Space Telescope Science Institute, which is operated by the Association of Universities for Research in Astronomy, Inc., under NASA contract NAS 5-26555.

Funding for SDSS-III has been provided by the Alfred P. Sloan Foundation, the Participating Institutions, the National Science Foundation, and the U.S. Department of Energy Office of Science. The SDSS-III web site is {\tt http://www.sdss3.org/}.
SDSS-III is managed by the Astrophysical Research Consortium for the Participating Institutions of the SDSS-III Collaboration including the University of Arizona, the Brazilian Participation Group, Brookhaven National Laboratory, Carnegie Mellon University, University of Florida, the French Participation Group, the German Participation Group, Harvard University, the Instituto de Astrofisica de Canarias, the Michigan State/Notre Dame/JINA Participation Group, Johns Hopkins University, Lawrence Berkeley National Laboratory, Max Planck Institute for Astrophysics, Max Planck Institute for Extraterrestrial Physics, New Mexico State University, New York University, Ohio State University, Pennsylvania State University, University of Portsmouth, Princeton University, the Spanish Participation Group, University of Tokyo, University of Utah, Vanderbilt University, University of Virginia, University of Washington, and Yale University.

This work is based in part on observations made with the {\sl Spitzer Space Telescope}, which is operated by the Jet Propulsion Laboratory, California Institute of Technology under a contract with NASA.
This publication makes use of data products from the {\sl Wide-field Infrared Survey Explorer}, which is a joint project of the University of California, Los Angeles, and the Jet Propulsion Laboratory/California Institute of Technology, funded by the National Aeronautics and Space Administration."
Herschel is an ESA space observatory with science instruments provided by European-led Principal Investigator consortia and with important participation from NASA.

This research has also made use of results from NASA's Astrophysics Data System. This research has made use of the NASA/IPAC Extragalactic Database (NED), which is operated by the Jet Propulsion Laboratory, California Institute of Technology, under contract with the National Aeronautics and Space Administration. This research has made use of the SIMBAD database, operated at CDS, Strasbourg, France. This research made use of Astropy, a community-developed core Python package for Astronomy \citep{2013A&A...558A..33A}.   We thank the people that support photutils \citep{larry_bradley_2018_1340699}. 

\section*{Data availability}

The data underlying this article are available in the article and in its online supplementary material.

%%%%%%%%%%%%%%%%%%%%%%%%%%%%%%%%%%%%%%%%%%%%%%%%%%



\bibliographystyle{mnras}
\bibliography{ManualBib} % if your bibtex file is called example.bib

%\begin{thebibliography}{99}
%\end{thebibliography}

%%%%%%%%%%%%%%%%%%%%%%%%%%%%%%%%%%%%%%%%%%%%%%%%%%

%%%%%%%%%%%%%%%%% APPENDICES %%%%%%%%%%%%%%%%%%%%%

\appendix

\section{Extra material}

Here we present all the information for the samples, photometry values, line emissions, examples of the \textsc{CIGALE} SED fitting, \textsc{CIGALE} derived parameters and histograms of the parameters analysed for all the 188 galaxies presented in this work. We include a table with the derived parameters for six AGN galaxies where a different viewing angle in \textsc{CIGALE} give different output parameters. In addition, as online material, we provide the SED fits of the 178 galaxies with good fits. For the remaining 10 galaxies where the SED fit is not good enough, we provide the AGN and no-AGN SED fits. 

% %% INFORMATION SAMPLES
\begin{table*}
\centering
\caption{Basic data for the SIGS sample galaxies.}
\label{tab:SIGSsample}
\newcommand{\0}{\phantom{0}}
\begin{tabular}{cccrlcccr}
\hline
Group & Galaxy & RA & Dec\ \ \  & Redshift & Sample & Interaction & Size & Angle$^{\rm a}$\\
 ID   & ID &\multicolumn{2}{c}{(J2000)} & \ \ \ ($z$) &&Stage&($\arcsec \times \arcsec$) & ($\degr$)\ \ \ \\
\hline
\01 & NGC 274$^{\rm b}$
                & 00:51:01.6 & $-$07:03:22.7 & 0.0058 & A & 4 &  33.8  $\times$  23.0 & 130.0 \\
    & NGC 275   & 00:51:04.8 & $-$07:03:59.8 & 0.0058 & A & 4 &  38.2  $\times$  28.1 &  25.0 \\
\02 & NGC 470   & 01:19:44.9 &    03:24:35.6 & 0.0079 & A & 2 &  90.0  $\times$  55.1 &  65.0 \\
    & NGC 474   & 01:20:06.7 &    03:24:55.4 & 0.0077 & A & 2 & 225.0  $\times$ 175.0 & 165.0 \\
\03 & NGC 520   & 01:24:35.1 &    03:47:32.7 & 0.0076 & A & 5 & 147.3  $\times$  97.2 & 235.0 \\
\04 &  IC 195$^{\rm b}$   & 02:03:44.6 &    14:42:33.5 & 0.0122 & A & 3 &  37.8  $\times$  21.2 &  39.8 \\
    &  IC 196   & 02:03:49.8 &    14:44:20.8 & 0.0122 & A & 3 &  95.0  $\times$  55.1 &  62.0 \\
\05 & NGC 833   & 02:09:20.8 & $-$10:07:59.2 & 0.0129 & A & 4 &  42.5  $\times$  23.0 & 175.0 \\
    & NGC 835   & 02:09:24.6 & $-$10:08:09.2 & 0.0136 & A & 4 &  42.8  $\times$  35.3 & 125.0 \\
    & NGC 838   & 02:09:38.5 & $-$10:08:48.1 & 0.0128 & A & 3 &  45.0  $\times$  25.9 & 175.0 \\
    & NGC 839   & 02:09:42.9 & $-$10:11:02.8 & 0.0129 & A & 2 &  45.0  $\times$  28.0 &   5.0 \\
\06 & NGC 935   & 02:28:10.6 &    19:36:05.4 & 0.0138 & A & 4 &  53.6  $\times$  33.1 &  65.0 \\
    & IC 1801   & 02:28:12.7 &    19:34:59.9 & 0.0134 & A & 4 &  34.2  $\times$  19.1 & 120.0 \\
\07 & NGC 1241  & 03:11:14.6 & $-$08:55:19.6 & 0.0135 & A & 3 & 100.1  $\times$  60.1 &  50.0 \\
    & NGC 1242  & 03:11:19.3 & $-$08:54:08.6 & 0.0134 & A & 3 &  28.1  $\times$  19.4 &  44.4 \\
\08 & NGC 1253  & 03:14:09.0 & $-$02:49:22.4 & 0.0057 & A & 3 & 168.1  $\times$  52.6 & 175.0 \\
    & NGC 1253A & 03:14:23.3 & $-$02:48:02.9 & 0.0061 & A & 3 &  55.1  $\times$  20.2 &   5.0 \\
\09 & NGC 2276  & 07:27:28.3 &    85:45:23.8 & 0.0081 & C & 2 &  85.0  $\times$  72.0 & 110.0 \\
 10 & NGC 2444  & 07:46:53.2 &    39:02:05.3 & 0.0135 & A & 4 &  54.4  $\times$  30.6 & 125.0 \\
    & NGC 2445  & 07:46:55.1 &    39:00:41.8 & 0.0133 & A & 4 &  52.2  $\times$  45.4 & 110.0 \\
 11 & NGC 2633  & 08:48:04.6 &    74:05:56.0 & 0.0072 & C & 2 &  79.9  $\times$  50.0 &  90.0 \\
    & NGC 2634$^{\rm b}$
                & 08:48:25.4 &    73:58:01.9 & 0.0075 & C & 2 &  40.7  $\times$  35.3 & 135.0 \\
    & NGC 2634A & 08:48:38.1 &    73:56:21.5 & 0.007  & C & 2 &  38.2  $\times$  18.0 & 155.0 \\
 12 & NGC 2719A & 09:00:15.5 &    35:43:09.5 & 0.0104 & A & 2 &  22.0  $\times$  15.1 &  35.0 \\
    & NGC 2719  & 09:00:15.6 &    35:43:41.9 & 0.0103 & A & 2 &  40.0  $\times$  13.0 &  40.0 \\
 13 & NGC 2805  & 09:20:20.4 &    64:06:10.1 & 0.0058 & C & 2 & 165.0  $\times$ 140.0 & 120.0 \\
    & NGC 2814  & 09:21:11.5 &    64:15:11.5 & 0.0053 & C & 2 &  50.0  $\times$  20.2 &  90.0 \\
 14 & NGC 2820A & 09:21:29.8 &    64:14:14.6 & 0.0051 & C & 3 &  15.5  $\times$  11.2 & 110.0 \\
    & NGC 2820  & 09:21:45.6 &    64:15:28.4 & 0.0053 & C & 3 & 119.9  $\times$  20.2 & 151.0 \\
 15 & NGC 2964  & 09:42:54.2 &    31:50:50.6 & 0.0044 & C & 2 &  78.5  $\times$  41.0 &   8.0 \\
    & NGC 2968$^{\rm b}$  & 09:43:12.0 &    31:55:43.3 & 0.0052 & C & 2 &  71.3  $\times$  54.4 & 145.0 \\
    & NGC 2970  & 09:43:31.1 &    31:58:37.2 & 0.0054 & C & 2 &  24.8  $\times$  24.8 &   0.0 \\
 16 & NGC 2976  & 09:47:15.5 &    67:54:59.0 & 0.0008 & C & 2 & 209.2  $\times$  84.6 &  50.6 \\
 17 & NGC 3031  & 09:55:33.2 &    69:03:55.1 & 0.0008 & C & 2 & 834.8  $\times$ 420.1 &  75.0 \\
    & NGC 3034  & 09:55:52.7 &    69:40:45.8 & 0.0007 & C & 2 & 285.1  $\times$ 159.8 & 155.0 \\
    & NGC 3077  & 10:03:19.1 &    68:44:02.0 & 0.0009 & C & 2 & 143.6  $\times$ 110.0 & 135.0 \\
 18 & NGC 3165  & 10:13:31.3 &    03:22:30.0 & 0.0045 & C & 3 &  70.9  $\times$  30.2 &  78.5 \\
    & NGC 3166  & 10:13:45.8 &    03:25:30.0 & 0.0045 & C & 3 & 141.1  $\times$  68.8 & 175.0 \\
    & NGC 3169  & 10:14:15.0 &    03:27:58.0 & 0.0041 & C & 2 & 150.1  $\times$ 110.2 & 145.0 \\
 19 & NGC 3185  & 10:17:38.6 &    21:41:17.9 & 0.0041 & C & 2 &  79.9  $\times$  55.1 &  35.0 \\
    & NGC 3187  & 10:17:47.9 &    21:52:23.9 & 0.0053 & C & 3 &  90.0  $\times$  40.0 & 170.0 \\
    & NGC 3190  & 10:18:05.6 &    21:49:56.3 & 0.0042 & C & 3 & 141.8  $\times$  50.0 &  32.0 \\
 20 & NGC 3226  & 10:23:27.0 &    19:53:54.6 & 0.0044 & C & 4 &  56.9  $\times$  44.6 & 110.0 \\
    & NGC 3227  & 10:23:30.6 &    19:51:54.0 & 0.0039 & C & 4 &  95.0  $\times$  60.1 &  55.0 \\
 21 & NGC 3395  & 10:49:49.3 &    32:58:45.5 & 0.0054 & C & 4 &  55.1  $\times$  32.0 & 120.0 \\
    & NGC 3396  & 10:49:55.6 &    32:59:24.7 & 0.0054 & C & 4 &  63.0  $\times$  29.9 &  10.0 \\
 22 & NGC 3424  & 10:51:46.3 &    32:54:02.9 & 0.005  & C & 2 &  93.6  $\times$  28.8 &  20.0 \\
    & NGC 3430  & 10:52:11.4 &    32:57:01.4 & 0.0053 & C & 2 & 119.2  $\times$  69.1 & 115.0 \\
 23 & UGC 6016$^{\rm b}$
                & 10:54:14.6 &    54:17:11.8 & 0.005  & A & 3 &  74.9  $\times$  47.9 & 150.0 \\
    & NGC 3448  & 10:54:38.6 &    54:18:22.3 & 0.0045 & A & 3 &  84.6  $\times$  40.3 & 155.0 \\
 24 &   IC 694  & 11:28:27.3 &    58:34:42.6 & 0.0132 & C & 4 &  11.9  $\times$  11.9 &   0.0 \\
    & NGC 3690  & 11:28:32.3 &    58:33:42.8 & 0.0104 & C & 4 &  65.9  $\times$  47.2 &  35.0 \\
 25 & NGC 3786  & 11:39:42.7 &    31:54:27.7 & 0.0089 & C & 3 &  62.3  $\times$  33.5 & 170.0 \\
    & NGC 3788  & 11:39:44.6 &    31:55:52.3 & 0.009  & C & 3 &  58.0  $\times$  25.9 &  85.0 \\
 26 & NGC 3799  & 11:40:09.4 &    15:19:38.3 & 0.011  & A & 3 &  25.2  $\times$  18.4 &  14.2 \\
    & NGC 3800  & 11:40:13.5 &    15:20:32.6 & 0.011  & A & 3 &  61.2  $\times$  24.5 & 142.0 \\
 27 &   IC 749  & 11:58:34.1 &    42:44:02.4 & 0.0027 & C & 2 &  79.9  $\times$  65.2 &  65.0 \\
    &   IC 750  & 11:58:52.2 &    42:43:21.0 & 0.0023 & C & 2 &  79.9  $\times$  45.4 & 130.0 \\
 28 & NGC 4038  & 12:01:54.3 & $-$18:53:03.1 & 0.0055 & A & 4 & 142.9  $\times$ 107.6 &  84.3 \\
 29 & NGC 4382$^{\rm b}$  & 12:25:24.1 &    18:11:29.4 & 0.0024 & C & 2 & 150.1  $\times$ 114.8 &  95.0 \\
    & NGC 4394  & 12:25:55.5 &    18:12:50.8 & 0.0031 & C & 2 & 110.2  $\times$  95.0 &  15.0 \\
\hline
\end{tabular}

\end{table*}

\begin{table*}
\contcaption{Basic data for the SIGS sample galaxies.}
\label{tab:SIGSsample2}
\begin{tabular}{cccrlcccr}
\hline
Group & Galaxy & RA & Dec\ \ \  & Redshift & Sample & Interaction & Size & Angle$^{\rm a}$\\
 ID   & ID &\multicolumn{2}{c}{(J2000)} & \ \ \ ($z$) &&Stage&($\arcsec \times \arcsec$) & ($\degr$)\ \ \ \\
\hline
30 & NGC 4567  & 12:36:31.5 &    11:15:43.6 & 0.0075 & C & 3 &  72.0  $\times$  43.2 & 145.0 \\
   & NGC 4568  & 12:36:34.5 &    11:14:12.5 & 0.0075 & C & 3 & 142.9  $\times$  41.8 & 122.0 \\
31 & NGC 4618  & 12:41:32.9 &    41:08:42.4 & 0.0018 & C & 3 & 135.0  $\times$ 105.1 & 115.0 \\
   & NGC 4625  & 12:41:52.7 &    41:16:26.4 & 0.0021 & C & 3 &  65.5  $\times$  57.2 &  56.7 \\
32 & NGC 4647  & 12:43:31.7 &    11:35:03.5 & 0.0047 & C & 3 &  68.0  $\times$  55.1 &  20.0 \\
   & NGC 4649  & 12:43:40.0 &    11:33:09.7 & 0.0037 & C & 3 & 119.9  $\times$  95.0 &  10.0 \\
33 & NGC 4933A$^{\rm b}$ & 13:03:53.9 & $-$11:30:23.8 & 0.0104 & A & 4 &  24.8  $\times$  20.2 & 155.0 \\
   & NGC 4933B & 13:03:57.2 & $-$11:29:43.8 & 0.0108 & A & 4 &  40.0  $\times$  38.2 & 130.0 \\
   & NGC 4933C & 13:04:01.1 & $-$11:29:26.2 & 0.0106 & A & 4 &  15.8  $\times$  14.0 &  20.0 \\
34 & M51A      & 13:29:51.6 &    47:10:34.7 & 0.0015 & C & 3 & 273.2  $\times$ 204.0 & 120.0 \\
   & M51B      & 13:29:59.6 &    47:15:58.0 & 0.0016 & C & 3 & 107.6  $\times$  65.5 &   5.0 \\
35 & NGC 5350  & 13:53:21.6 &    40:21:50.0 & 0.0077 & C & 3 & 100.1  $\times$  69.8 & 125.0 \\
35 & NGC 5353  & 13:53:26.7 &    40:16:58.8 & 0.0078 & C & 3 &  52.9  $\times$  28.1 &  52.0 \\
   & NGC 5354  & 13:53:26.7 &    40:18:10.1 & 0.0086 & C & 2 &  42.1  $\times$  36.0 &   0.0 \\
36 & NGC 5394  & 13:58:32.4 &    37:27:14.8 & 0.0115 & C & 4 &  50.0  $\times$  45.0 & 110.0 \\
   & NGC 5395  & 13:58:38.0 &    37:25:28.2 & 0.0117 & C & 4 &  92.2  $\times$  50.4 &  87.2 \\
37 & NGC 5457  & 14:03:12.5 &    54:20:56.4 & 0.0008 & C & 3 & 650.2  $\times$ 650.0 &   0.0 \\
   & NGC 5474  & 14:05:01.6 &    53:39:43.9 & 0.0009 & C & 3 & 142.6  $\times$ 124.0 & 282.5 \\
38 & NGC 5426  & 14:03:24.8 & $-$06:04:08.8 & 0.0086 & A & 4 &  79.9  $\times$  51.8 &  75.0 \\
   & NGC 5427  & 14:03:26.0 & $-$06:01:50.9 & 0.0087 & A & 4 &  77.0  $\times$  61.9 &   0.0 \\
39 & NGC 5480  & 14:06:21.6 &    50:43:30.4 & 0.0062 & C & 2 &  62.6  $\times$  50.0 &  85.0 \\
   & NGC 5481  & 14:06:41.3 &    50:43:23.9 & 0.0066 & C & 2 &  60.1  $\times$  42.1 &  15.0 \\
40 & NGC 5544  & 14:17:03.0 &    36:34:19.6 & 0.0101 & C & 3 &  69.8  $\times$  42.1 & 150.0 \\
41 & NGC 5614  & 14:24:07.6 &    34:51:31.7 & 0.013  & C & 4 &  79.2  $\times$  45.4 &  45.0 \\
42 & NGC 5846$^{\rm b}$  & 15:06:29.3 &    01:36:20.2 & 0.0057 & C & 2 &  92.2  $\times$  81.4 & 125.0 \\
   & NGC 5850  & 15:07:07.7 &    01:32:39.1 & 0.0085 & C & 2 & 138.0  $\times$ 110.0 &  10.0 \\
43 & NGC 5905  & 15:15:23.3 &    55:31:02.6 & 0.0113 & C & 3 & 160.0  $\times$  75.0 &  38.0 \\
   & NGC 5908  & 15:16:43.2 &    55:24:33.5 & 0.011  & C & 2 &  83.5  $\times$  41.8 &  63.2 \\
44 & NGC 5929  & 15:26:05.4 &    41:40:07.3 & 0.0083 & C & 4 &  22.7  $\times$  20.2 & 130.0 \\
   & NGC 5930  & 15:26:08.2 &    41:40:44.0 & 0.0087 & C & 4 &  67.0  $\times$  23.4 &  70.0 \\
45 & NGC 5953  & 15:34:31.9 &    15:11:38.8 & 0.0066 & A & 4 &  31.0  $\times$  28.1 & 160.0 \\
   & NGC 5954  & 15:34:34.8 &    15:12:05.8 & 0.0065 & A & 4 &  38.5  $\times$  20.9 &  90.0 \\
46 & NGC 5981  & 15:37:53.4 &    59:23:30.5 & 0.0059 & C & 2 &  95.0  $\times$  15.5 &  49.2 \\
   & NGC 5985  & 15:39:37.1 &    59:19:54.8 & 0.0084 & C & 2 & 169.9  $\times$  82.1 & 110.0 \\
47 & Arp 314A  & 22:58:02.2 & $-$03:46:10.9 & 0.0123 & A & 4 &  34.9  $\times$  28.1 & 115.0 \\
   & Arp 314C$^{\rm b}$
               & 22:58:07.4 & $-$03:48:41.4 & 0.0123 & A & 4 &  38.2  $\times$  31.0 &  70.0 \\
   & Arp 314B  & 22:58:07.9 & $-$03:47:19.7 & 0.0124 & A & 4 &  34.9  $\times$  32.0 &  90.0 \\
48 & NGC 7715  & 23:36:22.1 &    02:09:23.4 & 0.0092 & A & 4 &  52.9  $\times$  23.4 & 160.0 \\
   & NGC 7714  & 23:36:14.1 &    02:09:18.6 & 0.0093 & A & 4 &  81.7  $\times$  53.3 & 155.0 \\
\hline
\end{tabular}

\begin{flushleft}
\textbf{Note:}  Group IDs, Redshifts, Sample, and Interaction Stages are taken from 
\citetalias{2015ApJS..218....6B}, as described in Sec.~\ref{ssec:SIGS}.  A Sample of C or A indicates objects belonging to the Keel-Complete sample or the Arp sample, respectively.  The RA, Dec, Size, and Angle columns correspond to the centroids, semi-axis lengths, and position angles of the elliptical apertures used for the photometry as described in Sec.~\ref{ssec:apertures}.\\
$^{\rm a}$ Angles are given in degrees from the East as measured by \textit{photutils} (Sec.~\ref{ssec:background}), so Angle=PA$-$90 degrees. \\
$^{\rm b}$ This galaxy was not analyzed in the SED models described in Sec.~\ref{ssec:SED} because the photometry was too sparse to support reliable SED models.
\end{flushleft}
\end{table*}

\begin{table*}
\centering
\caption{Basic data for the SB sample galaxies.}
\label{tab:SBsample}
\begin{tabular}{ccrlccr}
\hline
Galaxy & RA & Dec\ \ \  & Redshift & Sample & Size & Angle$^{\rm a}$\\
ID &\multicolumn{2}{c}{(J2000)} & \ \ \ ($z$) & &($\arcsec \times \arcsec$) & ($\degr$)\ \ \ \\
\hline
NGC   23 & 00:09:53.4 &    25:55:25.6 & 0.0152 & S  &  86.3 $\times$  58.2 &  90.0 \\
NGC  253 & 00:47:32.4 & $-$25:17:44.0 & 0.0008 & S  & 820.4 $\times$ 226.5 & 140.0 \\
NGC  660 & 01:43:02.4 &    13:38:42.2 & 0.0028 & SB & 304.8 $\times$ 124.2 &  75.0 \\
NGC 1222 & 03:08:56.7 & $-$02:57:18.5 & 0.0081 & B  &  73.4 $\times$  60.0 &  70.0 \\
NGC 1365 & 03:33:36.4 & $-$36:08:28.2 & 0.0055 & B  & 353.8 $\times$ 221.6 & 128.0 \\
IC   342 & 03:46:48.5 &    68:05:46.9 & 0.0001 & B  & 716.5 $\times$ 598.5 &   0.0 \\
NGC 1614 & 04:33:59.8 & $-$08:34:44.0 & 0.0159 & B  &  82.9 $\times$  54.9 & 114.7 \\
NGC 1797 & 05:07:44.9 & $-$08:01:08.7 & 0.0149 & S  &  66.7 $\times$  41.0 & 162.9 \\
NGC 2146 & 06:18:37.7 &    78:21:25.3 & 0.003  & B  & 174.8 $\times$ 125.9 & 210.0 \\
NGC 2623 & 08:38:24.0 &    25:45:16.1 & 0.0185 & B  &  76.4 $\times$  45.8 & 160.0 \\
NGC 3256 & 10:27:51.3 & $-$43:54:13.5 & 0.0094 & SB & 224.4 $\times$ 135.8 &  10.0 \\
NGC 3310 & 10:38:45.9 &    53:30:12.2 & 0.0033 & B  & 126.4 $\times$ 104.4 &  93.3 \\
NGC 3556 & 11:11:31.0 &    55:40:26.8 & 0.0023 & B  & 262.2 $\times$  85.7 & 170.0 \\
NGC 3628 & 11:20:17.0 &    13:35:22.9 & 0.0028 & B  & 508.4 $\times$ 219.2 &  13.0 \\
NGC 4088 & 12:05:34.2 &    50:32:20.5 & 0.0025 & SB & 187.7 $\times$ 103.1 & 143.0 \\
NGC 4194 & 12:14:09.5 &    54:31:36.6 & 0.0083 & B  &  92.2 $\times$  53.4 &  90.0 \\
Mrk   52 & 12:25:42.8 &    00:34:21.4 & 0.0071 & B  &  62.7 $\times$  38.4 & 170.0 \\
NGC 4676 & 12:46:10.7 &    30:43:38.0 & 0.022  & B  & 155.1 $\times$  54.5 &  90.0 \\
NGC 4818 & 12:56:48.8 & $-$08:31:37.0 & 0.0036 & B  & 145.8 $\times$  66.0 &  90.0 \\
NGC 4945 & 13:05:27.5 & $-$49:28:05.6 & 0.0019 & SB & 571.9 $\times$ 150.0 & 133.0 \\
NGC 7252 & 22:20:44.7 & $-$24:40:41.7 & 0.016  & B  &  71.0 $\times$  62.7 &  80.0 \\
\hline
\end{tabular}

\begin{flushleft}
\textbf{Note:}  The RA, Dec, Size, and Angle columns define the centroids, semi-axis lengths, and angles of the elliptical apertures used for the photometry of the starburst sample galaxies. Redshifts were taken from NED.  Samples are B for \citet{2006ApJ...653.1129B} and S for added well known local starbursts.\\
$^{\rm a}$ Angles are given in degrees from the East as measured by \textit{photutils} (Sec.~\ref{ssec:background}), so Angle=PA$-$90 degrees. 

\end{flushleft}
\end{table*}

\begin{table*}
\centering
\caption{Basic data for the AGN sample galaxies.}
\label{tab:AGN}
\begin{tabular}{ccrlccr}
\hline
Galaxy & RA & Dec\ \ \  & Redshift & Sample & Size & Angle$^{\rm a}$\\
ID &\multicolumn{2}{c}{(J2000)} & \ \ \ ($z$) & &($\arcsec \times \arcsec$) & ($\degr$)\ \ \ \\
\hline
Mrk 335 & 00:06:19.5 & 20:12:10.5 & 0.0258 & S & 26.9 $\times$ 26.8 & 84.2 \\
Mrk 1502 & 00:53:34.9 & 12:41:36.2 & 0.0589 & S & 29.0 $\times$ 28.5 & 45.0 \\
NGC 931 & 02:28:14.5 & 31:18:42.0 & 0.0167 & S & 100.9 $\times$ 35.7 & 165.7 \\
NGC 1068 & 02:42:40.7 & $-$00:00:47.8 & 0.0038 & HRG & 215.1 $\times$ 174.7 & 170.0 \\
NGC 1194 & 03:03:49.1 & $-$01:06:13.5 & 0.0136 & S & 100.7 $\times$ 42.7 & 50.6 \\
NGC 1320 & 03:24:48.7 & $-$03:02:32.2 & 0.0089 & S & 63.5 $\times$ 33.9 & 47.2 \\
ESO 33$-$2 & 04:55:59.0 & $-$75:32:28.2 & 0.0181 & S & 31.0 $\times$ 29.0 & 45.0 \\
4U 0557$-$385 & 05:58:02.0 & $-$38:20:04.7 & 0.0339 & S & 25.0 $\times$ 23.0 & 229.3 \\
Mrk 3 & 06:15:36.4 & 71:02:15.1 & 0.0135 & S & 48.6 $\times$ 42.9 & 55.0 \\
ESO 428$-$14 & 07:16:31.2 & $-$29:19:29.0 & 0.0057 & S & 46.2 $\times$ 32.0 & 230.0 \\
NGC 3281 & 10:31:52.1 & $-$34:51:13.3 & 0.0107 & S & 114.6 $\times$ 54.7 & 55.0 \\
NGC 3516 & 11:06:47.5 & 72:34:06.9 & 0.0088 & S & 53.3 $\times$ 52.4 & 33.8 \\
NGC 4151 & 12:10:32.6 & 39:24:20.6 & 0.0033 & HR & 256.1 $\times$ 246.5 & 50.0 \\
NGC 4388 & 12:25:46.8 & 12:39:43.5 & 0.0084 & S & 200.7 $\times$ 54.2 & 179.4 \\
Mrk 771 & 12:32:03.6 & 20:09:29.2 & 0.063 & S & 27.8 $\times$ 26.6 & 21.1 \\
NGC 4941 & 13:04:13.1 & $-$05:33:05.8 & 0.0037 & S & 121.6 $\times$ 78.3 & 286.9 \\
MCG $-$03$-$34$-$064 & 13:22:24.5 & $-$16:43:42.5 & 0.0165 & G & 41.5 $\times$ 31.0 & 141.9 \\
ESO 383$-$35 & 13:35:53.7 & $-$34:17:43.9 & 0.0077 & S & 27.9 $\times$ 19.4 & 29.7 \\
ESO 445$-$50 & 13:49:19.3 & $-$30:18:34.0 & 0.0161 & S & 42.0 $\times$ 26.0 & 133.7 \\
NGC 5506 & 14:13:14.9 & $-$03:12:27.3 & 0.0062 & S & 87.3 $\times$ 34.9 & 179.8 \\
2XMM J141348.3$+$440014 & 14:13:48.3 & 44:00:14.0 & 0.0896 & S & 28.0 $\times$ 24.0 & 54.3 \\
NGC 5548 & 14:17:59.5 & 25:08:12.4 & 0.0172 & S & 54.2 $\times$ 44.6 & 190.2 \\
Mrk 1383 & 14:29:06.6 & 01:17:06.5 & 0.0866 & S & 29.2 $\times$ 28.5 & 3.9 \\
Mrk 841 & 15:04:01.2 & 10:26:16.1 & 0.0364 & S & 37.5 $\times$ 34.0 & 105.0 \\
ESO 141$-$55 & 19:21:14.1 & $-$58:40:13.1 & 0.0371 & S & 33.0 $\times$ 29.0 & 0.0 \\
IC 5063 & 20:52:02.3 & $-$57:04:07.6 & 0.0113 & S &79.8 $\times$ 62.6 & 205.1 \\
Mrk 1513 & 21:32:27.8 & 10:08:19.5 & 0.063 & S & 26.2 $\times$ 23.9 & 328.2 \\
Leda 68751 & 22:23:49.5 & $-$02:06:12.8 & 0.0559 & S & 20.0 $\times$ 18.0 & 15.0 \\
NGC 7674 & 23:27:57.0 & 08:46:43.3 & 0.0289 & HRG & 50.0 $\times$ 50.0 & 0.0 \\
\hline
\hline
\end{tabular}

\begin{flushleft}
\textbf{Note:} The RA, Dec, Size, and Angle columns define the centroids, semi-axis lengths, and angles of the elliptical apertures used for the photometry of the AGN sample galaxies. Redshifts were taken from NED. The Sample column indicates whether objects belong to the GOALS sample (G), \citet{2013RMxAA..49..301H} (HR) or taken from SIMBAD (S).\\
$^{\rm a}$ Angles are given in degrees from the East as measured by \textit{photutils} (Sec.~\ref{ssec:background}), so Angle=PA$-$90 degrees. 
\end{flushleft}
\end{table*}

\begin{table*}
\centering
\caption{Basic data for 38 of the 49 LSM sample galaxies.}
\label{tab:LSM}
\begin{tabular}{ccrlrlcr}
\hline
Galaxy & RA & Dec\ \ \  & Redshift & $\log(L_{IR})$ & Stage & Size & Angle$^{\rm a}$\\
ID &\multicolumn{2}{c}{(J2000)} & \ \ \ ($z$) & ($L_{\odot}$)\ \ \  & &($\arcsec \times \arcsec$) & ($\degr$)\ \ \ \\
\hline
NGC 0078                & 00:20:26.6 &    00:49:46.7 & 0.0183 &  9.98 & 2.0 & $60.0  \times 35.0 $ & 145.0 \\
UM 246                  & 00:29:45.1 &    00:10:09.0 & 0.0594 & 10.83 & 4.5 & $60.0  \times 30.0 $ &  45.0 \\
2MASX J01221811+0100262 & 01:22:17.8 &    01:00:27.5 & 0.0555 & 11.54 & 4.0 & $39.0  \times 36.0 $ &  30.0 \\
CGCG 087-046            & 07:54:31.8 &    16:48:26.3 & 0.0463 & 11.28 & 4.0 & $45.0  \times 35.0 $ & 115.0 \\
UGC 04383               & 08:23:33.5 &    21:20:34.7 & 0.0179 & 10.46 & 3.0 & $55.0  \times 38.0 $ & 125.0 \\
2MASX J08343370+1720462 & 08:34:33.7 &    17:20:46.4 & 0.0479 & 10.86 & 4.0 & $30.0  \times 28.0 $ & 115.0 \\
2MASX J08381760+3054533 & 08:38:17.6 &    30:54:53.5 & 0.0477 & 10.62 & 4.0 & $28.0  \times 22.0 $ & 110.0 \\
2MASX J08434495+3549421 & 08:43:45.0 &    35:49:42.0 & 0.054  & 10.45 & 5.0 & $28.0  \times 25.0 $ &  55.0 \\
UGC 05044               & 09:27:44.0 &    12:17:12.3 & 0.029  & 10.52 & 4.0 & $50.0  \times 45.0 $ & 110.0 \\
Arp 142                 & 09:37:44.0 &    02:45:15.1 & 0.0233 & 10.89 & 3.5 & $65.0  \times 55.0 $ & 130.0 \\
CGCG 266-026            & 10:10:00.8 &    54:40:19.8 & 0.0462 & 10.91 & 4.5 & $50.0  \times 32.0 $ & 155.0 \\
LSBCF 567-01            & 10:19:01.5 &    21:17:01.3 & 0.0036 &  8.38 & 5.0 & $65.0  \times 35.0 $ &  65.0 \\
2MASX J10225654+3446467 & 10:22:56.6 &    34:46:46.8 & 0.0561 & 10.66 & 3.5 & $40.0  \times 32.0 $ &  65.0 \\
UGC 05644               & 10:25:46.3 &    13:43:00.7 & 0.0323 & 10.52 & 3.0 & $75.0  \times 32.0 $ & 130.0 \\
CGCG 037-076            & 10:33:28.6 &    07:08:03.8 & 0.0445 & 10.79 & 4.0 & $33.0  \times 28.0 $ &  70.0 \\
UGC A219                & 10:49:05.0 &    52:20:07.8 & 0.008  &  8.82 & 5.0 & $30.0  \times 25.0 $ &  60.0 \\
NGC 3445                & 10:54:35.5 &    56:59:26.5 & 0.0068 &  9.76 & 3.0 & $50.0  \times 45.0 $ &  70.0 \\
2MASX J10591815+2432343 & 10:59:18.1 &    24:32:34.5 & 0.0431 & 12.18 & 3.5 & $50.0  \times 42.0 $ &  20.0 \\
VV 627                  & 11:00:59.8 &    57:47:04.0 & 0.0477 & 10.60 & 4.0 & $70.0  \times 40.0 $ & 145.0 \\
IC 0700                 & 11:29:15.5 &    20:35:05.7 & 0.0049 &  8.57 & 5.0 & $55.0  \times 30.0 $ & 155.0 \\
UGC 06665               & 11:42:12.4 &    00:20:02.5 & 0.0186 & 10.83 & 5.0 & $60.0  \times 48.0 $ & 125.0 \\
UGC 07388               & 12:20:15.7 &    33:39:38.9 & 0.0215 & 10.52 & 5.0 & $40.0  \times 35.0 $ & 125.0 \\
NGC 4320                & 12:22:57.7 &    10:32:54.0 & 0.0267 & 10.55 & 4.5 & $60.0  \times 45.0 $ & 115.0 \\
UGC 07936               & 12:46:00.1 &    45:12:00.0 & 0.0247 & 10.23 & 4.0 & $65.0  \times 50.0 $ & 130.0 \\
Mrk 0237                & 13:01:17.6 &    48:03:38.0 & 0.0298 & 10.87 & 4.0 & $50.0  \times 38.0 $ &  30.0 \\
UGC 08327               & 13:15:15.6 &    44:24:26.0 & 0.0367 & 11.17 & 3.5 & $60.0  \times 42.0 $ &   0.0 \\
UGC 08335               & 13:15:33.1 &    62:07:30.4 & 0.0308 & 11.70 & 4.0 & $60.0  \times 35.0 $ &  35.0 \\
NGC 5100                & 13:20:58.6 &    08:58:55.0 & 0.0319 & 11.21 & 3.5 & $60.0  \times 50.0 $ &  50.0 \\
CGCG 017-018            & 13:32:55.9 & $-$03:01:37.0 & 0.0465 & 10.86 & 2.0 & $38.0  \times 25.0 $ &  20.0 \\
NGC 5331                & 13:52:16.3 &    02:06:10.9 & 0.033  & 11.23 & 3.0 & $65.0  \times 50.0 $ &  80.0 \\
CGCG 076-015$^{\rm b}$  
                        & 14:44:27.1 &    12:15:25.8 & 0.0528 & 10.96 & 4.0 & $23.0  \times 23.0 $ &   0.0 \\
UGC 09618               & 14:57:00.5 &    24:36:49.9 & 0.0329 & 11.26 & 2.0 & $60.0  \times 40.0 $ &  95.0 \\
2MASX J15015015+2332536 & 15:01:50.2 &    23:32:53.7 & 0.0463 & 10.92 & 4.5 & $32.0  \times 27.0 $ &  60.0 \\
SBS 1509+583            & 15:10:17.8 &    58:10:37.5 & 0.0319 & 10.37 & 2.0 & $35.0  \times 25.0 $ &  30.0 \\
KUG 1553+200            & 15:55:56.9 &    19:56:58.0 & 0.0413 & 11.00 & 4.0 & $32.0  \times 32.0 $ &   0.0 \\
KUG 1556+326            & 15:58:37.8 &    32:27:42.2 & 0.0482 & 10.76 & 3.0 & $38.0  \times 32.0 $ & 110.0 \\
Mrk 0881                & 16:25:49.4 &    40:20:42.7 & 0.0288 & 10.90 & 5.0 & $35.0  \times 30.0 $ &  95.0 \\
2MASX J17045097+3449020 & 17:04:50.9 &    34:49:02.4 & 0.0563 & 11.31 & 2.0 & $35.0  \times 30.0 $ &  40.0 \\
\hline
\hline
\end{tabular}

\begin{flushleft}
\textbf{Note:} The RA, Dec, Size, and Angle columns define the centroids, semi-axis lengths, and angles of the elliptical apertures used for the photometry of the LSM sample galaxies.  This table presents data for 38 galaxies; corresponding quantities for the remainder of the 49-galaxy sample appear in Table~1 of \citet{2018MNRAS.480.3562D}.  Redshifts were taken from NED.  Infrared luminosities were taken from the Revised IRAS-FSC Redshift Catalog \citep[RIFSCz,][]{2014MNRAS.442.2739W}.  
The stages were determined by the entire team as described in Sec.~\ref{ssec:SIGS}.\\  $^{\rm a}$ Angles are given in degrees from the East as measured by \textit{photutils} (Sec.~\ref{ssec:background}), so Angle=PA$-$90 degrees. \\
$^{\rm b}$ This galaxy was not analyzed in the SED models described in Sec.~\ref{ssec:SED} because the photometry was too sparse to support reliable SED models.
\end{flushleft}
\end{table*}

% %% SPECTRAINFO

\begin{table*}
\centering
\caption{Integrated IR line intensities measured for SB sample galaxies using PAHFIT in units of $\mathrm{1 \times 10^{-21}\,W\,cm^{-2}}$.}
\label{tab:SHlines}
\begin{tabular}{ccccccccc}
\hline
Galaxy ID & [\ion{Ne}{ii}] & [\ion{Ne}{iii}] & [\ion{Ne}{v}] & [\ion{S}{iii}] & [\ion{S}{iv}] & [\ion{Fe}{ii}] & H$_{2}$ S(2) & H$_{2}$ S(1)\\
&  12.81~\micron & 15.56~\micron & 14.32~\micron & 18.71~\micron & 10.51~\micron & 12.64~\micron & 12.28~\micron & 17.03~\micron \\
\hline
NGC 23 & 49.37$\pm$0.14 & 6.95$\pm$0.04 & $\cdots$ & 16.08$\pm$0.14 & 1.62$\pm$0.06 & 0.48$\pm$0.06 & 5.71$\pm$0.11 & 6.65$\pm$0.07 \\
NGC 253 & 2689.15$\pm$1.29 & 183.27$\pm$0.57 & $\cdots$ & 576.23$\pm$1.08 & $\cdots$ & 77.21$\pm$0.83 & 93.53$\pm$0.80 & 79.39$\pm$0.67 \\
NGC 520 & 46.18$\pm$0.21 & 6.84$\pm$0.09 & 0.23$\pm$0.08 & 6.71$\pm$0.11 & 0.13$\pm$0.09 & $\cdots$ & 5.24$\pm$0.12 & 8.89$\pm$0.10 \\
NGC 660 & 286.20$\pm$0.34 & 27.60$\pm$0.19 & 1.30$\pm$0.14 & 54.78$\pm$0.24 & 1.47$\pm$0.22 & 3.40$\pm$0.14 & 6.30$\pm$0.13 & 17.82$\pm$0.20 \\
NGC 1222 & 82.21$\pm$0.16 & 78.18$\pm$0.13 & 0.47$\pm$0.06 & 49.33$\pm$0.17 & 25.16$\pm$0.12 & 0.69$\pm$0.09 & 4.79$\pm$0.09 & 7.92$\pm$0.08 \\
NGC 1365 & 178.44$\pm$0.37 & 55.36$\pm$0.23 & 18.74$\pm$0.20 & 55.25$\pm$0.30 & 28.31$\pm$0.29 & 4.35$\pm$0.17 & 12.61$\pm$0.18 & 20.18$\pm$0.18 \\
IC 342 & 639.63$\pm$0.38 & 34.58$\pm$0.16 & 2.31$\pm$0.16 & 255.47$\pm$0.39 & 5.20$\pm$0.21 & 11.80$\pm$0.21 & 12.01$\pm$0.26 & 9.30$\pm$0.18 \\
NGC 1614 & 265.07$\pm$0.30 & 66.59$\pm$0.19 & 1.63$\pm$0.10 & 70.91$\pm$0.30 & 11.65$\pm$0.17 & 1.97$\pm$0.14 & 6.84$\pm$0.15 & 10.05$\pm$0.14 \\
NGC 1797 & 67.82$\pm$0.19 & 5.33$\pm$0.07 & 0.34$\pm$0.12 & 18.63$\pm$0.16 & 1.11$\pm$0.09 & 0.53$\pm$0.10 & 3.50$\pm$0.14 & 3.92$\pm$0.10 \\
NGC 2146 & 803.04$\pm$1.52 & 121.59$\pm$0.43 & 2.99$\pm$0.31 & 253.62$\pm$1.17 & 21.55$\pm$0.82 & 10.90$\pm$0.30 & 10.07$\pm$0.28 & 37.91$\pm$0.35 \\
NGC 2623 & 73.10$\pm$0.57 & 20.68$\pm$0.26 & 3.54$\pm$0.11 & 13.55$\pm$0.32 & 7.31$\pm$0.42 & 1.31$\pm$0.21 & 5.51$\pm$0.17 & 14.12$\pm$0.31 \\
NGC 3256 & 495.57$\pm$0.50 & 61.04$\pm$0.18 & 0.92$\pm$0.15 & 138.36$\pm$0.35 & 10.49$\pm$0.17 & 8.29$\pm$0.20 & 20.34$\pm$0.17 & 26.12$\pm$0.16 \\
NGC 3310 & 43.45$\pm$0.22 & 28.62$\pm$0.15 & 0.25$\pm$0.10 & 19.36$\pm$0.26 & 5.56$\pm$0.15 & 0.46$\pm$0.09 & 0.84$\pm$0.16 & 2.04$\pm$0.13 \\
NGC 3556 & 21.51$\pm$0.10 & 3.01$\pm$0.05 & 0.06$\pm$0.06 & 10.28$\pm$0.12 & 0.42$\pm$0.10 & 0.24$\pm$0.08 & 1.67$\pm$0.12 & 2.46$\pm$0.06 \\
NGC 3628 & 170.98$\pm$0.66 & 13.28$\pm$0.12 & 0.69$\pm$0.09 & 32.91$\pm$0.31 & $\cdots$ & 2.92$\pm$0.18 & 10.02$\pm$0.21 & 28.46$\pm$0.29 \\
NGC 4088 & 35.01$\pm$0.14 & 2.24$\pm$0.06 & 0.43$\pm$0.16 & 9.31$\pm$0.08 & 0.28$\pm$0.13 & 0.28$\pm$0.06 & 2.45$\pm$0.09 & 3.97$\pm$0.06 \\
NGC 4194 & 162.22$\pm$0.26 & 50.57$\pm$0.17 & 1.99$\pm$0.09 & 56.92$\pm$0.22 & 15.11$\pm$0.21 & 2.50$\pm$0.14 & 6.63$\pm$0.15 & 7.26$\pm$0.13 \\
Mrk 52 & 24.39$\pm$0.09 & 3.34$\pm$0.07 & $\cdots$ & 17.71$\pm$0.09 & 1.11$\pm$0.07 & 0.41$\pm$0.06 & 1.00$\pm$0.06 & 1.54$\pm$0.07 \\
NGC 4676 & 34.68$\pm$0.23 & 6.87$\pm$0.12 & 0.44$\pm$0.05 & 16.18$\pm$0.22 & 2.63$\pm$0.20 & 0.65$\pm$0.08 & 3.07$\pm$0.09 & 8.86$\pm$0.15 \\
NGC 4818 & 154.06$\pm$0.22 & 11.12$\pm$0.12 & 0.97$\pm$0.12 & 47.40$\pm$0.24 & 2.35$\pm$0.10 & 4.34$\pm$0.17 & 6.11$\pm$0.32 & 13.59$\pm$0.15 \\
NGC 4945 & 978.73$\pm$2.18 & 114.10$\pm$0.49 & 9.49$\pm$0.27 & 106.69$\pm$0.77 & 31.63$\pm$2.62 & 17.47$\pm$0.50 & 80.48$\pm$0.50 & 228.37$\pm$1.41 \\
NGC 7252 & 36.31$\pm$0.13 & 2.99$\pm$0.06 & 0.34$\pm$0.06 & 8.28$\pm$0.09 & 0.12$\pm$0.07 & 0.08$\pm$0.05 & 3.12$\pm$0.13 & 3.56$\pm$0.07 \\
NGC 7714 & 94.55$\pm$0.17 & 63.77$\pm$0.11 & $\cdots$ & 56.86$\pm$0.14 & 16.38$\pm$0.15 & 2.26$\pm$0.11 & 3.44$\pm$0.10 & 3.19$\pm$0.09 \\
\hline
\end{tabular}
\end{table*}

% %% PHOTOMETRY

\begin{landscape}
\begin{table*}
\centering
\caption{{\sl GALEX} and SDSS DR12 photometry for the four study samples.}
\label{tab:all_phot_galex_sdss}
\begin{tabular}{cccccccc}
\hline
Galaxy & \multicolumn{2}{c}{\sl GALEX} & \multicolumn{5}{c}{SDSS DR12} \\
ID     & FUV & NUV & $u$ & $g$ & $r$ & $i$ & $z$ \\
       & (mJy) & (mJy) & (mJy) & (mJy) & (mJy) & (mJy) & (mJy) \\
\hline  
\multicolumn{7}{c}{Photometry for the SIGS sample} \\
\hline
NGC274 & $\cdots$ & $\cdots$ & 4.53 $\pm$ 0.13 & 21.15 $\pm$ 0.42 & 43.97 $\pm$ 0.88 & 64.74 $\pm$ 1.30 & 81.14 $\pm$ 1.66 \\
NGC275 & 3.30 $\pm$ 0.33 & 4.56 $\pm$ 0.46 & 10.80 $\pm$ 0.24 & 23.35 $\pm$ 0.47 & 36.91 $\pm$ 0.74 & 43.79 $\pm$ 0.88 & 50.87 $\pm$ 1.08 \\
NGC470 & 3.18 $\pm$ 0.32 & 4.67 $\pm$ 0.47 & 16.59 $\pm$ 0.37 & 44.43 $\pm$ 0.89 & 78.42 $\pm$ 1.57 & 107.23 $\pm$ 2.15 & 131.44 $\pm$ 2.67 \\
NGC474 & 1.44 $\pm$ 0.15 & 1.98 $\pm$ 0.20 & 20.66 $\pm$ 0.61 & 67.87 $\pm$ 1.36 & 125.69 $\pm$ 2.52 & 179.48 $\pm$ 3.61 & 243.94 $\pm$ 5.05 \\
NGC520 & 1.69 $\pm$ 0.17 & 3.12 $\pm$ 0.31 & 4.25 $\pm$ 0.22 & 62.00 $\pm$ 1.25 & 113.32 $\pm$ 2.27 & 160.68 $\pm$ 3.22 & 192.31 $\pm$ 3.92 \\
IC195 & 0.05 $\pm$ 0.01 & 0.10 $\pm$ 0.01 & 1.96 $\pm$ 0.06 & 9.48 $\pm$ 0.19 & 19.99 $\pm$ 0.40 & 29.45 $\pm$ 0.59 & 37.98 $\pm$ 0.77 \\
IC196 & 0.65 $\pm$ 0.06 & 0.82 $\pm$ 0.08 & 4.33 $\pm$ 0.14 & 18.03 $\pm$ 0.36 & 36.51 $\pm$ 0.73 & 53.72 $\pm$ 1.08 & 68.04 $\pm$ 1.41 \\
NGC833 & $\cdots$ & 0.29 $\pm$ 0.03 & 3.33 $\pm$ 0.09 & 16.49 $\pm$ 0.33 & 35.86 $\pm$ 0.72 & 53.35 $\pm$ 1.07 & 68.14 $\pm$ 1.39 \\
NGC835 & $\cdots$ & 1.67 $\pm$ 0.17 & 7.74 $\pm$ 0.17 & 29.07 $\pm$ 0.58 & 57.24 $\pm$ 1.15 & 81.33 $\pm$ 1.63 & 102.36 $\pm$ 2.07 \\
NGC838 & $\cdots$ & 2.04 $\pm$ 0.20 & 6.39 $\pm$ 0.14 & 16.92 $\pm$ 0.34 & 30.38 $\pm$ 0.61 & 39.34 $\pm$ 0.79 & 48.92 $\pm$ 1.02 \\
\hline
\end{tabular}

\begin{flushleft}
\textbf{Note:} Photometry expressed in mJy in the UV and optical bands for the SIGS,
SB, AGN, and LSM samples described in Sec.~\ref{Sec:Obs}.  The full table is available
in the online version of this paper.  A portion is shown here for guidance regarding its
form and content. \\
\end{flushleft}
\end{table*}
% \end{landscape}

% \begin{landscape}
\begin{table*}
\centering
\caption{2MASS and {\sl Spitzer}/IRAC photometry for the four study samples.}
\label{tab:all_phot_2mass_irac}
\begin{tabular}{cccccccc}
\hline
Galaxy & \multicolumn{3}{c}{2MASS} & \multicolumn{4}{c}{{\sl Spitzer}/IRAC} \\
ID     & $J$ & $H$ & $K_s$ & 3.6\,$\mu$m & 4.5\,$\mu$m & 5.8\,$\mu$m & 8.0\,$\mu$m \\
       & (mJy) & (mJy) & (mJy) & (mJy) & (mJy) & (mJy) & (mJy) \\
\hline
\multicolumn{7}{c}{Photometry for the SIGS sample} \\
\hline
NGC274 & 111.29 $\pm$ 2.28 & 132.95 $\pm$ 2.76 & 102.24 $\pm$ 2.28 & 50.25 $\pm$ 1.51 & 31.65 $\pm$ 0.95 & 23.19 $\pm$ 0.70 & 17.06 $\pm$ 0.52 \\
NGC275 & 64.31 $\pm$ 1.41 & 73.18 $\pm$ 1.68 & 59.58 $\pm$ 1.68 & 39.06 $\pm$ 1.17 & 27.29 $\pm$ 0.82 & 70.69 $\pm$ 2.13 & 170.52 $\pm$ 5.12 \\
NGC470 & 186.08 $\pm$ 3.96 & 199.56 $\pm$ 4.58 & 184.86 $\pm$ 4.47 & 108.22 $\pm$ 3.25 & 73.47 $\pm$ 2.20 & 163.96 $\pm$ 4.93 & 417.70 $\pm$ 12.53 \\
NGC474 & 296.05 $\pm$ 7.02 & 314.31 $\pm$ 8.98 & 271.30 $\pm$ 8.35 & 149.83 $\pm$ 4.50 & 86.50 $\pm$ 2.60 & 141.57 $\pm$ 4.34 & 100.29 $\pm$ 3.14 \\
NGC520 & 288.70 $\pm$ 6.23 & 357.34 $\pm$ 8.33 & 312.68 $\pm$ 7.30 & 182.73 $\pm$ 5.48 & 138.65 $\pm$ 4.16 & 357.79 $\pm$ 10.74 & 916.85 $\pm$ 27.51 \\
IC195 & 54.38 $\pm$ 1.22 & 66.50 $\pm$ 1.60 & 53.22 $\pm$ 1.47 & 25.29 $\pm$ 0.76 & 15.80 $\pm$ 0.47 & 10.82 $\pm$ 0.34 & 6.30 $\pm$ 0.21 \\
IC196 & 95.34 $\pm$ 2.39 & 129.52 $\pm$ 3.48 & 105.90 $\pm$ 3.34 & 49.53 $\pm$ 1.49 & 32.17 $\pm$ 0.97 & 29.94 $\pm$ 0.92 & 42.85 $\pm$ 1.30 \\
NGC833 & 101.56 $\pm$ 2.08 & 125.16 $\pm$ 2.62 & 106.44 $\pm$ 2.32 & 48.47 $\pm$ 1.45 & 30.78 $\pm$ 0.92 & 25.01 $\pm$ 0.76 & 25.52 $\pm$ 0.77 \\
NGC835 & 155.86 $\pm$ 3.17 & 194.36 $\pm$ 4.01 & 166.35 $\pm$ 3.51 & 89.23 $\pm$ 2.68 & 60.95 $\pm$ 1.83 & 117.65 $\pm$ 3.53 & 291.88 $\pm$ 8.76 \\
NGC838 & 73.96 $\pm$ 1.57 & 93.24 $\pm$ 2.04 & 86.14 $\pm$ 1.97 & 67.69 $\pm$ 2.03 & 50.08 $\pm$ 1.50 & 212.29 $\pm$ 6.37 & 589.03 $\pm$ 17.67 \\
\hline
\end{tabular}

\begin{flushleft}
\textbf{Note:} Photometry in seven near- and mid-IR bands for the SIGS,
SB, AGN, and LSM samples described in Sec.~\ref{Sec:Obs}.  The full table is available
in the online version of this paper.  A portion is shown here for guidance regarding its
form and content. \\
\end{flushleft}
\end{table*}
\end{landscape}

\begin{landscape}
\begin{table*}
\centering
\caption{{\sl WISE} and {\sl Spitzer}/MIPS photometry for the four study samples.}
\label{tab:all_phot_wise_mips}
\begin{tabular}{cccccccc}
\hline
Galaxy & \multicolumn{4}{c}{\sl WISE} & \multicolumn{3}{c}{{\sl Spitzer}/MIPS} \\
ID     & 3.4\,$\mu$m & 4.6\,$\mu$m & 12\,$\mu$m & 22\,$\mu$m & 24\,$\mu$m & 70\,$\mu$m & 160\,$\mu$m \\
       & (mJy) & (mJy) & (mJy) & (mJy) & (mJy) & (Jy) & (Jy) \\
\hline
\multicolumn{7}{c}{Photometry for the SIGS sample} \\
\hline
NGC274 & 52.6 $\pm$ 3.2 & 28.3 $\pm$ 1.7 & 11.3 $\pm$ 0.7 & $\cdots$ & $\cdots$ & $\cdots$ & $\cdots$ \\
NGC275 & 38.1 $\pm$ 2.3 & 24.5 $\pm$ 1.5 & 146.5 $\pm$ 8.8 & 384.6 $\pm$ 23.1 & 459.32 $\pm$ 18.39 & 5.40 $\pm$ 0.22 & 7.34 $\pm$ 0.32 \\
NGC470 & 112.5 $\pm$ 6.8 & 69.5 $\pm$ 4.2 & 334.5 $\pm$ 20.1 & 802.2 $\pm$ 48.1 & 799.55 $\pm$ 32.02 & 9.59 $\pm$ 0.39 & 13.72 $\pm$ 0.55 \\
NGC474 & 151.8 $\pm$ 9.1 & 83.3 $\pm$ 5.0 & $\cdots$ & $\cdots$ & 152.39 $\pm$ 8.99 & $\cdots$ & $\cdots$ \\
NGC520 & 183.0 $\pm$ 11.0 & 131.4 $\pm$ 7.9 & 738.3 $\pm$ 44.3 & 2233.4 $\pm$ 134.0 & 2347.72 $\pm$ 93.94 & 33.44 $\pm$ 1.34 & $\cdots$ \\
IC195 & 26.0 $\pm$ 1.6 & 13.9 $\pm$ 0.8 & 4.2 $\pm$ 0.3 & 3.5 $\pm$ 0.3 & $\cdots$ & $\cdots$ & $\cdots$ \\
IC196 & 52.6 $\pm$ 3.2 & 28.4 $\pm$ 1.7 & 37.5 $\pm$ 2.3 & 36.0 $\pm$ 2.2 & $\cdots$ & $\cdots$ & $\cdots$ \\
NGC833 & 51.8 $\pm$ 3.1 & 29.2 $\pm$ 1.7 & 26.7 $\pm$ 1.6 & 49.6 $\pm$ 3.0 & $\cdots$ & $\cdots$ & $\cdots$ \\
NGC835 & 89.9 $\pm$ 5.4 & 56.5 $\pm$ 3.4 & 242.0 $\pm$ 14.5 & 412.1 $\pm$ 24.7 & 434.53 $\pm$ 17.43 & 6.82 $\pm$ 0.27 & 9.33 $\pm$ 0.38 \\
NGC838 & 62.1 $\pm$ 3.7 & 46.2 $\pm$ 2.8 & 459.2 $\pm$ 27.6 & 1344.6 $\pm$ 80.7 & 1458.64 $\pm$ 58.35 & 11.50 $\pm$ 0.46 & 9.60 $\pm$ 0.41 \\
\hline
\end{tabular}

\begin{flushleft}
\textbf{Note:} Photometry in seven mid-IR bands for the SIGS,
SB, AGN, and LSM samples described in Sec.~\ref{Sec:Obs}.  The full table is available
in the online version of this paper.  A portion is shown here for guidance regarding its
form and content. \\
\end{flushleft}
\end{table*}
% \end{landscape}

% \begin{landscape}
\begin{table*}
\centering
\caption{{\sl Herschel}/PACS and SPIRE photometry for the four study samples.}
\label{tab:all_phot_herschel}
\begin{tabular}{ccccccc}
\hline
Galaxy & \multicolumn{3}{c}{{\sl Herschel}/PACS} & \multicolumn{3}{c}{{\sl Herschel}/SPIRE} \\
ID     & 70\,$\mu$m & 100\,$\mu$m & 160\,$\mu$m & 250\,$\mu$m & 350\,$\mu$m & 500\,$\mu$m \\
       & (Jy) & (Jy) & (Jy) & (Jy) & (Jy) & (Jy) \\
\hline
\multicolumn{7}{c}{Photometry for the SIGS sample} \\
\hline
NGC274 & $\cdots$ & $\cdots$ & $\cdots$ & $\cdots$ & $\cdots$ & $\cdots$ \\
NGC275 & 5.96 $\pm$ 0.60 & 8.71 $\pm$ 0.87 & 7.85 $\pm$ 0.79 & $\cdots$ & $\cdots$ & $\cdots$ \\
NGC470 & $\cdots$ & $\cdots$ & $\cdots$ & $\cdots$ & $\cdots$ & $\cdots$ \\
NGC474 & $\cdots$ & $\cdots$ & $\cdots$ & $\cdots$ & $\cdots$ & $\cdots$ \\
NGC520 & 41.68 $\pm$ 4.17 & 51.86 $\pm$ 5.19 & 39.13 $\pm$ 3.91 & $\cdots$ & $\cdots$ & $\cdots$ \\
IC195 & $\cdots$ & $\cdots$ & $\cdots$ & $\cdots$ & $\cdots$ & $\cdots$ \\
IC196 & $\cdots$ & $\cdots$ & $\cdots$ & $\cdots$ & $\cdots$ & $\cdots$ \\
NGC833 & 0.39 $\pm$ 0.07 & 0.79 $\pm$ 0.10 & 1.14 $\pm$ 0.12 & 0.58 $\pm$ 0.04 & 0.27 $\pm$ 0.02 & 0.10 $\pm$ 0.01 \\
NGC835 & 7.44 $\pm$ 0.75 & 11.72 $\pm$ 1.17 & 10.58 $\pm$ 1.06 & 4.04 $\pm$ 0.28 & 1.49 $\pm$ 0.10 & 0.46 $\pm$ 0.03 \\
NGC838 & 14.98 $\pm$ 1.50 & 17.84 $\pm$ 1.78 & 12.99 $\pm$ 1.30 & 3.91 $\pm$ 0.27 & 1.37 $\pm$ 0.10 & 0.41 $\pm$ 0.03 \\
\hline
\end{tabular}

\begin{flushleft}
\textbf{Note:} Photometry in the FIR bands for the SIGS,
SB, AGN, and LSM samples described in Sec.~\ref{Sec:Obs}.  The full table is available
in the online version of this paper.  A portion is shown here for guidance regarding its
form and content. \\
\end{flushleft}
\end{table*}
\end{landscape}

%% SED Fittings

\begin{figure*}
\includegraphics[width=0.98\textwidth]{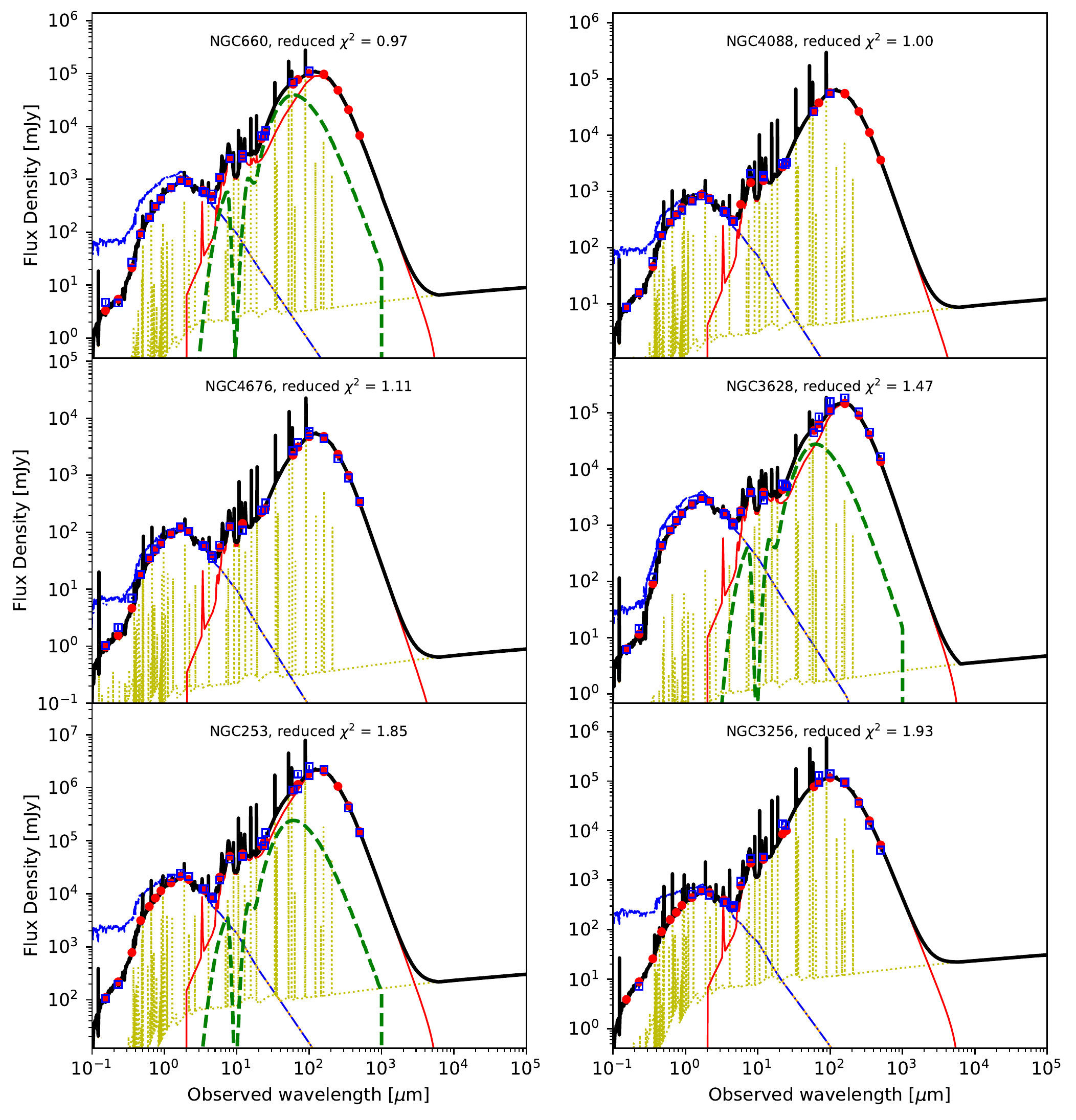}
\caption{Best-fit SED models for 6 galaxies in the SB sample containing the nebular emission (gold dotted lines), both attenuated stellar emission (orange) and non-attenuated stellar emission (blue dot-dashed), dust emission (red solid), and AGN emission (green
dashed). The red dots are the best model flux densities and the blue squares mark the observed flux densities with 1$\sigma$ error bars.}
\label{fig:SED_SB}
\end{figure*}

\begin{figure*}
\includegraphics[width=0.98\textwidth]{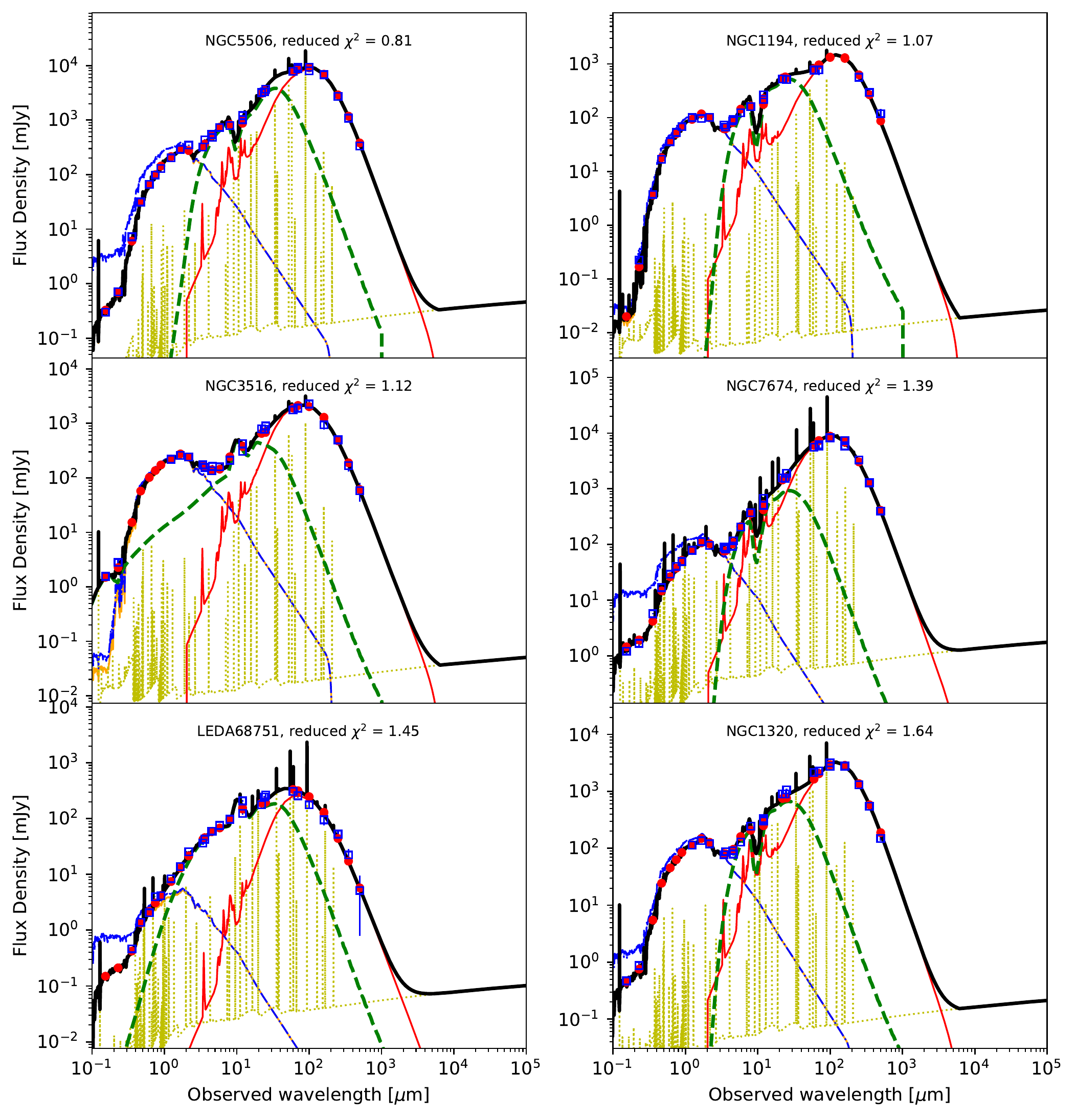}
\caption{Best-fit SED models for 6 galaxies in the AGN sample. The colors and lines are identical to Figure~\ref{fig:SED_SB}.}
\label{fig:SED_AGN}
\end{figure*}

\begin{figure*}
\includegraphics[width=0.98\textwidth]{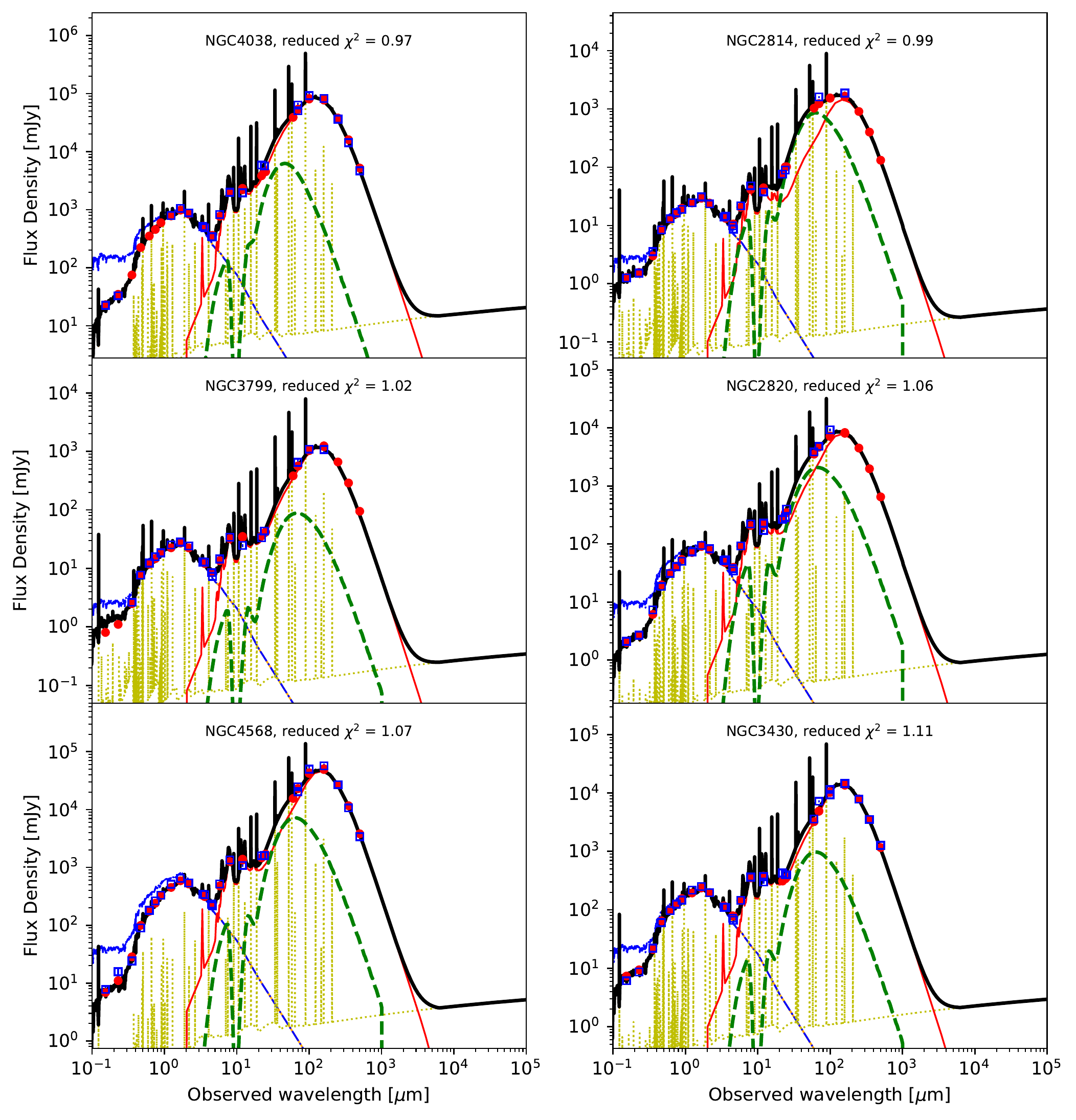}
\caption{Best-fit SED models for 6 galaxies in the SIGS sample. The colors and lines are identical to Figure~\ref{fig:SED_SB}.}
\label{fig:SED_SIGS}
\end{figure*}

\begin{figure*}
\includegraphics[width=0.98\textwidth]{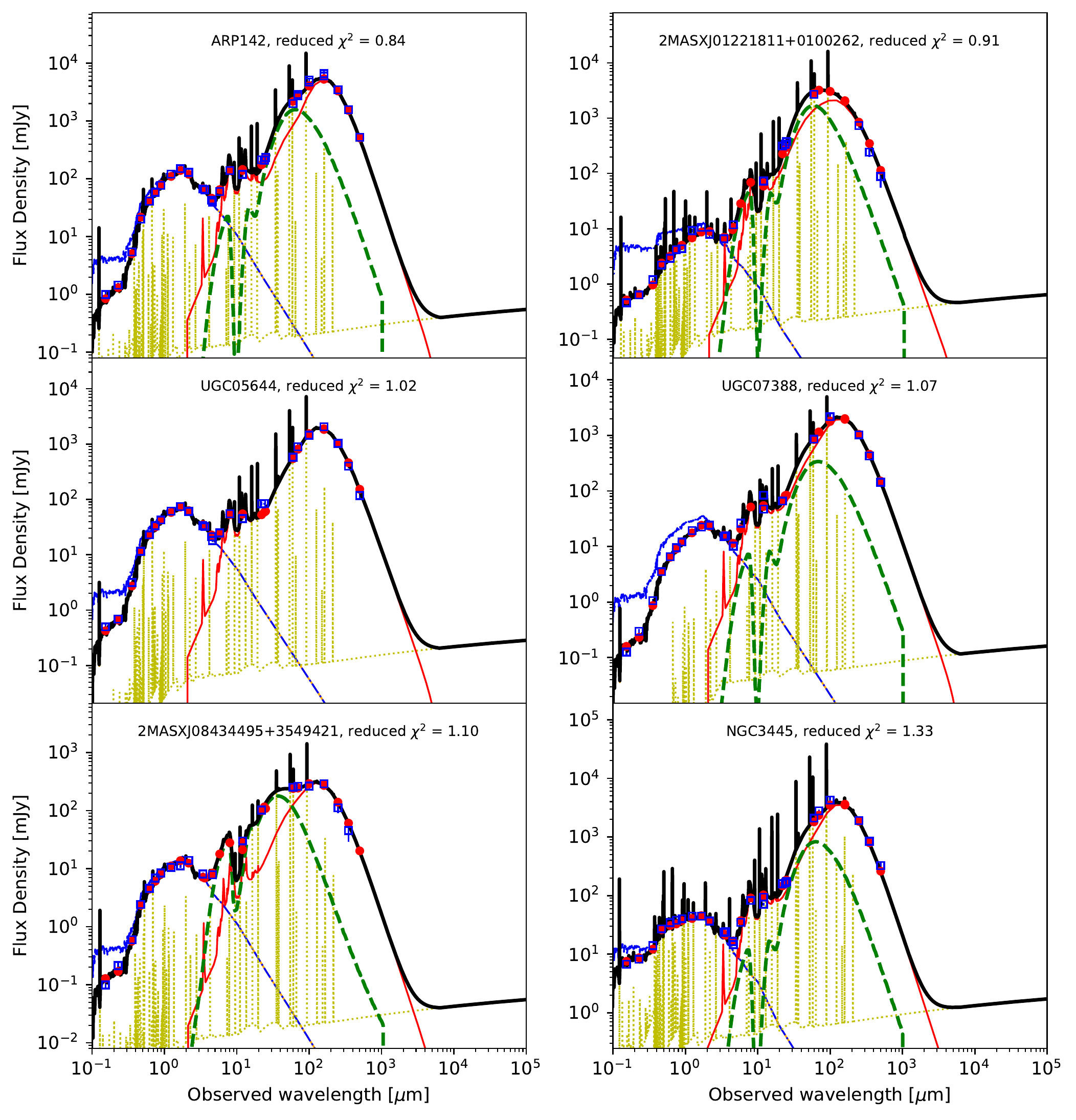}
\caption{Best-fit SED models for 6 galaxies in the LSM sample. The colors and lines are identical to Figure~\ref{fig:SED_SB}.}
\label{fig:SED_LSM}
\end{figure*}

% PARAMETERS

\begin{table*}
\centering
\caption{\textsc{CIGALE}-derived parameters for the AGN sample.}
\label{tab:CIGALE1AGN}
\begin{tabular}{ccccccc}
\hline
Galaxy &$f_{\rm{AGN}}$ & $L_{\rm{AGN}}$ & Old Att. & Young Att. &Dust $\alpha$ & $L_{\rm{dust}}$ \\
ID&&$\log(L_\odot)$&$\log(L_\odot)$& $\log(L_\odot)$&&$\log(L_\odot)$ \\
\hline
Mrk335 & 0.77 $\pm$ 0.04 & $11.03 \pm 0.02$ & $9.42 \pm 0.15$ & $9.78 \pm 0.07$ & 1.03 $\pm$ 0.08 & $9.99 \pm 0.05$ \\
Mrk1502 & 0.49 $\pm$ 0.03 & $11.72 \pm 0.02$ & $11.28 \pm 0.05$ & $11.31 \pm 0.04$ & 1.52 $\pm$ 0.08 & $11.64 \pm 0.03$ \\
NGC931 & 0.51 $\pm$ 0.04 & $10.68 \pm 0.04$ & $10.50 \pm 0.04$ & $9.95 \pm 0.07$ & 2.68 $\pm$ 0.21 & $10.62 \pm 0.04$ \\
NGC1068 & 0.42 $\pm$ 0.07 & $10.93 \pm 0.08$ & $10.86 \pm 0.07$ & $10.56 \pm 0.11$ & 2.01 $\pm$ 0.10 & $11.06 \pm 0.05$ \\
NGC1194 & 0.66 $\pm$ 0.07 & $10.17 \pm 0.05$ & $9.83 \pm 0.08$ & $8.06 \pm 0.44$ & 2.20 $\pm$ 0.24 & $9.84 \pm 0.08$ \\
NGC1320 & 0.43 $\pm$ 0.06 & $9.85 \pm 0.05$ & $9.87 \pm 0.06$ & $9.08 \pm 0.30$ & 1.89 $\pm$ 0.12 & $9.95 \pm 0.07$ \\
ESO33-2 & 0.63 $\pm$ 0.03 & $10.33 \pm 0.02$ & $10.02 \pm 0.03$ & $7.99 \pm 0.59$ & 1.75 $\pm$ 0.09 & $10.03 \pm 0.03$ \\
4U0557-385 & 0.90 $\pm$ 0.04 & $11.25 \pm 0.02$ & $10.14 \pm 0.04$ & $9.32 \pm 0.19$ & 1.84 $\pm$ 0.12 & $10.21 \pm 0.02$ \\
Mrk3 & 0.64 $\pm$ 0.06 & $10.62 \pm 0.03$ & $10.34 \pm 0.08$ & $8.75 \pm 0.04$ & 1.63 $\pm$ 0.13 & $10.36 \pm 0.08$ \\
ESO428-14 & 0.31 $\pm$ 0.04 & $9.50 \pm 0.06$ & $9.77 \pm 0.07$ & $8.89 \pm 0.44$ & 1.74 $\pm$ 0.09 & $9.83 \pm 0.03$ \\
NGC3281 & 0.40 $\pm$ 0.05 & $10.37 \pm 0.07$ & $10.52 \pm 0.03$ & $8.25 \pm 0.99$ & 1.75 $\pm$ 0.09 & $10.52 \pm 0.03$ \\
NGC3516 & 0.45 $\pm$ 0.03 & $10.12 \pm 0.05$ & $9.85 \pm 0.04$ & $7.98 \pm 1.06$ & 1.44 $\pm$ 0.11 & $9.86 \pm 0.03$ \\
NGC4151 & 0.67 $\pm$ 0.07 & $9.86 \pm 0.08$ & $9.25 \pm 0.19$ & $8.92 \pm 0.20$ & 2.18 $\pm$ 0.19 & $9.44 \pm 0.07$ \\
NGC4388 & 0.24 $\pm$ 0.04 & $10.14 \pm 0.08$ & $10.49 \pm 0.03$ & $10.00 \pm 0.05$ & 2.01 $\pm$ 0.10 & $10.63 \pm 0.02$ \\
Mrk771 & 0.53 $\pm$ 0.03 & $11.47 \pm 0.04$ & $10.63 \pm 0.07$ & $9.06 \pm 1.66$ & 1.68 $\pm$ 0.11 & $10.64 \pm 0.04$ \\
NGC4941 & $\cdots$ & $\cdots$ & $9.39 \pm 0.02$ & $8.40 \pm 0.02$ & 2.71 $\pm$ 0.20 & $9.44 \pm 0.02$ \\
MCG-03-34-064 & 0.46 $\pm$ 0.04 & $10.69 \pm 0.06$ & $10.67 \pm 0.05$ & $9.82 \pm 0.29$ & 1.49 $\pm$ 0.07 & $10.74 \pm 0.03$ \\
ESO383-35 & 0.65 $\pm$ 0.04 & $9.85 \pm 0.03$ & $9.47 \pm 0.05$ & $8.31 \pm 0.30$ & 1.33 $\pm$ 0.12 & $9.51 \pm 0.06$ \\
ESO445-50 & 0.81 $\pm$ 0.04 & $10.99 \pm 0.02$ & $10.28 \pm 0.05$ & $8.18 \pm 0.56$ & 1.55 $\pm$ 0.10 & $10.29 \pm 0.05$ \\
NGC5506 & 0.56 $\pm$ 0.03 & $10.25 \pm 0.03$ & $10.05 \pm 0.02$ & $8.94 \pm 0.02$ & 1.75 $\pm$ 0.09 & $10.09 \pm 0.02$ \\
2XMMJ141348.3+440014 & 0.85 $\pm$ 0.04 & $11.89 \pm 0.13$ & $10.09 \pm 0.27$ & $10.36 \pm 0.16$ & 2.02 $\pm$ 0.10 & $10.59 \pm 0.04$ \\
NGC5548 & 0.54 $\pm$ 0.08 & $10.46 \pm 0.09$ & $10.13 \pm 0.15$ & $9.47 \pm 0.33$ & 1.92 $\pm$ 0.15 & $10.23 \pm 0.07$ \\
Mrk1383 & 0.65 $\pm$ 0.03 & $12.11 \pm 0.02$ & $11.04 \pm 0.03$ & $9.33 \pm 1.13$ & 1.75 $\pm$ 0.09 & $11.05 \pm 0.02$ \\
Mrk841 & 0.64 $\pm$ 0.03 & $11.08 \pm 0.02$ & $9.78 \pm 0.03$ & $10.31 \pm 0.02$ & 1.00 $\pm$ 0.05 & $10.48 \pm 0.02$ \\
ESO141-55 & 0.52 $\pm$ 0.03 & $11.33 \pm 0.10$ & $10.20 \pm 0.25$ & $10.57 \pm 0.09$ & 2.00 $\pm$ 0.10 & $10.77 \pm 0.02$ \\
IC5063 & 0.70 $\pm$ 0.03 & $10.67 \pm 0.02$ & $10.28 \pm 0.02$ & $8.59 \pm 0.26$ & 2.00 $\pm$ 0.10 & $10.29 \pm 0.02$ \\
Mrk1513 & 0.66 $\pm$ 0.03 & $11.69 \pm 0.04$ & $10.48 \pm 0.11$ & $10.78 \pm 0.05$ & 1.52 $\pm$ 0.08 & $11.00 \pm 0.02$ \\
LEDA68751 & 0.80 $\pm$ 0.04 & $11.35 \pm 0.03$ & $10.32 \pm 0.14$ & $10.13 \pm 0.19$ & 1.21 $\pm$ 0.15 & $10.57 \pm 0.04$ \\
NGC7674 & 0.27 $\pm$ 0.05 & $11.03 \pm 0.08$ & $11.23 \pm 0.05$ & $10.92 \pm 0.04$ & 1.77 $\pm$ 0.09 & $11.43 \pm 0.04$ \\
\hline
\end{tabular}

\begin{flushleft}
\textbf{Note:} Galaxy ID is the common identifier used in the same order as in Table~\ref{tab:AGN}, $f_{\rm{AGN}}$ is the fraction of AGN contribution (from both torus and accretion) to the IR, or $L_{\rm{IR}}^{\rm{AGN}} = f_{\rm{AGN}} \times L_{\rm{IR}}^{\rm{Tot}}$ as defined by \citet{2015A&A...576A..10C}, $L_{\rm{AGN}}$ is the AGN luminosity of the three AGN components by \citet{2006MNRAS.366..767F} , Old Att. is the attenuation from the old stellar population, Young Att. is the attenuation from the young stellar population, Dust $\alpha$ is the parameter that defines the contribution of the local heating intensity in the dust (eq.~\ref{eq:alpha}) and $L_{\rm{dust}}$ is the dust luminosity.\\
\end{flushleft}
\end{table*}

\begin{table*}
\centering
\contcaption{\textsc{CIGALE}-derived parameters for the AGN sample.}
\label{tab:CIGALE2AGN}
\begin{tabular}{cccccc}
\hline
Galaxy &  SFR & $\tau_{\rm{main}}$ & Stellar Age & $M_{\rm{gas}}$ & $M_{\star}$ \\
ID& $\log(M_\odot /yr)$ & $\log(yr)$ & $\log(yr)$ & $\log(M_\odot)$& $\log(M_\odot)$\\
\hline
Mrk335 & $0.49 \pm 0.09$ & $9.60 \pm 0.28$ & $8.34 \pm 0.44$ & $8.29 \pm 0.64$ & $8.94 \pm 0.37$ \\
Mrk1502 & $1.86 \pm 0.03$ & $9.56 \pm 0.33$ & $8.92 \pm 0.20$ & $10.40 \pm 0.17$ & $10.93 \pm 0.13$ \\
NGC931 & $0.49 \pm 0.05$ & $8.99 \pm 0.04$ & $9.61 \pm 0.02$ & $10.77 \pm 0.02$ & $11.16 \pm 0.02$ \\
NGC1068 & $1.07 \pm 0.11$ & $8.95 \pm 0.10$ & $9.36 \pm 0.07$ & $10.70 \pm 0.09$ & $11.13 \pm 0.08$ \\
NGC1194 & $-1.31 \pm 0.42$ & $8.47 \pm 0.33$ & $9.55 \pm 0.09$ & $10.45 \pm 0.07$ & $10.84 \pm 0.06$ \\
NGC1320 & $-0.37 \pm 0.25$ & $8.81 \pm 0.15$ & $9.52 \pm 0.06$ & $10.15 \pm 0.03$ & $10.56 \pm 0.02$ \\
ESO33-2 & $-1.46 \pm 0.58$ & $8.39 \pm 0.40$ & $9.56 \pm 0.12$ & $10.48 \pm 0.10$ & $10.88 \pm 0.09$ \\
4U0557-385 & $-0.06 \pm 0.11$ & $8.72 \pm 0.12$ & $9.48 \pm 0.04$ & $10.61 \pm 0.07$ & $11.02 \pm 0.07$ \\
Mrk3 & $-0.74 \pm 0.03$ & $8.70 \pm 0.02$ & $9.60 \pm 0.02$ & $10.77 \pm 0.02$ & $11.16 \pm 0.02$ \\
ESO428-14 & $-0.59 \pm 0.43$ & $8.72 \pm 0.39$ & $9.40 \pm 0.27$ & $9.85 \pm 0.18$ & $10.28 \pm 0.13$ \\
NGC3281 & $-1.19 \pm 0.93$ & $8.21 \pm 0.52$ & $9.40 \pm 0.16$ & $10.71 \pm 0.11$ & $11.12 \pm 0.09$ \\
NGC3516 & $-1.35 \pm 1.11$ & $8.26 \pm 0.49$ & $9.52 \pm 0.17$ & $10.40 \pm 0.13$ & $10.80 \pm 0.11$ \\
NGC4151 & $-0.36 \pm 0.20$ & $8.92 \pm 0.18$ & $9.57 \pm 0.09$ & $9.98 \pm 0.06$ & $10.38 \pm 0.05$ \\
NGC4388 & $0.54 \pm 0.04$ & $8.75 \pm 0.13$ & $9.35 \pm 0.11$ & $10.51 \pm 0.09$ & $10.94 \pm 0.07$ \\
Mrk771 & $-0.28 \pm 1.89$ & $8.57 \pm 1.11$ & $9.62 \pm 0.12$ & $10.78 \pm 0.13$ & $11.17 \pm 0.12$ \\
NGC4941 & $-0.90 \pm 0.02$ & $8.70 \pm 0.02$ & $9.48 \pm 0.02$ & $9.82 \pm 0.02$ & $10.22 \pm 0.02$ \\
MCG-03-34-064 & $0.33 \pm 0.29$ & $8.69 \pm 0.28$ & $9.34 \pm 0.20$ & $10.48 \pm 0.14$ & $10.91 \pm 0.11$ \\
ESO383-35 & $-1.15 \pm 0.26$ & $8.70 \pm 0.02$ & $9.54 \pm 0.06$ & $9.85 \pm 0.09$ & $10.25 \pm 0.08$ \\
ESO445-50 & $-1.28 \pm 0.53$ & $8.53 \pm 0.28$ & $9.67 \pm 0.07$ & $10.83 \pm 0.08$ & $11.21 \pm 0.07$ \\
NGC5506 & $-0.55 \pm 0.02$ & $8.70 \pm 0.02$ & $9.48 \pm 0.02$ & $10.17 \pm 0.02$ & $10.57 \pm 0.02$ \\
2XMMJ141348.3+440014 & $1.06 \pm 0.20$ & $9.54 \pm 0.33$ & $8.91 \pm 0.53$ & $9.65 \pm 0.77$ & $10.13 \pm 0.61$ \\
NGC5548 & $0.14 \pm 0.32$ & $8.85 \pm 0.25$ & $9.59 \pm 0.08$ & $10.63 \pm 0.08$ & $11.03 \pm 0.07$ \\
Mrk1383 & $-0.16 \pm 1.13$ & $8.40 \pm 0.47$ & $9.60 \pm 0.13$ & $11.18 \pm 0.14$ & $11.57 \pm 0.12$ \\
Mrk841 & $1.04 \pm 0.02$ & $9.59 \pm 0.29$ & $9.01 \pm 0.14$ & $9.71 \pm 0.13$ & $10.22 \pm 0.11$ \\
ESO141-55 & $1.30 \pm 0.09$ & $9.61 \pm 0.29$ & $9.11 \pm 0.26$ & $10.18 \pm 0.45$ & $10.66 \pm 0.37$ \\
IC5063 & $-0.71 \pm 0.15$ & $8.70 \pm 0.02$ & $9.60 \pm 0.02$ & $10.77 \pm 0.02$ & $11.16 \pm 0.02$ \\
Mrk1513 & $1.49 \pm 0.05$ & $9.60 \pm 0.28$ & $8.39 \pm 0.45$ & $9.41 \pm 0.75$ & $10.03 \pm 0.54$ \\
LEDA68751 & $0.66 \pm 0.18$ & $9.39 \pm 0.42$ & $9.22 \pm 0.30$ & $10.00 \pm 0.45$ & $10.44 \pm 0.41$ \\
NGC7674 & $1.45 \pm 0.04$ & $9.01 \pm 0.09$ & $9.36 \pm 0.03$ & $10.96 \pm 0.05$ & $11.40 \pm 0.04$ \\
\hline
\end{tabular}

\begin{flushleft}
\textbf{Note:} Galaxy ID is the common identifier used in the same order as in Table~\ref{tab:AGN}, SFR is the star formation rate, $\tau_{\rm{main}}$ is the e-folding time of the main stellar population model, Stellar Age is the age of the oldest stars in the galaxy, $M_{\rm{gas}}$ is the gas mass and $M_{\star}$ is the stellar mass.\\
\end{flushleft}
\end{table*}

\begin{table*}
\centering
\caption{\textsc{CIGALE}-derived parameters for the SB sample. Units as Table~\ref{tab:CIGALE1AGN}.}
\label{tab:CIGALE1SB}
\begin{tabular}{ccccccc}
\hline
Galaxy &$f_{\rm{AGN}}$ & $L_{\rm{AGN}}$ & Old Att. & Young Att. &Dust $\alpha$ & $L_{\rm{dust}}$ \\
ID&&$\log(L_\odot)$&$\log(L_\odot)$& $\log(L_\odot)$&&$\log(L_\odot)$ \\
\hline
NGC23 & 0.32 $\pm$ 0.06 & $10.60 \pm 0.10$ & $10.73 \pm 0.04$ & $10.39 \pm 0.06$ & 2.02 $\pm$ 0.10 & $10.92 \pm 0.04$ \\
NGC253 & 0.23 $\pm$ 0.12 & $9.96 \pm 0.22$ & $10.32 \pm 0.04$ & $9.95 \pm 0.15$ & 2.38 $\pm$ 0.21 & $10.50 \pm 0.07$ \\
NGC660 & 0.43 $\pm$ 0.08 & $10.14 \pm 0.09$ & $10.14 \pm 0.04$ & $9.59 \pm 0.13$ & 2.32 $\pm$ 0.25 & $10.26 \pm 0.05$ \\
NGC1222 & $\cdots$ & $\cdots$ & $10.35 \pm 0.02$ & $10.31 \pm 0.02$ & 1.50 $\pm$ 0.07 & $10.66 \pm 0.02$ \\
NGC1365 & 0.39 $\pm$ 0.12 & $10.87 \pm 0.14$ & $10.83 \pm 0.10$ & $10.60 \pm 0.16$ & 2.58 $\pm$ 0.34 & $11.06 \pm 0.08$ \\
IC342 & $\cdots$ & $\cdots$ & $8.39 \pm 0.03$ & $7.49 \pm 0.24$ & 2.26 $\pm$ 0.11 & $8.45 \pm 0.02$ \\
NGC1614 & $\cdots$ & $\cdots$ & $11.31 \pm 0.04$ & $11.24 \pm 0.05$ & 1.37 $\pm$ 0.12 & $11.61 \pm 0.04$ \\
NGC1797 & $\cdots$ & $\cdots$ & $10.63 \pm 0.07$ & $10.68 \pm 0.05$ & 1.75 $\pm$ 0.09 & $11.00 \pm 0.02$ \\
NGC2146 & 0.38 $\pm$ 0.04 & $10.41 \pm 0.06$ & $10.60 \pm 0.04$ & $9.09 \pm 0.77$ & 2.00 $\pm$ 0.10 & $10.62 \pm 0.02$ \\
NGC2623 & 0.27 $\pm$ 0.06 & $10.72 \pm 0.11$ & $10.83 \pm 0.04$ & $10.78 \pm 0.05$ & 1.73 $\pm$ 0.09 & $11.14 \pm 0.04$ \\
NGC3256 & $\cdots$ & $\cdots$ & $11.30 \pm 0.02$ & $11.21 \pm 0.02$ & 1.75 $\pm$ 0.09 & $11.59 \pm 0.02$ \\
NGC3310 & 0.20 $\pm$ 0.07 & $9.61 \pm 0.16$ & $9.63 \pm 0.05$ & $10.00 \pm 0.04$ & 1.94 $\pm$ 0.11 & $10.20 \pm 0.03$ \\
NGC3556 & 0.28 $\pm$ 0.09 & $9.53 \pm 0.15$ & $9.71 \pm 0.05$ & $9.47 \pm 0.09$ & 2.65 $\pm$ 0.29 & $9.93 \pm 0.06$ \\
NGC3628 & 0.22 $\pm$ 0.03 & $9.85 \pm 0.07$ & $10.34 \pm 0.02$ & $9.29 \pm 0.06$ & 2.93 $\pm$ 0.15 & $10.39 \pm 0.02$ \\
NGC4088 & $\cdots$ & $\cdots$ & $9.85 \pm 0.03$ & $9.64 \pm 0.02$ & 2.01 $\pm$ 0.10 & $10.09 \pm 0.02$ \\
NGC4194 & $\cdots$ & $\cdots$ & $10.59 \pm 0.02$ & $10.50 \pm 0.02$ & 1.50 $\pm$ 0.07 & $10.88 \pm 0.02$ \\
Mrk52 & $\cdots$ & $\cdots$ & $9.87 \pm 0.04$ & $9.91 \pm 0.03$ & 1.75 $\pm$ 0.09 & $10.23 \pm 0.03$ \\
NGC4676 & $\cdots$ & $\cdots$ & $10.72 \pm 0.02$ & $10.36 \pm 0.02$ & 2.00 $\pm$ 0.10 & $10.90 \pm 0.02$ \\
NGC4818 & 0.24 $\pm$ 0.09 & $9.53 \pm 0.19$ & $9.89 \pm 0.05$ & $9.33 \pm 0.09$ & 1.82 $\pm$ 0.15 & $10.02 \pm 0.05$ \\
NGC4945 & $\cdots$ & $\cdots$ & $10.77 \pm 0.17$ & $10.60 \pm 0.21$ & 2.27 $\pm$ 0.11 & $11.02 \pm 0.02$ \\
NGC7252 & $\cdots$ & $\cdots$ & $10.60 \pm 0.03$ & $10.23 \pm 0.04$ & 2.00 $\pm$ 0.10 & $10.78 \pm 0.02$ \\
\hline
\end{tabular}

\end{table*}

\begin{table*}
\centering
\contcaption{\textsc{CIGALE}-derived parameters for the SB sample. Units as Table~\ref{tab:CIGALE2AGN}.}
\label{tab:CIGALE2SB}
\begin{tabular}{cccccc}
\hline
Galaxy &  SFR & $\tau_{\rm{main}}$ & Stellar Age & $M_{\rm{gas}}$ & $M_{\star}$ \\
ID& $\log(M_\odot /yr)$ & $\log(yr)$ & $\log(yr)$ & $\log(M_\odot)$& $\log(M_\odot)$\\
\hline
NGC23 & $0.91 \pm 0.06$ & $8.96 \pm 0.09$ & $9.47 \pm 0.06$ & $10.81 \pm 0.04$ & $11.23 \pm 0.04$ \\
NGC253 & $0.47 \pm 0.15$ & $8.92 \pm 0.12$ & $9.36 \pm 0.06$ & $10.14 \pm 0.06$ & $10.57 \pm 0.05$ \\
NGC660 & $0.10 \pm 0.13$ & $8.97 \pm 0.09$ & $9.49 \pm 0.09$ & $10.02 \pm 0.08$ & $10.44 \pm 0.06$ \\
NGC1222 & $0.82 \pm 0.02$ & $9.68 \pm 0.22$ & $9.08 \pm 0.09$ & $9.54 \pm 0.07$ & $10.05 \pm 0.05$ \\
NGC1365 & $1.14 \pm 0.14$ & $8.96 \pm 0.11$ & $9.39 \pm 0.10$ & $10.82 \pm 0.09$ & $11.25 \pm 0.07$ \\
IC342 & $-1.99 \pm 0.24$ & $8.84 \pm 0.15$ & $9.53 \pm 0.07$ & $8.49 \pm 0.02$ & $8.89 \pm 0.02$ \\
NGC1614 & $1.75 \pm 0.05$ & $9.60 \pm 0.27$ & $8.85 \pm 0.12$ & $10.19 \pm 0.08$ & $10.74 \pm 0.06$ \\
NGC1797 & $1.24 \pm 0.10$ & $9.68 \pm 0.22$ & $9.12 \pm 0.16$ & $9.99 \pm 0.12$ & $10.48 \pm 0.09$ \\
NGC2146 & $-0.42 \pm 0.80$ & $7.97 \pm 0.92$ & $8.69 \pm 0.35$ & $9.74 \pm 0.11$ & $10.30 \pm 0.06$ \\
NGC2623 & $1.29 \pm 0.05$ & $9.58 \pm 0.30$ & $9.00 \pm 0.09$ & $9.94 \pm 0.06$ & $10.45 \pm 0.04$ \\
NGC3256 & $1.72 \pm 0.02$ & $9.66 \pm 0.24$ & $9.09 \pm 0.09$ & $10.47 \pm 0.06$ & $10.97 \pm 0.05$ \\
NGC3310 & $0.56 \pm 0.03$ & $9.67 \pm 0.22$ & $8.93 \pm 0.10$ & $9.08 \pm 0.08$ & $9.61 \pm 0.06$ \\
NGC3556 & $0.00 \pm 0.08$ & $8.97 \pm 0.08$ & $9.47 \pm 0.06$ & $9.88 \pm 0.05$ & $10.29 \pm 0.04$ \\
NGC3628 & $-0.21 \pm 0.06$ & $8.70 \pm 0.03$ & $9.48 \pm 0.02$ & $10.51 \pm 0.02$ & $10.91 \pm 0.02$ \\
NGC4088 & $0.17 \pm 0.02$ & $8.99 \pm 0.06$ & $9.35 \pm 0.04$ & $9.68 \pm 0.04$ & $10.12 \pm 0.04$ \\
NGC4194 & $1.01 \pm 0.02$ & $9.42 \pm 0.39$ & $9.01 \pm 0.08$ & $9.73 \pm 0.04$ & $10.24 \pm 0.03$ \\
Mrk52 & $0.43 \pm 0.03$ & $9.59 \pm 0.19$ & $9.33 \pm 0.06$ & $9.52 \pm 0.04$ & $9.97 \pm 0.03$ \\
NGC4676 & $0.92 \pm 0.02$ & $9.00 \pm 0.03$ & $9.50 \pm 0.02$ & $10.82 \pm 0.02$ & $11.23 \pm 0.02$ \\
NGC4818 & $-0.16 \pm 0.09$ & $8.79 \pm 0.15$ & $9.36 \pm 0.09$ & $9.77 \pm 0.06$ & $10.20 \pm 0.05$ \\
NGC4945 & $1.22 \pm 0.25$ & $9.32 \pm 0.45$ & $9.23 \pm 0.21$ & $10.45 \pm 0.13$ & $10.91 \pm 0.09$ \\
NGC7252 & $0.76 \pm 0.03$ & $8.99 \pm 0.05$ & $9.49 \pm 0.04$ & $10.66 \pm 0.03$ & $11.07 \pm 0.02$ \\
\hline
\end{tabular}

\end{table*}

\begin{table*}
\centering
\caption{\textsc{CIGALE}-derived parameters for the SIGS sample. Units as Table~\ref{tab:CIGALE1AGN}.}
\label{tab:CIGALE1SIGS1}
\begin{tabular}{ccccccc}
\hline
Galaxy &$f_{\rm{AGN}}$ & $L_{\rm{AGN}}$ & Old Att. & Young Att. &Dust $\alpha$ & $L_{\rm{dust}}$ \\
ID&&$\log(L_\odot)$&$\log(L_\odot)$& $\log(L_\odot)$&&$\log(L_\odot)$ \\
\hline
NGC275 & 0.20 $\pm$ 0.14 & $9.26 \pm 0.30$ & $9.42 \pm 0.09$ & $9.57 \pm 0.07$ & 2.09 $\pm$ 0.18 & $9.84 \pm 0.07$ \\
NGC470 & $\cdots$ & $\cdots$ & $10.22 \pm 0.02$ & $10.04 \pm 0.03$ & 2.00 $\pm$ 0.10 & $10.47 \pm 0.02$ \\
NGC474 & $\cdots$ & $\cdots$ & $9.65 \pm 0.05$ & $8.94 \pm 0.06$ & 2.04 $\pm$ 0.10 & $9.74 \pm 0.04$ \\
NGC520 & 0.58 $\pm$ 0.08 & $10.69 \pm 0.07$ & $10.45 \pm 0.06$ & $9.81 \pm 0.16$ & 2.37 $\pm$ 0.28 & $10.56 \pm 0.08$ \\
IC196 & $\cdots$ & $\cdots$ & $9.66 \pm 0.02$ & $8.82 \pm 0.02$ & 2.32 $\pm$ 0.12 & $9.73 \pm 0.02$ \\
NGC833 & 0.27 $\pm$ 0.07 & $9.09 \pm 0.15$ & $9.50 \pm 0.02$ & $8.11 \pm 0.17$ & 2.87 $\pm$ 0.17 & $9.52 \pm 0.02$ \\
NGC835 & $\cdots$ & $\cdots$ & $10.60 \pm 0.04$ & $10.12 \pm 0.07$ & 2.00 $\pm$ 0.10 & $10.74 \pm 0.02$ \\
NGC838 & $\cdots$ & $\cdots$ & $10.71 \pm 0.02$ & $10.48 \pm 0.04$ & 1.75 $\pm$ 0.09 & $10.94 \pm 0.02$ \\
NGC839 & $\cdots$ & $\cdots$ & $10.70 \pm 0.02$ & $10.48 \pm 0.03$ & 1.50 $\pm$ 0.07 & $10.93 \pm 0.02$ \\
NGC935 & $\cdots$ & $\cdots$ & $10.54 \pm 0.02$ & $10.11 \pm 0.03$ & 2.25 $\pm$ 0.11 & $10.70 \pm 0.02$ \\
IC1801 & $\cdots$ & $\cdots$ & $10.06 \pm 0.02$ & $9.77 \pm 0.03$ & 2.25 $\pm$ 0.11 & $10.27 \pm 0.02$ \\
NGC1241 & 0.21 $\pm$ 0.09 & $10.13 \pm 0.18$ & $10.57 \pm 0.05$ & $10.09 \pm 0.07$ & 2.54 $\pm$ 0.27 & $10.71 \pm 0.05$ \\
NGC1242 & 0.41 $\pm$ 0.08 & $9.44 \pm 0.09$ & $9.40 \pm 0.07$ & $9.09 \pm 0.14$ & 2.68 $\pm$ 0.23 & $9.60 \pm 0.05$ \\
NGC1253 & 0.21 $\pm$ 0.09 & $9.20 \pm 0.18$ & $9.61 \pm 0.14$ & $9.19 \pm 0.22$ & 2.57 $\pm$ 0.27 & $9.77 \pm 0.05$ \\
NGC1253A & $\cdots$ & $\cdots$ & $8.22 \pm 0.03$ & $8.58 \pm 0.02$ & 1.75 $\pm$ 0.09 & $8.79 \pm 0.02$ \\
NGC2276 & 0.20 $\pm$ 0.08 & $10.13 \pm 0.18$ & $10.40 \pm 0.06$ & $10.38 \pm 0.06$ & 2.44 $\pm$ 0.24 & $10.73 \pm 0.04$ \\
NGC2444 & $\cdots$ & $\cdots$ & $9.05 \pm 0.02$ & $8.00 \pm 0.09$ & 2.00 $\pm$ 0.10 & $9.10 \pm 0.02$ \\
NGC2445 & $\cdots$ & $\cdots$ & $10.28 \pm 0.05$ & $10.24 \pm 0.05$ & 1.97 $\pm$ 0.10 & $10.59 \pm 0.02$ \\
NGC2633 & $\cdots$ & $\cdots$ & $10.40 \pm 0.02$ & $10.32 \pm 0.02$ & 1.75 $\pm$ 0.09 & $10.70 \pm 0.02$ \\
NGC2634 & $\cdots$ & $\cdots$ & $8.64 \pm 0.02$ & $7.04 \pm 0.36$ & 2.96 $\pm$ 0.15 & $8.65 \pm 0.02$ \\
NGC2719A & $\cdots$ & $\cdots$ & $8.65 \pm 0.06$ & $9.19 \pm 0.02$ & 1.50 $\pm$ 0.07 & $9.35 \pm 0.02$ \\
NGC2719 & 0.26 $\pm$ 0.11 & $8.88 \pm 0.22$ & $8.80 \pm 0.14$ & $9.10 \pm 0.05$ & 1.88 $\pm$ 0.13 & $9.32 \pm 0.07$ \\
NGC2805 & $\cdots$ & $\cdots$ & $9.38 \pm 0.04$ & $9.51 \pm 0.02$ & 2.47 $\pm$ 0.12 & $9.79 \pm 0.02$ \\
NGC2814 & 0.42 $\pm$ 0.11 & $8.86 \pm 0.12$ & $8.69 \pm 0.08$ & $8.63 \pm 0.13$ & 2.54 $\pm$ 0.30 & $8.99 \pm 0.08$ \\
NGC2820A & $\cdots$ & $\cdots$ & $7.69 \pm 0.02$ & $8.21 \pm 0.02$ & 1.69 $\pm$ 0.11 & $8.37 \pm 0.02$ \\
NGC2820 & 0.25 $\pm$ 0.07 & $9.24 \pm 0.13$ & $9.46 \pm 0.04$ & $9.29 \pm 0.05$ & 2.47 $\pm$ 0.20 & $9.72 \pm 0.04$ \\
NGC2964 & 0.22 $\pm$ 0.08 & $9.57 \pm 0.16$ & $9.82 \pm 0.06$ & $9.71 \pm 0.03$ & 2.23 $\pm$ 0.16 & $10.10 \pm 0.04$ \\
NGC2970 & $\cdots$ & $\cdots$ & $7.89 \pm 0.02$ & $6.86 \pm 0.02$ & 1.17 $\pm$ 0.18 & $7.94 \pm 0.02$ \\
NGC2976 & 0.29 $\pm$ 0.04 & $8.32 \pm 0.06$ & $8.42 \pm 0.05$ & $8.31 \pm 0.05$ & 2.91 $\pm$ 0.15 & $8.70 \pm 0.03$ \\
NGC3031 & $\cdots$ & $\cdots$ & $9.46 \pm 0.02$ & $8.76 \pm 0.02$ & 2.96 $\pm$ 0.15 & $9.55 \pm 0.02$ \\
NGC3034 & 0.48 $\pm$ 0.03 & $10.19 \pm 0.05$ & $10.20 \pm 0.02$ & $8.46 \pm 0.02$ & 1.75 $\pm$ 0.09 & $10.21 \pm 0.02$ \\
NGC3077 & 0.30 $\pm$ 0.07 & $8.30 \pm 0.11$ & $8.45 \pm 0.09$ & $8.18 \pm 0.11$ & 2.00 $\pm$ 0.10 & $8.66 \pm 0.03$ \\
NGC3165 & $\cdots$ & $\cdots$ & $7.86 \pm 0.09$ & $8.04 \pm 0.05$ & 2.01 $\pm$ 0.10 & $8.30 \pm 0.02$ \\
NGC3166 & 0.21 $\pm$ 0.04 & $9.06 \pm 0.09$ & $9.60 \pm 0.02$ & $8.36 \pm 0.02$ & 2.56 $\pm$ 0.14 & $9.63 \pm 0.02$ \\
NGC3169 & $\cdots$ & $\cdots$ & $10.04 \pm 0.06$ & $9.38 \pm 0.27$ & 2.71 $\pm$ 0.19 & $10.14 \pm 0.03$ \\
NGC3185 & $\cdots$ & $\cdots$ & $9.12 \pm 0.02$ & $8.81 \pm 0.02$ & 2.25 $\pm$ 0.11 & $9.32 \pm 0.02$ \\
NGC3187 & 0.21 $\pm$ 0.05 & $8.64 \pm 0.11$ & $8.91 \pm 0.04$ & $8.86 \pm 0.02$ & 2.95 $\pm$ 0.15 & $9.22 \pm 0.02$ \\
NGC3190 & $\cdots$ & $\cdots$ & $9.73 \pm 0.02$ & $8.22 \pm 0.03$ & 2.50 $\pm$ 0.13 & $9.74 \pm 0.02$ \\
NGC3226 & $\cdots$ & $\cdots$ & $8.79 \pm 0.02$ & $7.49 \pm 0.24$ & 2.26 $\pm$ 0.11 & $8.82 \pm 0.02$ \\
NGC3227 & 0.20 $\pm$ 0.03 & $9.32 \pm 0.08$ & $9.76 \pm 0.04$ & $9.30 \pm 0.03$ & 2.26 $\pm$ 0.11 & $9.91 \pm 0.02$ \\
NGC3395 & $\cdots$ & $\cdots$ & $9.52 \pm 0.08$ & $9.65 \pm 0.07$ & 2.00 $\pm$ 0.10 & $9.93 \pm 0.02$ \\
NGC3396 & $\cdots$ & $\cdots$ & $9.52 \pm 0.04$ & $9.70 \pm 0.02$ & 1.75 $\pm$ 0.09 & $9.96 \pm 0.02$ \\
NGC3424 & $\cdots$ & $\cdots$ & $9.89 \pm 0.02$ & $9.62 \pm 0.02$ & 2.00 $\pm$ 0.10 & $10.10 \pm 0.02$ \\
NGC3430 & $\cdots$ & $\cdots$ & $9.61 \pm 0.03$ & $9.67 \pm 0.02$ & 2.39 $\pm$ 0.12 & $9.98 \pm 0.02$ \\
NGC3448 & 0.40 $\pm$ 0.05 & $9.46 \pm 0.07$ & $9.20 \pm 0.05$ & $9.36 \pm 0.04$ & 2.27 $\pm$ 0.11 & $9.63 \pm 0.03$ \\
IC694 & $\cdots$ & $\cdots$ & $9.02 \pm 0.06$ & $7.54 \pm 0.03$ & 1.24 $\pm$ 0.19 & $9.03 \pm 0.06$ \\
NGC3690 & $\cdots$ & $\cdots$ & $11.47 \pm 0.02$ & $11.51 \pm 0.03$ & 1.25 $\pm$ 0.06 & $11.83 \pm 0.02$ \\
NGC3786 & 0.22 $\pm$ 0.08 & $9.39 \pm 0.16$ & $9.75 \pm 0.05$ & $9.36 \pm 0.03$ & 2.22 $\pm$ 0.11 & $9.92 \pm 0.04$ \\
NGC3788 & $\cdots$ & $\cdots$ & $9.86 \pm 0.02$ & $9.45 \pm 0.07$ & 2.29 $\pm$ 0.11 & $10.02 \pm 0.02$ \\
NGC3799 & $\cdots$ & $\cdots$ & $9.27 \pm 0.08$ & $9.26 \pm 0.08$ & 2.25 $\pm$ 0.11 & $9.60 \pm 0.02$ \\
NGC3800 & $\cdots$ & $\cdots$ & $10.36 \pm 0.03$ & $10.09 \pm 0.02$ & 2.23 $\pm$ 0.11 & $10.57 \pm 0.02$ \\
IC749 & $\cdots$ & $\cdots$ & $8.61 \pm 0.05$ & $8.74 \pm 0.04$ & 2.25 $\pm$ 0.11 & $9.02 \pm 0.02$ \\
IC750 & 0.33 $\pm$ 0.07 & $9.32 \pm 0.11$ & $9.52 \pm 0.04$ & $8.88 \pm 0.04$ & 2.65 $\pm$ 0.26 & $9.62 \pm 0.04$ \\
NGC4038 & $\cdots$ & $\cdots$ & $10.61 \pm 0.04$ & $10.55 \pm 0.04$ & 2.00 $\pm$ 0.10 & $10.91 \pm 0.02$ \\
NGC4394 & $\cdots$ & $\cdots$ & $9.08 \pm 0.03$ & $8.26 \pm 0.02$ & 2.59 $\pm$ 0.21 & $9.15 \pm 0.02$ \\
\hline
\end{tabular}

\begin{flushleft}
\textbf{Note:} The full table is available
in the online version of this paper. A portion is shown here for guidance regarding its
form and content. \\
\end{flushleft}
\end{table*}

\begin{table*}
\centering
\contcaption{\textsc{CIGALE}-derived parameters for the SIGS sample.  
Units as Table~\ref{tab:CIGALE1AGN}.}
\label{tab:CIGALE1SIGS2}
\begin{tabular}{cccccc}
\hline
Galaxy &  SFR & $\tau_{\rm{main}}$ & Stellar Age & $M_{\rm{gas}}$ & $M_{\star}$ \\
ID& $\log(M_\odot /yr)$ & $\log(yr)$ & $\log(yr)$ & $\log(M_\odot)$& $\log(M_\odot)$\\
\hline
NGC275 & $0.11 \pm 0.06$ & $9.51 \pm 0.25$ & $9.31 \pm 0.09$ & $9.22 \pm 0.06$ & $9.67 \pm 0.05$ \\
NGC470 & $0.57 \pm 0.02$ & $9.00 \pm 0.02$ & $9.36 \pm 0.02$ & $10.09 \pm 0.02$ & $10.53 \pm 0.02$ \\
NGC474 & $-0.36 \pm 0.02$ & $8.70 \pm 0.02$ & $9.48 \pm 0.02$ & $10.35 \pm 0.02$ & $10.76 \pm 0.02$ \\
NGC520 & $0.35 \pm 0.15$ & $9.00 \pm 0.02$ & $9.58 \pm 0.05$ & $10.49 \pm 0.04$ & $10.89 \pm 0.04$ \\
IC196 & $-0.46 \pm 0.02$ & $8.70 \pm 0.02$ & $9.48 \pm 0.02$ & $10.26 \pm 0.02$ & $10.67 \pm 0.02$ \\
NGC833 & $-1.16 \pm 0.17$ & $8.67 \pm 0.10$ & $9.61 \pm 0.04$ & $10.42 \pm 0.04$ & $10.81 \pm 0.04$ \\
NGC835 & $0.65 \pm 0.07$ & $8.86 \pm 0.15$ & $9.40 \pm 0.10$ & $10.57 \pm 0.05$ & $11.00 \pm 0.04$ \\
NGC838 & $0.99 \pm 0.04$ & $9.04 \pm 0.37$ & $9.18 \pm 0.11$ & $10.14 \pm 0.05$ & $10.61 \pm 0.04$ \\
NGC839 & $0.99 \pm 0.03$ & $9.42 \pm 0.25$ & $9.31 \pm 0.07$ & $10.12 \pm 0.03$ & $10.57 \pm 0.03$ \\
NGC935 & $0.62 \pm 0.03$ & $8.99 \pm 0.05$ & $9.49 \pm 0.03$ & $10.53 \pm 0.02$ & $10.94 \pm 0.02$ \\
IC1801 & $0.29 \pm 0.03$ & $9.00 \pm 0.02$ & $9.36 \pm 0.02$ & $9.81 \pm 0.02$ & $10.25 \pm 0.02$ \\
NGC1241 & $0.62 \pm 0.05$ & $8.91 \pm 0.13$ & $9.52 \pm 0.13$ & $10.78 \pm 0.11$ & $11.18 \pm 0.09$ \\
NGC1242 & $-0.21 \pm 0.08$ & $8.99 \pm 0.05$ & $9.49 \pm 0.05$ & $9.67 \pm 0.05$ & $10.08 \pm 0.04$ \\
NGC1253 & $-0.10 \pm 0.23$ & $8.88 \pm 0.24$ & $9.38 \pm 0.21$ & $9.81 \pm 0.16$ & $10.24 \pm 0.13$ \\
NGC1253A & $-0.69 \pm 0.02$ & $8.94 \pm 0.38$ & $9.16 \pm 0.19$ & $8.52 \pm 0.09$ & $9.00 \pm 0.06$ \\
NGC2276 & $0.91 \pm 0.05$ & $9.40 \pm 0.26$ & $9.32 \pm 0.09$ & $10.06 \pm 0.06$ & $10.51 \pm 0.05$ \\
NGC2444 & $-1.28 \pm 0.09$ & $8.70 \pm 0.02$ & $9.61 \pm 0.03$ & $10.26 \pm 0.03$ & $10.65 \pm 0.03$ \\
NGC2445 & $0.76 \pm 0.04$ & $9.01 \pm 0.33$ & $9.20 \pm 0.18$ & $9.97 \pm 0.10$ & $10.44 \pm 0.07$ \\
NGC2633 & $0.84 \pm 0.02$ & $9.68 \pm 0.16$ & $9.33 \pm 0.06$ & $9.88 \pm 0.05$ & $10.34 \pm 0.04$ \\
NGC2634 & $-2.23 \pm 0.36$ & $8.69 \pm 0.07$ & $9.69 \pm 0.03$ & $9.94 \pm 0.04$ & $10.31 \pm 0.03$ \\
NGC2719A & $-0.08 \pm 0.02$ & $9.49 \pm 0.36$ & $8.79 \pm 0.10$ & $8.32 \pm 0.05$ & $8.88 \pm 0.03$ \\
NGC2719 & $-0.18 \pm 0.05$ & $8.86 \pm 0.25$ & $9.12 \pm 0.09$ & $8.99 \pm 0.04$ & $9.47 \pm 0.03$ \\
NGC2805 & $0.22 \pm 0.02$ & $8.75 \pm 0.25$ & $9.08 \pm 0.10$ & $9.41 \pm 0.04$ & $9.89 \pm 0.03$ \\
NGC2814 & $-0.68 \pm 0.06$ & $9.03 \pm 0.15$ & $9.36 \pm 0.05$ & $8.82 \pm 0.04$ & $9.25 \pm 0.03$ \\
NGC2820A & $-1.06 \pm 0.02$ & $9.33 \pm 0.41$ & $9.02 \pm 0.10$ & $7.67 \pm 0.06$ & $8.18 \pm 0.04$ \\
NGC2820 & $-0.16 \pm 0.04$ & $9.02 \pm 0.12$ & $9.36 \pm 0.03$ & $9.35 \pm 0.02$ & $9.79 \pm 0.02$ \\
NGC2964 & $0.23 \pm 0.02$ & $9.00 \pm 0.02$ & $9.36 \pm 0.02$ & $9.75 \pm 0.02$ & $10.18 \pm 0.02$ \\
NGC2970 & $-2.41 \pm 0.02$ & $8.70 \pm 0.02$ & $9.60 \pm 0.02$ & $9.09 \pm 0.02$ & $9.48 \pm 0.02$ \\
NGC2976 & $-1.13 \pm 0.04$ & $8.97 \pm 0.08$ & $9.48 \pm 0.05$ & $8.77 \pm 0.03$ & $9.19 \pm 0.02$ \\
NGC3031 & $-0.52 \pm 0.02$ & $8.70 \pm 0.02$ & $9.48 \pm 0.02$ & $10.20 \pm 0.02$ & $10.61 \pm 0.02$ \\
NGC3034 & $-1.06 \pm 0.02$ & $7.70 \pm 0.02$ & $8.60 \pm 0.02$ & $9.30 \pm 0.02$ & $9.87 \pm 0.02$ \\
NGC3077 & $-1.14 \pm 0.12$ & $8.85 \pm 0.17$ & $9.40 \pm 0.11$ & $8.80 \pm 0.07$ & $9.22 \pm 0.06$ \\
NGC3165 & $-1.23 \pm 0.05$ & $8.99 \pm 0.06$ & $9.49 \pm 0.05$ & $8.67 \pm 0.04$ & $9.09 \pm 0.03$ \\
NGC3166 & $-1.08 \pm 0.02$ & $8.70 \pm 0.02$ & $9.60 \pm 0.02$ & $10.42 \pm 0.02$ & $10.81 \pm 0.02$ \\
NGC3169 & $-0.07 \pm 0.22$ & $8.87 \pm 0.15$ & $9.55 \pm 0.07$ & $10.37 \pm 0.02$ & $10.77 \pm 0.02$ \\
NGC3185 & $-0.67 \pm 0.02$ & $9.00 \pm 0.02$ & $9.61 \pm 0.02$ & $9.61 \pm 0.02$ & $10.01 \pm 0.02$ \\
NGC3187 & $-0.42 \pm 0.02$ & $8.78 \pm 0.31$ & $9.09 \pm 0.13$ & $8.77 \pm 0.06$ & $9.26 \pm 0.04$ \\
NGC3190 & $-1.22 \pm 0.02$ & $8.70 \pm 0.02$ & $9.60 \pm 0.02$ & $10.28 \pm 0.02$ & $10.67 \pm 0.02$ \\
NGC3226 & $-1.79 \pm 0.24$ & $8.63 \pm 0.17$ & $9.60 \pm 0.05$ & $9.84 \pm 0.05$ & $10.23 \pm 0.04$ \\
NGC3227 & $-0.17 \pm 0.02$ & $9.00 \pm 0.03$ & $9.61 \pm 0.03$ & $10.10 \pm 0.03$ & $10.49 \pm 0.03$ \\
NGC3395 & $0.31 \pm 0.02$ & $9.55 \pm 0.31$ & $9.07 \pm 0.09$ & $9.07 \pm 0.05$ & $9.57 \pm 0.04$ \\
NGC3396 & $0.24 \pm 0.02$ & $9.43 \pm 0.37$ & $9.05 \pm 0.11$ & $8.98 \pm 0.06$ & $9.49 \pm 0.04$ \\
NGC3424 & $0.13 \pm 0.02$ & $9.00 \pm 0.02$ & $9.36 \pm 0.02$ & $9.65 \pm 0.02$ & $10.09 \pm 0.02$ \\
NGC3430 & $0.21 \pm 0.02$ & $9.00 \pm 0.02$ & $9.36 \pm 0.02$ & $9.74 \pm 0.02$ & $10.17 \pm 0.02$ \\
NGC3448 & $-0.08 \pm 0.03$ & $8.75 \pm 0.24$ & $9.08 \pm 0.09$ & $9.10 \pm 0.04$ & $9.59 \pm 0.03$ \\
IC694 & $-1.96 \pm 0.02$ & $7.70 \pm 0.02$ & $8.60 \pm 0.02$ & $8.40 \pm 0.02$ & $8.97 \pm 0.02$ \\
NGC3690 & $2.02 \pm 0.03$ & $9.12 \pm 0.59$ & $8.41 \pm 0.17$ & $9.95 \pm 0.18$ & $10.59 \pm 0.14$ \\
NGC3786 & $-0.10 \pm 0.02$ & $8.99 \pm 0.03$ & $9.60 \pm 0.03$ & $10.17 \pm 0.03$ & $10.57 \pm 0.03$ \\
NGC3788 & $-0.01 \pm 0.07$ & $8.98 \pm 0.07$ & $9.58 \pm 0.07$ & $10.20 \pm 0.08$ & $10.60 \pm 0.07$ \\
NGC3799 & $-0.09 \pm 0.02$ & $9.00 \pm 0.02$ & $9.36 \pm 0.02$ & $9.43 \pm 0.02$ & $9.87 \pm 0.02$ \\
NGC3800 & $0.64 \pm 0.02$ & $9.00 \pm 0.02$ & $9.36 \pm 0.02$ & $10.16 \pm 0.02$ & $10.59 \pm 0.02$ \\
IC749 & $-0.70 \pm 0.03$ & $8.97 \pm 0.08$ & $9.33 \pm 0.08$ & $8.78 \pm 0.08$ & $9.22 \pm 0.07$ \\
IC750 & $-0.61 \pm 0.05$ & $8.71 \pm 0.08$ & $9.32 \pm 0.05$ & $9.33 \pm 0.04$ & $9.77 \pm 0.04$ \\
NGC4038 & $1.08 \pm 0.03$ & $9.23 \pm 0.24$ & $9.33 \pm 0.16$ & $10.34 \pm 0.10$ & $10.79 \pm 0.07$ \\
NGC4394 & $-1.02 \pm 0.02$ & $8.70 \pm 0.02$ & $9.48 \pm 0.02$ & $9.69 \pm 0.02$ & $10.10 \pm 0.02$ \\
\hline
\end{tabular}

\begin{flushleft}
\textbf{Note:} The full table is available
in the online version of this paper. A portion is shown here for guidance regarding its
form and content. \\
\end{flushleft}
\end{table*}

\begin{table*}
\centering
\caption{\textsc{CIGALE}-derived parameters for the LSM sample. 
Units as Table~\ref{tab:CIGALE1AGN}.}
\label{tab:CIGALELSM}
\begin{tabular}{ccccccc}
\hline
Galaxy &$f_{\rm{AGN}}$ & $L_{\rm{AGN}}$ & Old Att. & Young Att. &Dust $\alpha$ & $L_{\rm{dust}}$ \\
ID&&$\log(L_\odot)$&$\log(L_\odot)$& $\log(L_\odot)$&&$\log(L_\odot)$ \\
\hline
NGC0078 & $\cdots$ & $\cdots$ & $10.03 \pm 0.02$ & $9.27 \pm 0.02$ & 2.00 $\pm$ 0.10 & $10.11 \pm 0.02$ \\
UM246 & 0.25 $\pm$ 0.09 & $10.36 \pm 0.16$ & $10.48 \pm 0.09$ & $10.49 \pm 0.07$ & 2.20 $\pm$ 0.22 & $10.82 \pm 0.07$ \\
2MASXJ01221811+0100262 & 0.42 $\pm$ 0.10 & $11.27 \pm 0.12$ & $11.06 \pm 0.06$ & $11.05 \pm 0.07$ & 1.75 $\pm$ 0.09 & $11.40 \pm 0.06$ \\
CGCG087-046 & 0.51 $\pm$ 0.06 & $11.07 \pm 0.06$ & $10.86 \pm 0.06$ & $10.54 \pm 0.08$ & 2.78 $\pm$ 0.20 & $11.06 \pm 0.04$ \\
UGC04383 & 0.22 $\pm$ 0.04 & $9.89 \pm 0.11$ & $10.09 \pm 0.05$ & $10.11 \pm 0.04$ & 2.00 $\pm$ 0.10 & $10.44 \pm 0.02$ \\
2MASXJ08343370+1720462 & 0.52 $\pm$ 0.06 & $10.55 \pm 0.06$ & $10.34 \pm 0.06$ & $9.95 \pm 0.09$ & 2.29 $\pm$ 0.12 & $10.51 \pm 0.05$ \\
2MASXJ08381760+3054533 & 0.22 $\pm$ 0.10 & $10.03 \pm 0.20$ & $10.51 \pm 0.04$ & $9.64 \pm 0.27$ & 2.12 $\pm$ 0.13 & $10.57 \pm 0.05$ \\
2MASXJ08434495+3549421 & 0.49 $\pm$ 0.10 & $10.53 \pm 0.10$ & $10.34 \pm 0.10$ & $9.98 \pm 0.05$ & 1.87 $\pm$ 0.15 & $10.52 \pm 0.08$ \\
UGC05044 & 0.50 $\pm$ 0.09 & $10.59 \pm 0.08$ & $10.40 \pm 0.09$ & $10.05 \pm 0.06$ & 2.50 $\pm$ 0.32 & $10.59 \pm 0.08$ \\
ARP142 & 0.29 $\pm$ 0.05 & $10.41 \pm 0.08$ & $10.65 \pm 0.04$ & $10.22 \pm 0.02$ & 2.82 $\pm$ 0.19 & $10.80 \pm 0.03$ \\
CGCG266-026 & $\cdots$ & $\cdots$ & $10.81 \pm 0.04$ & $10.44 \pm 0.08$ & 2.37 $\pm$ 0.13 & $10.99 \pm 0.02$ \\
LSBCF567-01 & $\cdots$ & $\cdots$ & $7.63 \pm 0.02$ & $8.13 \pm 0.02$ & 1.75 $\pm$ 0.09 & $8.30 \pm 0.02$ \\
2MASXJ10225654+3446467 & 0.54 $\pm$ 0.12 & $10.51 \pm 0.11$ & $10.36 \pm 0.11$ & $9.53 \pm 0.16$ & 2.48 $\pm$ 0.31 & $10.43 \pm 0.11$ \\
UGC05644 & $\cdots$ & $\cdots$ & $10.57 \pm 0.04$ & $10.21 \pm 0.02$ & 2.23 $\pm$ 0.11 & $10.75 \pm 0.02$ \\
CGCG037-076 & $\cdots$ & $\cdots$ & $10.44 \pm 0.05$ & $10.68 \pm 0.04$ & 1.26 $\pm$ 0.08 & $10.93 \pm 0.03$ \\
UGCA219 & $\cdots$ & $\cdots$ & $8.19 \pm 0.02$ & $8.70 \pm 0.02$ & 1.26 $\pm$ 0.06 & $8.87 \pm 0.02$ \\
NGC3445 & 0.24 $\pm$ 0.05 & $9.12 \pm 0.10$ & $8.94 \pm 0.03$ & $9.46 \pm 0.02$ & 2.24 $\pm$ 0.11 & $9.62 \pm 0.02$ \\
2MASXJ10591815+2432343 & 0.24 $\pm$ 0.04 & $11.27 \pm 0.09$ & $11.47 \pm 0.02$ & $11.40 \pm 0.02$ & 1.75 $\pm$ 0.09 & $11.78 \pm 0.02$ \\
VV627 & $\cdots$ & $\cdots$ & $10.63 \pm 0.04$ & $10.28 \pm 0.02$ & 2.25 $\pm$ 0.11 & $10.82 \pm 0.02$ \\
IC0700 & $\cdots$ & $\cdots$ & $7.98 \pm 0.02$ & $8.41 \pm 0.02$ & 1.75 $\pm$ 0.09 & $8.60 \pm 0.02$ \\
UGC06665 & $\cdots$ & $\cdots$ & $10.45 \pm 0.04$ & $10.60 \pm 0.03$ & 1.50 $\pm$ 0.07 & $10.88 \pm 0.02$ \\
UGC07388 & 0.21 $\pm$ 0.08 & $9.78 \pm 0.18$ & $10.23 \pm 0.05$ & $9.68 \pm 0.08$ & 2.37 $\pm$ 0.17 & $10.35 \pm 0.04$ \\
NGC4320 & 0.46 $\pm$ 0.14 & $10.20 \pm 0.13$ & $10.03 \pm 0.13$ & $9.80 \pm 0.10$ & 2.73 $\pm$ 0.34 & $10.26 \pm 0.11$ \\
UGC07936 & 0.38 $\pm$ 0.08 & $9.94 \pm 0.10$ & $9.74 \pm 0.11$ & $9.87 \pm 0.06$ & 2.75 $\pm$ 0.26 & $10.15 \pm 0.06$ \\
MRK0237 & 0.30 $\pm$ 0.06 & $10.38 \pm 0.10$ & $10.31 \pm 0.04$ & $10.47 \pm 0.04$ & 2.00 $\pm$ 0.10 & $10.74 \pm 0.03$ \\
UGC08327 & 0.28 $\pm$ 0.07 & $10.67 \pm 0.10$ & $10.81 \pm 0.05$ & $10.59 \pm 0.11$ & 1.77 $\pm$ 0.09 & $11.04 \pm 0.04$ \\
UGC08335 & $\cdots$ & $\cdots$ & $11.38 \pm 0.05$ & $11.41 \pm 0.07$ & 1.40 $\pm$ 0.12 & $11.74 \pm 0.06$ \\
NGC5100 & 0.28 $\pm$ 0.07 & $10.63 \pm 0.11$ & $10.89 \pm 0.05$ & $10.40 \pm 0.08$ & 2.30 $\pm$ 0.12 & $11.03 \pm 0.04$ \\
CGCG017-018 & $\cdots$ & $\cdots$ & $10.60 \pm 0.02$ & $10.52 \pm 0.02$ & 2.00 $\pm$ 0.10 & $10.89 \pm 0.02$ \\
NGC5331 & 0.24 $\pm$ 0.13 & $10.97 \pm 0.24$ & $11.25 \pm 0.08$ & $10.97 \pm 0.09$ & 2.13 $\pm$ 0.13 & $11.46 \pm 0.08$ \\
UGC09618 & $\cdots$ & $\cdots$ & $11.42 \pm 0.09$ & $10.95 \pm 0.26$ & 2.25 $\pm$ 0.11 & $11.57 \pm 0.02$ \\
2MASXJ15015015+2332536 & 0.33 $\pm$ 0.08 & $10.41 \pm 0.12$ & $10.54 \pm 0.05$ & $10.16 \pm 0.11$ & 2.30 $\pm$ 0.15 & $10.72 \pm 0.05$ \\
SBS1509+583 & $\cdots$ & $\cdots$ & $10.22 \pm 0.04$ & $9.77 \pm 0.03$ & 2.25 $\pm$ 0.11 & $10.37 \pm 0.02$ \\
KUG1553+200 & $\cdots$ & $\cdots$ & $10.76 \pm 0.03$ & $10.70 \pm 0.03$ & 1.75 $\pm$ 0.09 & $11.07 \pm 0.02$ \\
KUG1556+326 & 0.26 $\pm$ 0.11 & $10.23 \pm 0.21$ & $10.49 \pm 0.06$ & $10.14 \pm 0.09$ & 2.28 $\pm$ 0.13 & $10.67 \pm 0.05$ \\
MRK0881 & 0.32 $\pm$ 0.06 & $10.35 \pm 0.09$ & $10.23 \pm 0.05$ & $10.40 \pm 0.03$ & 2.00 $\pm$ 0.10 & $10.67 \pm 0.03$ \\
2MASXJ17045097+3449020 & $\cdots$ & $\cdots$ & $11.19 \pm 0.02$ & $11.07 \pm 0.02$ & 1.75 $\pm$ 0.09 & $11.47 \pm 0.02$ \\
\hline
\end{tabular}

\end{table*}

\begin{table*}
\centering
\contcaption{\textsc{CIGALE}-derived parameters for the LSM sample. Units as Table~\ref{tab:CIGALE2AGN}.}
\label{tab:CIGALELSM2}
\begin{tabular}{cccccccc}
\hline
Galaxy &  SFR & $\tau_{\rm{main}}$ & Stellar Age & $M_{\rm{gas}}$ & $M_{\star}$ \\
ID& $\log(M_\odot /yr)$ & $\log(yr)$ & $\log(yr)$ & $\log(M_\odot)$& $\log(M_\odot)$\\
\hline
NGC0078 & $-0.21 \pm 0.02$ & $8.70 \pm 0.02$ & $9.48 \pm 0.02$ & $10.50 \pm 0.02$ & $10.91 \pm 0.02$ \\
UM246 & $1.03 \pm 0.06$ & $9.00 \pm 0.20$ & $9.32 \pm 0.11$ & $10.47 \pm 0.09$ & $10.91 \pm 0.08$ \\
2MASXJ01221811+0100262 & $1.58 \pm 0.06$ & $9.69 \pm 0.21$ & $8.95 \pm 0.16$ & $10.12 \pm 0.13$ & $10.65 \pm 0.10$ \\
CGCG087-046 & $1.07 \pm 0.08$ & $8.84 \pm 0.15$ & $9.39 \pm 0.10$ & $11.00 \pm 0.05$ & $11.42 \pm 0.04$ \\
UGC04383 & $0.64 \pm 0.04$ & $8.70 \pm 0.08$ & $9.06 \pm 0.04$ & $9.82 \pm 0.02$ & $10.31 \pm 0.02$ \\
2MASXJ08343370+1720462 & $0.47 \pm 0.08$ & $8.91 \pm 0.13$ & $9.45 \pm 0.09$ & $10.40 \pm 0.07$ & $10.82 \pm 0.05$ \\
2MASXJ08381760+3054533 & $0.17 \pm 0.25$ & $8.77 \pm 0.14$ & $9.50 \pm 0.05$ & $10.77 \pm 0.02$ & $11.17 \pm 0.02$ \\
2MASXJ08434495+3549421 & $0.53 \pm 0.05$ & $8.92 \pm 0.12$ & $9.53 \pm 0.12$ & $10.71 \pm 0.11$ & $11.11 \pm 0.09$ \\
UGC05044 & $0.56 \pm 0.05$ & $8.84 \pm 0.15$ & $9.45 \pm 0.15$ & $10.64 \pm 0.12$ & $11.06 \pm 0.09$ \\
ARP142 & $0.75 \pm 0.02$ & $9.00 \pm 0.02$ & $9.61 \pm 0.02$ & $11.03 \pm 0.02$ & $11.42 \pm 0.02$ \\
CGCG266-026 & $0.96 \pm 0.07$ & $8.99 \pm 0.05$ & $9.47 \pm 0.05$ & $10.81 \pm 0.07$ & $11.23 \pm 0.06$ \\
LSBCF567-01 & $-1.14 \pm 0.02$ & $9.64 \pm 0.19$ & $9.29 \pm 0.06$ & $7.86 \pm 0.05$ & $8.32 \pm 0.04$ \\
2MASXJ10225654+3446467 & $0.11 \pm 0.11$ & $8.70 \pm 0.05$ & $9.48 \pm 0.02$ & $10.82 \pm 0.02$ & $11.23 \pm 0.02$ \\
UGC05644 & $0.75 \pm 0.02$ & $9.00 \pm 0.02$ & $9.61 \pm 0.02$ & $11.03 \pm 0.02$ & $11.42 \pm 0.02$ \\
CGCG037-076 & $1.22 \pm 0.03$ & $9.53 \pm 0.32$ & $8.57 \pm 0.10$ & $9.32 \pm 0.09$ & $9.93 \pm 0.07$ \\
UGCA219 & $-0.57 \pm 0.02$ & $9.10 \pm 0.37$ & $8.97 \pm 0.07$ & $8.15 \pm 0.04$ & $8.66 \pm 0.03$ \\
NGC3445 & $0.19 \pm 0.02$ & $9.57 \pm 0.30$ & $9.14 \pm 0.14$ & $9.02 \pm 0.10$ & $9.51 \pm 0.07$ \\
2MASXJ10591815+2432343 & $1.91 \pm 0.02$ & $9.28 \pm 0.55$ & $8.72 \pm 0.10$ & $10.28 \pm 0.04$ & $10.85 \pm 0.03$ \\
VV627 & $0.81 \pm 0.02$ & $9.00 \pm 0.02$ & $9.61 \pm 0.02$ & $11.09 \pm 0.02$ & $11.49 \pm 0.02$ \\
IC0700 & $-0.86 \pm 0.02$ & $9.03 \pm 0.15$ & $9.20 \pm 0.05$ & $8.27 \pm 0.02$ & $8.74 \pm 0.02$ \\
UGC06665 & $1.13 \pm 0.04$ & $9.36 \pm 0.44$ & $8.61 \pm 0.09$ & $9.35 \pm 0.12$ & $9.94 \pm 0.10$ \\
UGC07388 & $0.19 \pm 0.08$ & $8.81 \pm 0.15$ & $9.37 \pm 0.09$ & $10.12 \pm 0.05$ & $10.54 \pm 0.04$ \\
NGC4320 & $0.36 \pm 0.08$ & $8.97 \pm 0.08$ & $9.57 \pm 0.09$ & $10.57 \pm 0.07$ & $10.97 \pm 0.06$ \\
UGC07936 & $0.60 \pm 0.07$ & $9.21 \pm 0.23$ & $9.37 \pm 0.11$ & $9.97 \pm 0.08$ & $10.41 \pm 0.07$ \\
MRK0237 & $1.00 \pm 0.03$ & $9.64 \pm 0.20$ & $9.30 \pm 0.07$ & $10.03 \pm 0.05$ & $10.49 \pm 0.04$ \\
UGC08327 & $1.12 \pm 0.10$ & $8.98 \pm 0.09$ & $9.36 \pm 0.06$ & $10.69 \pm 0.06$ & $11.12 \pm 0.05$ \\
UGC08335 & $1.92 \pm 0.07$ & $9.61 \pm 0.29$ & $8.66 \pm 0.14$ & $10.11 \pm 0.09$ & $10.70 \pm 0.07$ \\
NGC5100 & $0.94 \pm 0.08$ & $8.89 \pm 0.14$ & $9.42 \pm 0.09$ & $10.85 \pm 0.05$ & $11.28 \pm 0.04$ \\
CGCG017-018 & $1.04 \pm 0.02$ & $9.00 \pm 0.02$ & $9.36 \pm 0.02$ & $10.56 \pm 0.02$ & $10.99 \pm 0.02$ \\
NGC5331 & $1.48 \pm 0.09$ & $8.86 \pm 0.18$ & $9.21 \pm 0.16$ & $10.81 \pm 0.11$ & $11.27 \pm 0.08$ \\
UGC09618 & $1.45 \pm 0.26$ & $8.65 \pm 0.43$ & $9.00 \pm 0.21$ & $10.75 \pm 0.07$ & $11.26 \pm 0.04$ \\
2MASXJ15015015+2332536 & $0.70 \pm 0.11$ & $8.98 \pm 0.07$ & $9.44 \pm 0.07$ & $10.48 \pm 0.09$ & $10.90 \pm 0.08$ \\
SBS1509+583 & $0.29 \pm 0.02$ & $9.00 \pm 0.02$ & $9.61 \pm 0.02$ & $10.57 \pm 0.02$ & $10.97 \pm 0.02$ \\
KUG1553+200 & $1.23 \pm 0.02$ & $9.70 \pm 0.21$ & $9.16 \pm 0.11$ & $10.06 \pm 0.09$ & $10.54 \pm 0.07$ \\
KUG1556+326 & $0.66 \pm 0.08$ & $8.79 \pm 0.15$ & $9.36 \pm 0.09$ & $10.60 \pm 0.06$ & $11.03 \pm 0.04$ \\
MRK0881 & $0.94 \pm 0.03$ & $9.63 \pm 0.26$ & $9.09 \pm 0.10$ & $9.69 \pm 0.07$ & $10.19 \pm 0.05$ \\
2MASXJ17045097+3449020 & $1.59 \pm 0.02$ & $9.65 \pm 0.23$ & $9.23 \pm 0.10$ & $10.51 \pm 0.08$ & $10.99 \pm 0.06$ \\
\hline
\end{tabular}

\end{table*}

\begin{table*}
\centering
\caption{\textsc{CIGALE}-derived parameters for six AGN galaxies, where a Type 1 AGN ($\psi=70$) give better $\chi^2$. Units as Table~\ref{tab:CIGALE1AGN}.}
\label{tab:CIGALE1Psi}
\begin{tabular}{ccccccc}
\hline
Galaxy &$f_{\rm{AGN}}$ & $L_{\rm{AGN}}$ & Old Att. & Young Att. &Dust $\alpha$ & $L_{\rm{dust}}$ \\
ID&&$\log(L_\odot)$&$\log(L_\odot)$& $\log(L_\odot)$&&$\log(L_\odot)$ \\
\hline
Mrk335 & 0.86 $\pm$ 0.04 & $11.14 \pm 0.02$ & $9.88 \pm 0.07$ & $8.53 \pm 0.90$ & 1.06 $\pm$ 0.16 & $9.90 \pm 0.06$ \\
Mrk771 & 0.81 $\pm$ 0.04 & $11.43 \pm 0.02$ & $10.18 \pm 0.13$ & $9.84 \pm 0.23$ & 2.24 $\pm$ 0.29 & $10.37 \pm 0.09$ \\
2XMMJ141348.3+440014 & 0.90 $\pm$ 0.04 & $11.96 \pm 0.02$ & $10.30 \pm 0.18$ & $9.95 \pm 0.35$ & 2.33 $\pm$ 0.25 & $10.49 \pm 0.04$ \\
Mrk1383 & 0.80 $\pm$ 0.04 & $12.00 \pm 0.04$ & $10.52 \pm 0.25$ & $10.59 \pm 0.20$ & 2.05 $\pm$ 0.16 & $10.89 \pm 0.05$ \\
ESO141-55 & 0.74 $\pm$ 0.04 & $11.56 \pm 0.02$ & $10.52 \pm 0.10$ & $9.55 \pm 0.64$ & 2.56 $\pm$ 0.21 & $10.57 \pm 0.06$ \\
Mrk1513 & 0.86 $\pm$ 0.04 & $11.87 \pm 0.02$ & $10.61 \pm 0.11$ & $9.65 \pm 0.63$ & 2.13 $\pm$ 0.23 & $10.66 \pm 0.07$ \\
\hline
\end{tabular}
\end{table*}

\begin{table*}
\centering
\contcaption{\textsc{CIGALE}-derived parameters for six AGN galaxies, where a Type 1 AGN ($\psi=70$) give better $\chi^2$. Units as Table~\ref{tab:CIGALE1AGN}.}
\label{tab:CIGALE2Psi}
\begin{tabular}{cccccc}
\hline
Galaxy &  SFR & $\tau_{\rm{main}}$ & Stellar Age & $M_{\rm{gas}}$ & $M_{\star}$ \\
ID& $\log(M_\odot /yr)$ & $\log(yr)$ & $\log(yr)$ & $\log(M_\odot)$& $\log(M_\odot)$\\
\hline
Mrk335 & $-0.94 \pm 0.91$ & $8.51 \pm 0.53$ & $9.58 \pm 0.14$ & $10.08 \pm 0.17$ & $10.47 \pm 0.16$ \\
Mrk771 & $0.51 \pm 0.24$ & $8.94 \pm 0.22$ & $9.47 \pm 0.12$ & $10.47 \pm 0.16$ & $10.88 \pm 0.14$ \\
2XMMJ141348.3+440014 & $0.56 \pm 0.40$ & $9.31 \pm 0.52$ & $9.34 \pm 0.30$ & $10.27 \pm 0.45$ & $10.68 \pm 0.41$ \\
Mrk1383 & $1.31 \pm 0.21$ & $9.00 \pm 0.39$ & $9.24 \pm 0.22$ & $10.68 \pm 0.19$ & $11.14 \pm 0.15$ \\
ESO141-55 & $0.12 \pm 0.68$ & $8.60 \pm 0.47$ & $9.49 \pm 0.21$ & $10.66 \pm 0.22$ & $11.06 \pm 0.19$ \\
Mrk1513 & $0.18 \pm 0.65$ & $8.72 \pm 0.67$ & $9.51 \pm 0.19$ & $10.68 \pm 0.24$ & $11.08 \pm 0.22$ \\
\hline
\end{tabular}

\end{table*}

%% HISTOGRAMS

\begin{figure*}
\includegraphics[width=0.49\textwidth]{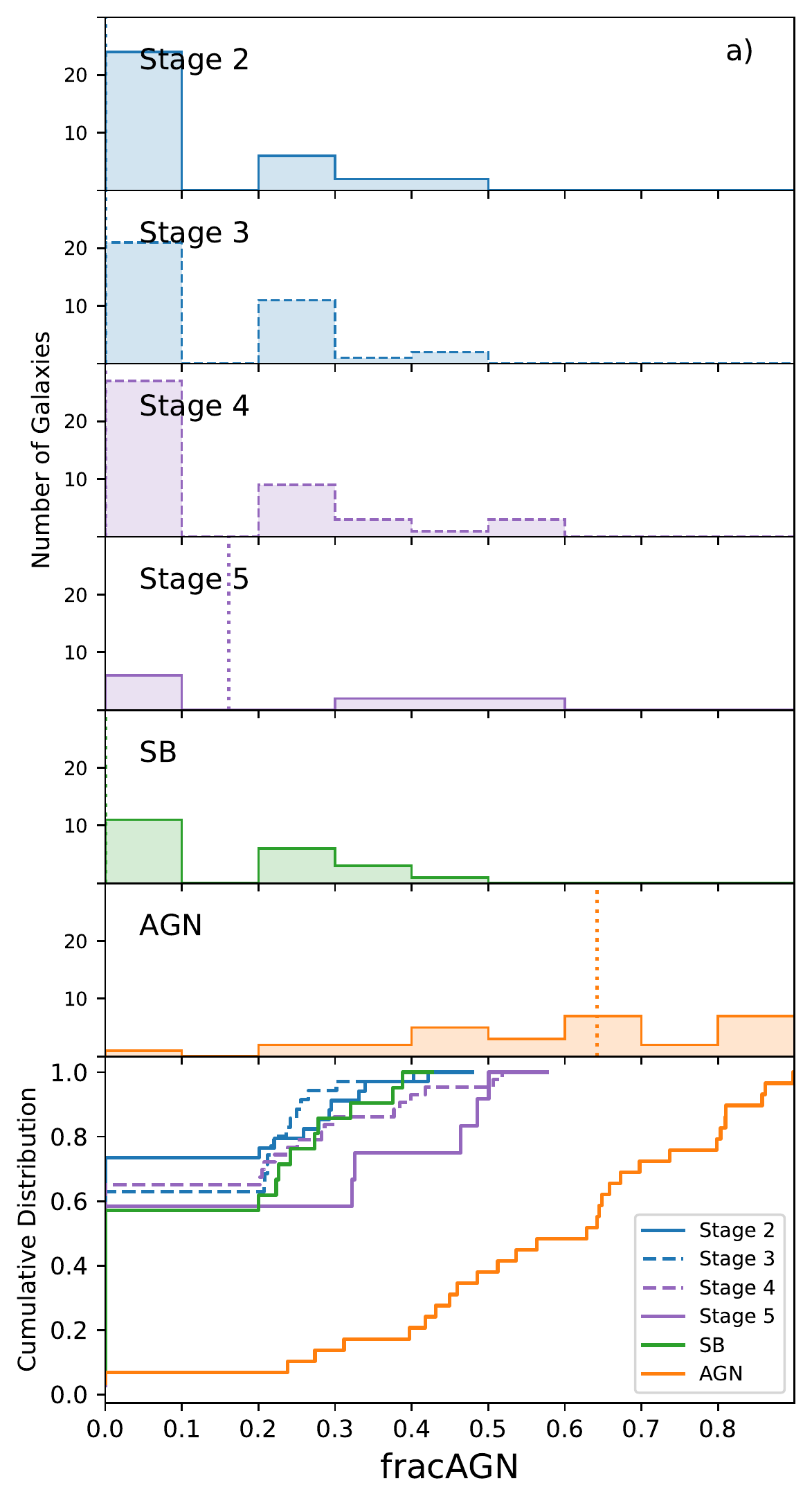}
\includegraphics[width=0.49\textwidth]{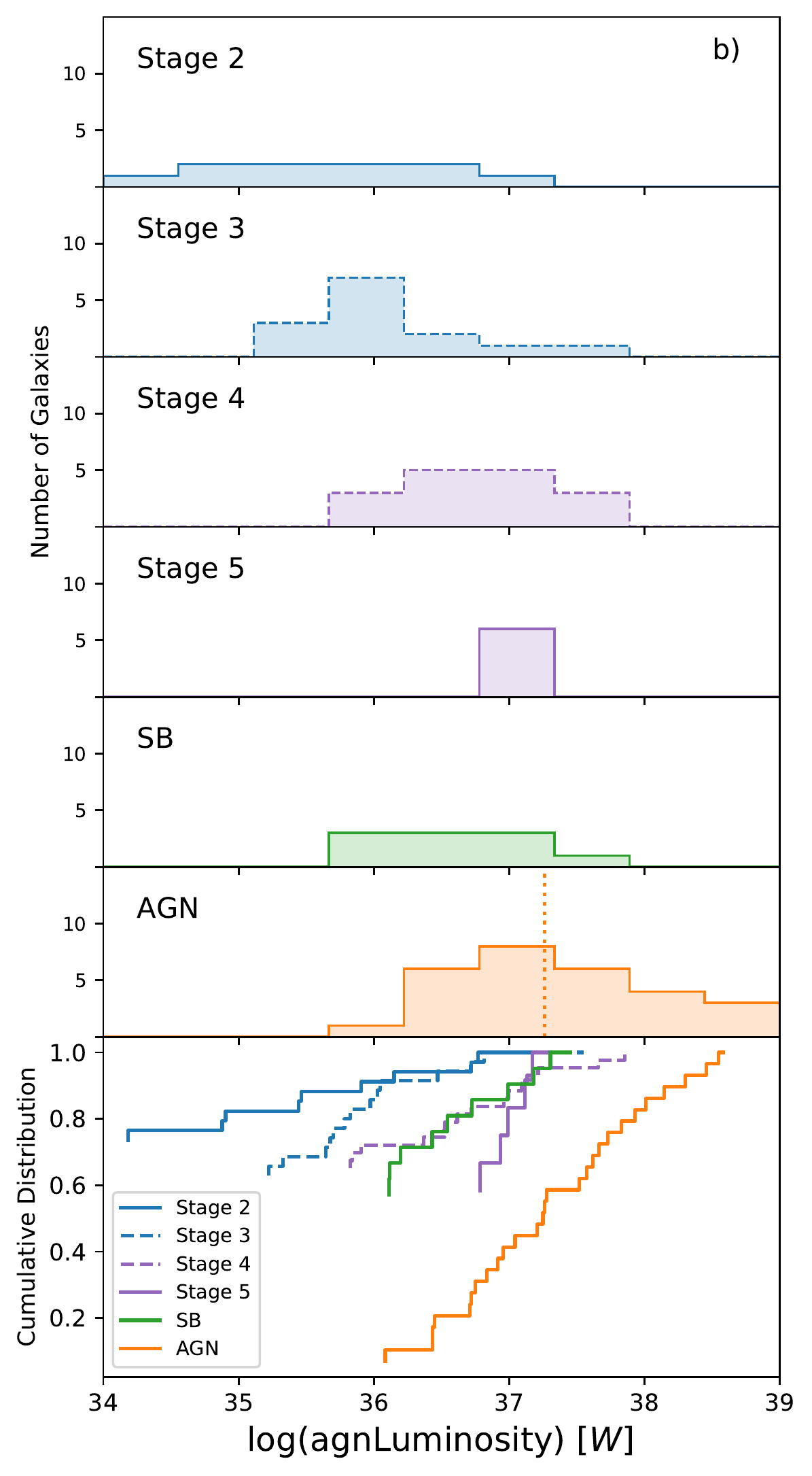}
\caption{Histograms (top) for the stages of the SIGS+LSM, AGN and SB samples and the normalized cumulative distributions (bottom) for the AGN fraction (a) and luminosity (b). The colors and lines are identical to Figure~\ref{fig:Hist}.}
\label{fig:Hist2}
\end{figure*}

\begin{figure*}
\includegraphics[width=0.49\textwidth]{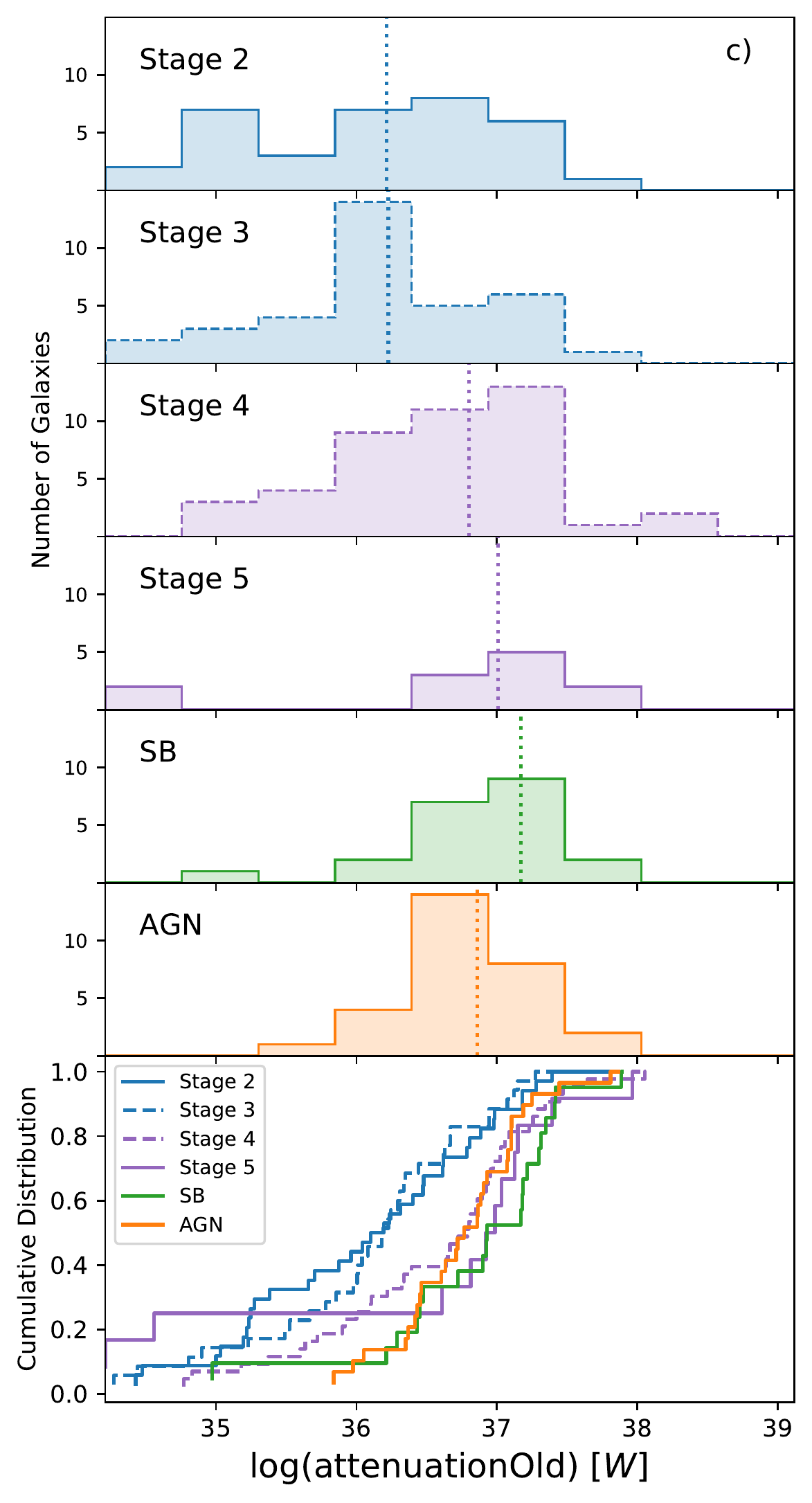}
\includegraphics[width=0.49\textwidth]{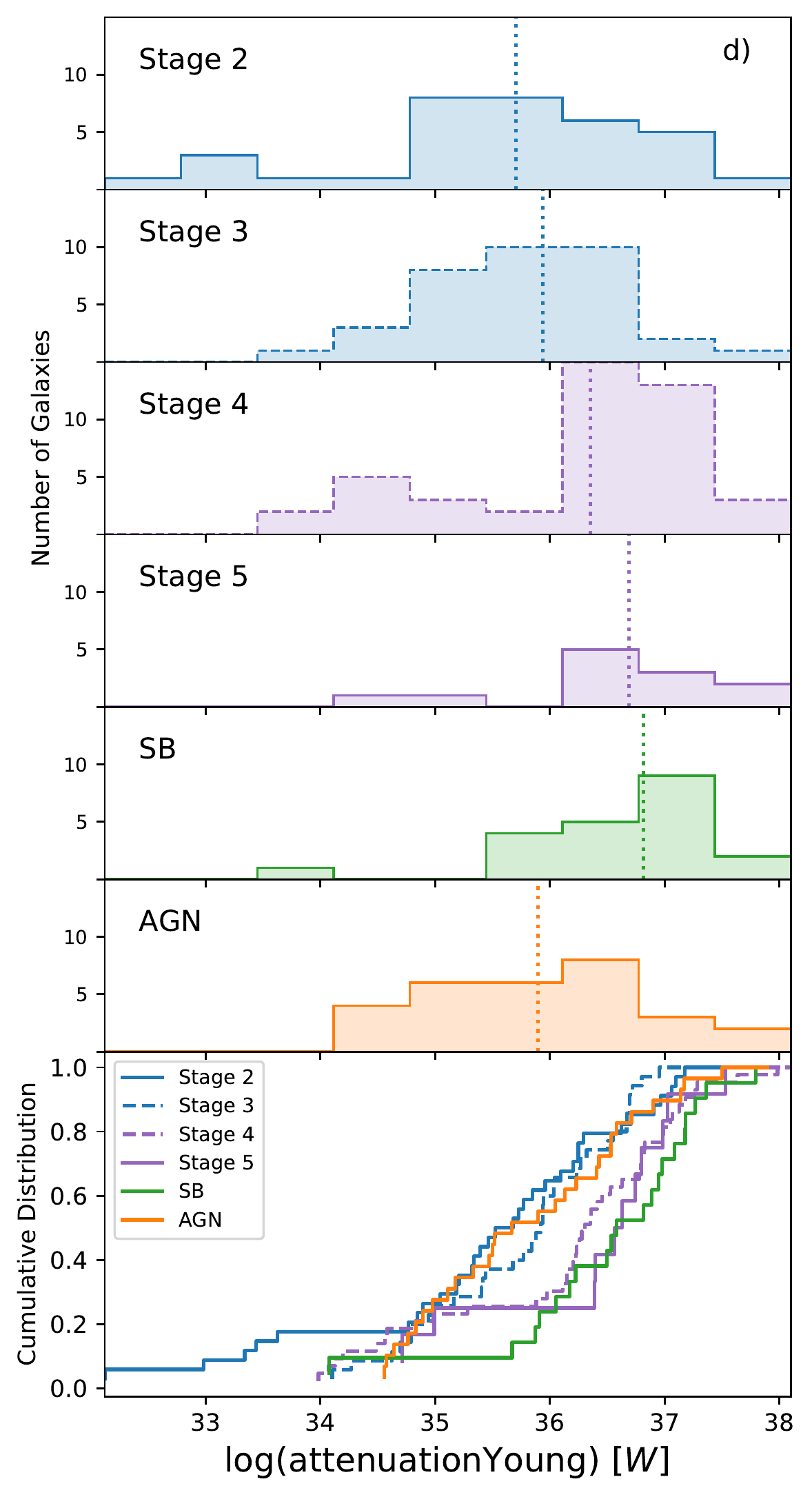}
\contcaption{Histograms (top) for the stages of the SIGS+LSM, AGN and SB samples and the normalized cumulative distributions (bottom) for the attenuation of old (c) and young stars (d). The colors and lines are identical to Figure~\ref{fig:Hist}.}
\end{figure*}

\begin{figure*}
\includegraphics[width=0.49\textwidth]{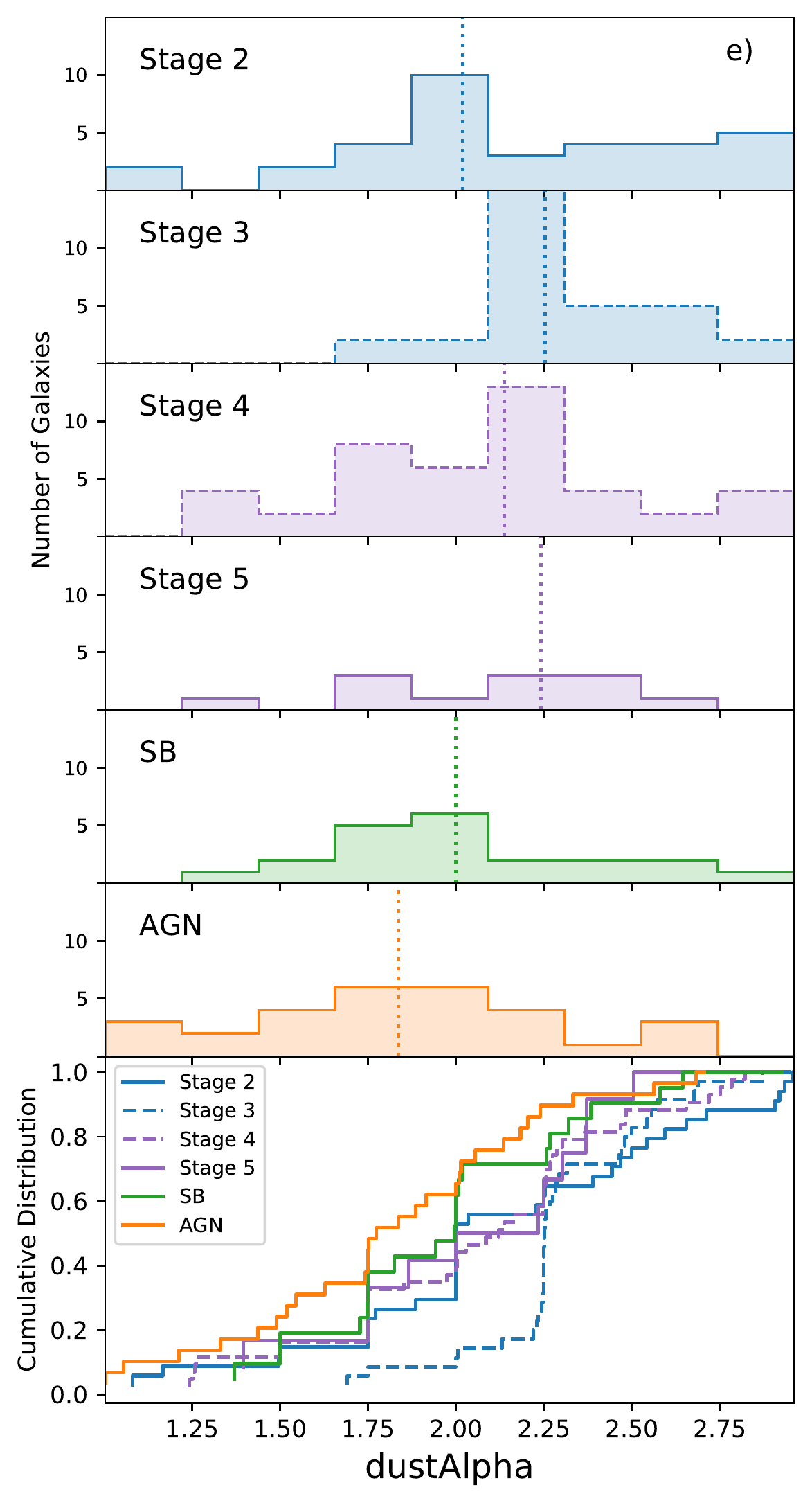}
\includegraphics[width=0.49\textwidth]{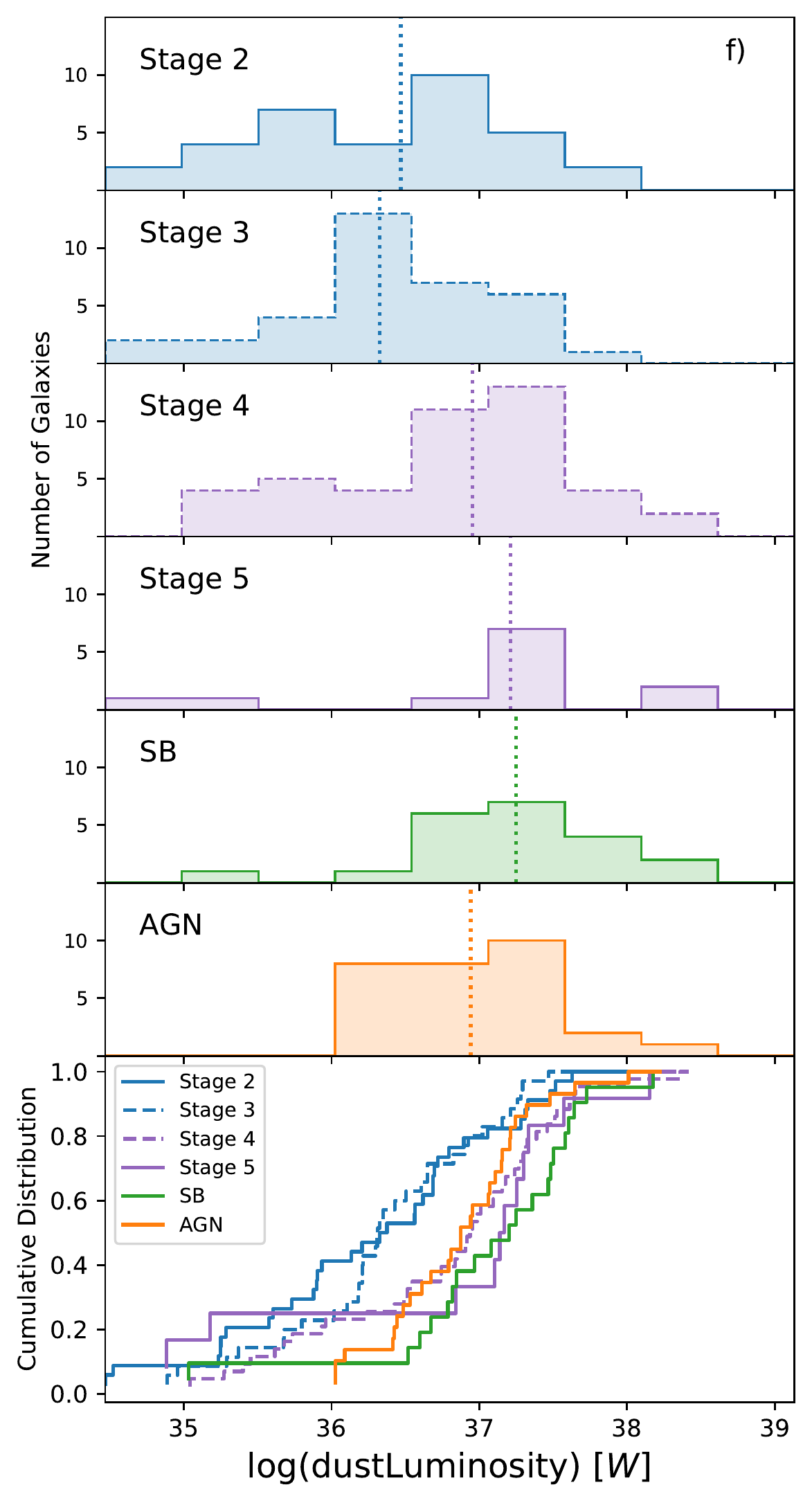}
\contcaption{Histograms (top) for the stages of the SIGS+LSM, AGN and SB samples and the normalized cumulative distributions (bottom) for the dust parameter $\alpha$ (e) and luminosity (f). The colors and lines are identical to Figure~\ref{fig:Hist}.}
\end{figure*}

\begin{figure*}
\includegraphics[width=0.49\textwidth]{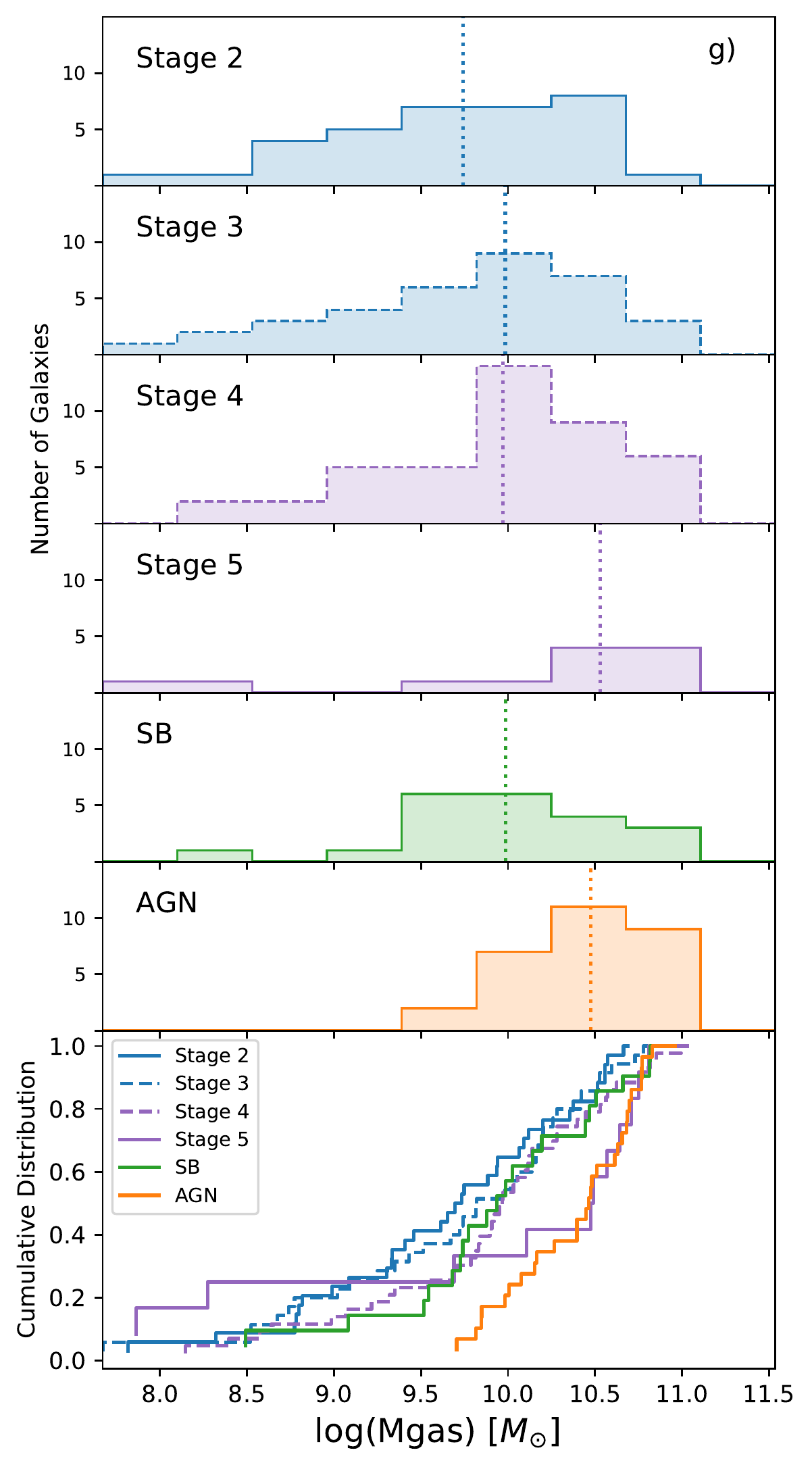}
\includegraphics[width=0.49\textwidth]{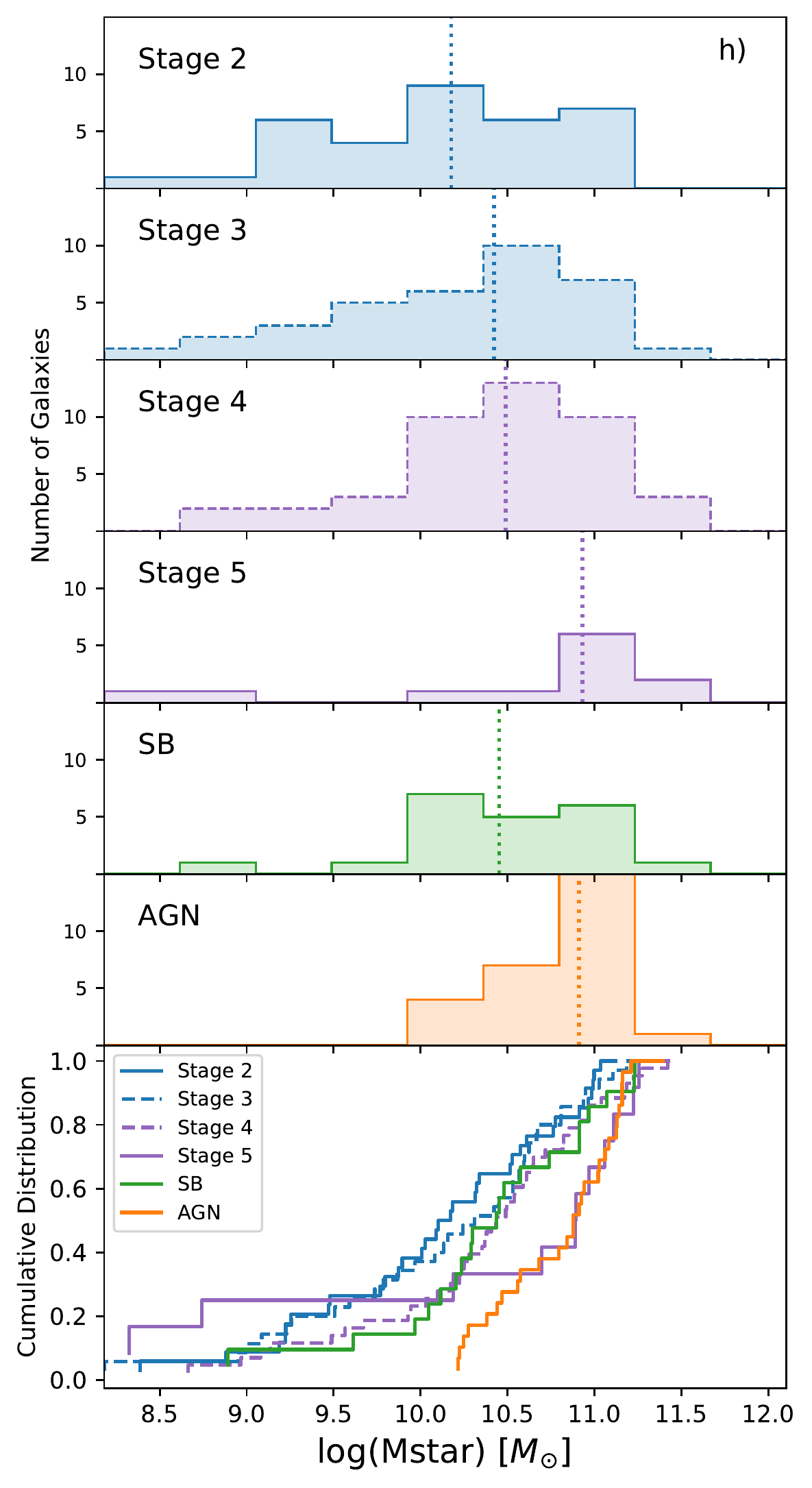}
\contcaption{Histograms (top) for the stages of the SIGS+LSM, AGN and SB samples and the normalized cumulative distributions (bottom) for the gas (g) and stellar mass (h). The colors and lines are identical to Figure~\ref{fig:Hist}.}
\end{figure*}

\begin{figure*}
\includegraphics[width=0.49\textwidth]{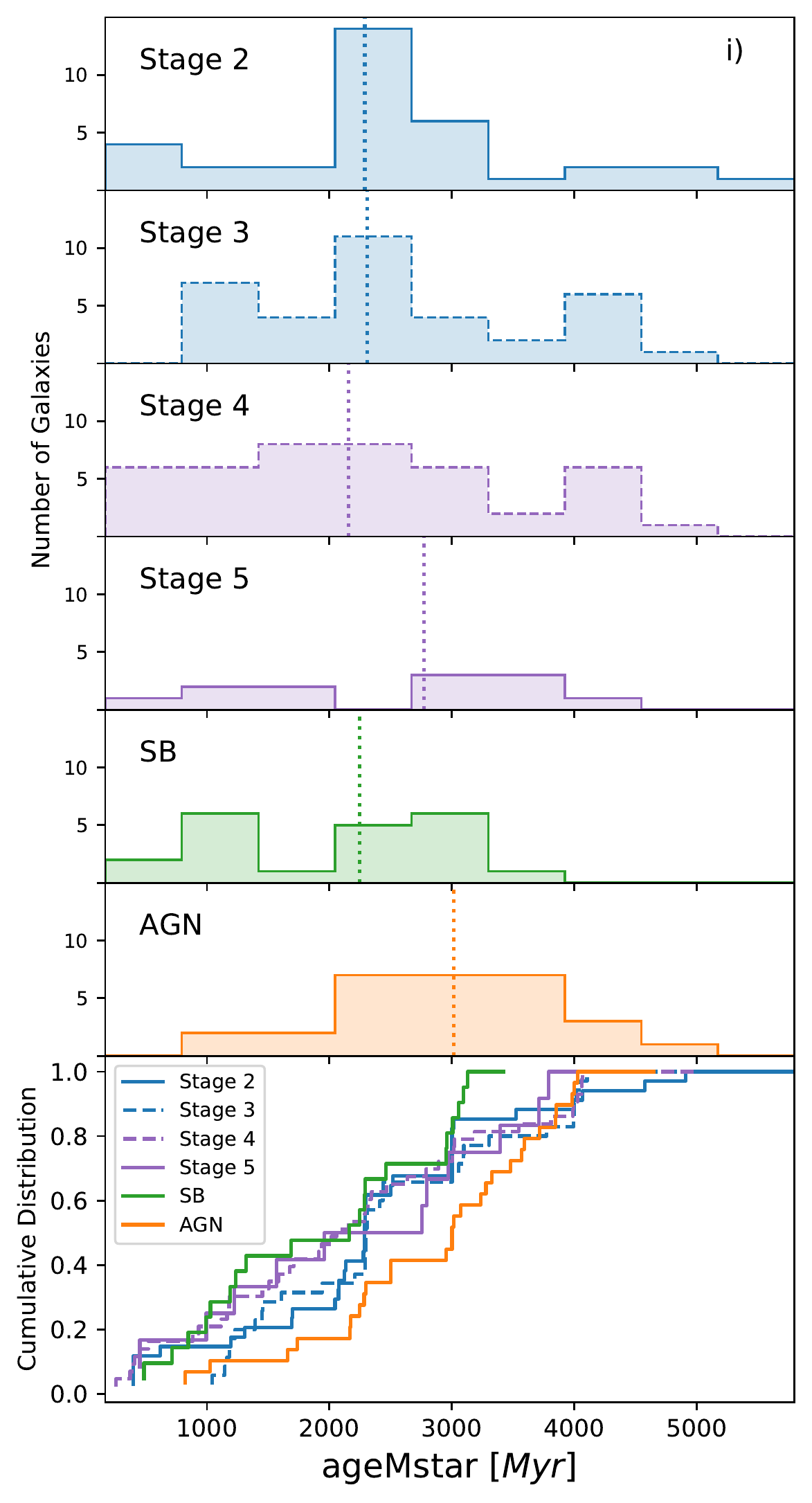}
\contcaption{Histogram for the stages of the SIGS+LSM, AGN and SB samples and the normalized cumulative distributions (bottom) for weighted age of the stars (i). The colors and lines are identical to Figure~\ref{fig:Hist}.}
\end{figure*}

%%%%%%%%%%%%%%%%%%%%%%%%%%%%%%%%%%%%%%%%%%%%%%%%%%

% Don't change these lines
\bsp	% typesetting comment
\label{lastpage}
\end{document}